\documentclass[twocolumn]{aastex61}

\usepackage{soul} 
\usepackage{amsmath}
\usepackage{amssymb}
\usepackage{xspace}
\usepackage{xifthen}

\mathchardef\mhyphen="2D

\newlength{\dhatheight}

\newcommand{\unit}[1]{\ensuremath{\mathrm{\,#1}}\xspace}

\providecommand{\deg}{}
\renewcommand{\deg}{\unit{deg}}

\usepackage[outdir=./]{epstopdf}
\usepackage{graphicx,verbatim}
\usepackage{xspace}
\usepackage[figuresright]{rotating}
\usepackage{tabularx}
\graphicspath{{./}{./figures/}}
\newcommand{\nv}{\hat{\bf n}}

\newcommand{\ccl}{{\tt CCL}\xspace}
\newcommand{\halofit}{{\tt halofit}\xspace}
\newcommand{\class}{{\tt CLASS}\xspace}
\newcommand{\camb}{{\tt CAMB}\xspace}
\newcommand{\hiclass}{{\tt hi$\_$CLASS\xspace}}

\begin{document}

\title{Core Cosmology Library: Precision Cosmological Predictions for LSST}


\author{Nora Elisa Chisari}
\affiliation{Department of Physics, University of Oxford, Denys Wilkinson Building, Keble Road, Oxford OX1 3RH, United Kingdom}
\author{David Alonso}
\affiliation{School of Physics and Astronomy, Cardiff University, The Parade, Cardiff, CF24 3AA, United Kingdom}
\affiliation{Department of Physics, University of Oxford, Denys Wilkinson Building, Keble Road, Oxford OX1 3RH, United Kingdom}
\author{Elisabeth Krause}
\affiliation{Steward Observatory/Department of Astronomy, University of Arizona, 933 North Cherry Avenue, Tucson, AZ 85721, USA}
\affiliation{Kavli Institute for Particle Astrophysics and Cosmology, Stanford, CA 94305-4085, USA}
\author{C. Danielle Leonard}
\affiliation{McWilliams Center for Cosmology, Department of Physics, Carnegie Mellon University, Pittsburgh, PA 15213, USA}
\author{Philip Bull}
\affiliation{School of Physics \& Astronomy, Queen Mary University of London, 327 Mile End Road, London E1 4NS, UK}
\affiliation{Department of Astronomy, University of California Berkeley, Berkeley, CA 94720, USA}
\author{J\'er\'emy Neveu}
\affiliation{Laboratoire de l'Acc\'el\'erateur Lin\'eaire, Universit\'e Paris-Sud, CNRS/IN2P3, Universit\'e Paris-Saclay, Orsay, France}
\author{Antonio Villarreal}
\affiliation{Argonne National Laboratory, Argonne, IL 60439, USA}
\author{Sukhdeep Singh}
\affiliation{Berkeley Center for Cosmological Physics and Department of Physics, University of California, Berkeley, California}
\affiliation{McWilliams Center for Cosmology, Department of Physics, Carnegie Mellon University, Pittsburgh, PA 15213, USA}
\author{Thomas McClintock}
\affiliation{Department of Physics, University of Arizona, Tucson, AZ 85721, USA}
\author{John Ellison}
\affiliation{Department of Physics and Astronomy, University of California, Riverside, CA 92521, USA}
\author{Zilong Du}
\affiliation{Department of Physics and Astronomy, University of California, Riverside, CA 92521, USA}
\author{Joe Zuntz}
\affiliation{Institute for Astronomy, Royal Observatory Edinburgh, Edinburgh EH9 3HJ, UK}
\author{Alexander Mead}
\affiliation{Department of Physics and Astronomy, University of British Columbia, 6224 Agricultural Road, Vancouver, BC V6T 1Z1, Canada}
\author{Shahab Joudaki}
\affiliation{Department of Physics, University of Oxford, Denys Wilkinson Building, Keble Road, Oxford OX1 3RH, United Kingdom}
\author{Christiane S. Lorenz}
\affiliation{Department of Physics, University of Oxford, Denys Wilkinson Building, Keble Road, Oxford OX1 3RH, United Kingdom}
\author{Tilman Tr\"oster}
\affiliation{Institute for Astronomy, Royal Observatory Edinburgh, Edinburgh EH9 3HJ, UK}
\author{Javier Sanchez}
\affiliation{Department of Physics and Astronomy, University of California, Irvine, CA 92697, USA}
\author{Francois Lanusse}
\affiliation{McWilliams Center for Cosmology, Department of Physics, Carnegie Mellon University, Pittsburgh, PA 15213, USA}
\author{Mustapha Ishak}
\affiliation{Department of Physics, The University of Texas at Dallas, Richardson, TX 75083, USA}
\author{Ren\'ee Hlozek}
\affiliation{Dunlap Institute for Astronomy and Astrophysics \& Department for Astronomy and Astrophysics, University of Toronto, ON M5S 3H4}
\author{Jonathan Blazek}
\affiliation{Center for Cosmology and Astroparticle Physics, Ohio State, Columbus, OH 43210, USA}
\affiliation{SNSF Ambizione, Laboratory of Astrophysics, \'Ecole Polytechnique F\'ed\'erale de Lausanne (EPFL), 1290 Versoix, Switzerland}
\author{Jean-Eric Campagne}
\affiliation{Laboratoire de l'Acc\'el\'erateur Lin\'eaire, Universit\'e Paris-Sud, CNRS/IN2P3, Universit\'e Paris-Saclay, Orsay, France}
\author{Husni Almoubayyed}
\affiliation{McWilliams Center for Cosmology, Department of Physics, Carnegie Mellon University, Pittsburgh, PA 15213, USA}
\author{Tim Eifler}
\affiliation{Steward Observatory/Department of Astronomy, University of Arizona, 933 North Cherry Avenue, Tucson, AZ 85721, USA}
\affiliation{Jet Propulsion Laboratory, California Institute of Technology, Pasadena, CA 91109, USA}
\author{Matthew Kirby}
\affiliation{Department of Physics, University of Arizona, Tucson, AZ 85721, USA}
\author{David Kirkby}
\affiliation{Department of Physics and Astronomy, University of California, Irvine, CA 92697, USA}
\author{St\'ephane Plaszczynski}
\affiliation{Laboratoire de l'Acc\'el\'erateur Lin\'eaire, Universit\'e Paris-Sud, CNRS/IN2P3, Universit\'e Paris-Saclay, Orsay, France}
\author{An\v{z}e Slosar}
\affiliation{Brookhaven National Laboratory, Physics Department, Upton, NY 11973, USA}
\author{Michal Vrastil}
\affiliation{Institute of Physics CAS, Prague, 182 21, CZ}
\author{Erika L. Wagoner}
\affiliation{Department of Physics, University of Arizona, Tucson, AZ 85721, USA}

\collaboration{(LSST Dark Energy Science Collaboration)}

\begin{abstract}

The Core Cosmology Library (\ccl) provides routines to compute basic cosmological observables to a high degree of accuracy, which have been verified with an extensive suite of validation tests. Predictions are provided for many cosmological quantities, including distances, angular power spectra, correlation functions, halo bias and the halo mass function through state-of-the-art modeling prescriptions available in the literature. Fiducial specifications for the expected galaxy distributions for the Large Synoptic Survey Telescope (LSST) are also included, together with the capability of computing redshift distributions for a user-defined photometric redshift model. A rigorous validation procedure, based on comparisons between \ccl and independent software packages, allows us to establish a well-defined numerical accuracy for each predicted quantity. As a result, predictions for correlation functions of galaxy clustering, galaxy-galaxy lensing and cosmic shear are demonstrated to be within a fraction of the expected statistical uncertainty of the observables for the models and in the range of scales of interest to LSST. \ccl is an open source software package written in C, with a {\tt python} interface and publicly available at \url{https://github.com/LSSTDESC/CCL}. 

\end{abstract}


\section{Introduction}
\label{sec:intro}

Starting in the next decade, large-scale galaxy surveys will drive a new era of high precision cosmology \citep{DESCWhite,green11,Laureijs11}. One of their main goals is to answer the question of the origin of cosmic acceleration, in other words, to elucidate the nature of ``dark energy'', broadly understood as a family of potential models: from a cosmological constant to a dynamical field and modifications of gravity (see for example  \citealt{Carroll2001CC,Peebles2003,Padmanabhan2003,Copeland2006,Ishak2007,Weinberg13} and references therein). These data will also allow us to shed light on a number of open questions in fundamental physics, such as the nature of dark matter \citep{Feng10,Porter11}, the mass of neutrinos \citep{Wong11,Lesgourgues12,Allison15} or the level of primordial non-Gaussianity \citep{Dalal08,Desjacques10}.

High precision constraints on dark energy models will be achieved by probing at the same time the expansion and growth history of the Universe over a long redshift baseline. For this purpose, it will be crucial to combine measurements of multiple cosmological probes: weak and strong gravitational lensing, the clustering of galaxies, distances to supernovae, and the abundance, clustering and gravitational lensing of galaxy clusters. Current weak lensing surveys, such as the Dark Energy Survey\footnote{\url{https://www.darkenergysurvey.org}} and the Kilo-Degree Survey\footnote{\url{http://kids.strw.leidenuniv.nl}}, have started to take this approach already \citep{Joudaki18,vanUitert18,DEScombined,krause17}. From a theoretical perspective, there are two challenges faced by the next generation of galaxy surveys. 

First, we need to ensure that all probes are modeled accurately from a {\it physical} point of view, including cosmological, astrophysical, and observational effects, to avoid potential biases in the final cosmological results. In the context of weak gravitational lensing, for example, phenomena that can lead to biases include the impact of baryons on the distribution of matter and the intrinsic alignments of galaxies \citep[e.g.][]{vanDaalen11,Semboloni11,Troxel14,Krause15,Blazek17,Chisari18}. In the context of galaxy clustering, the most relevant astrophysical systematic is the galaxy-matter bias relation on small scales \citep{2013MNRAS.436.2038Z,2016arXiv161109787D}. Effects such as magnification of number counts and redshift space distortions need to be included in the models as well \citep{Alonso15,ghosh18}. 

Second, even standard cosmological quantities in the simplest models, such as distances in a $\Lambda$CDM cosmology, have to be predicted to a validated high degree of {\it numerical} accuracy. Achieving this objective is not trivial, as computing these quantities generally requires, for example, numerical integration or interpolation, both of which are prone to numerical error.

Commonly used, publicly available, cosmological prediction tools are: {\tt astropy}\footnote{\url{http://www.astropy.org}} \citep{astropy}, {\tt NumCosmo}\footnote{\url{https://numcosmo.github.io}} \citep{numcosmo} and {\tt CAMB}\footnote{\url{https://camb.info}} \citep{camb}. However, none of these meets {\it all} the necessary capabilities for cosmological analysis with the next generation of dark energy experiments.

Faced with these challenges, the Dark Energy Science Collaboration (DESC), one of the science collaborations of the Large Synoptic Survey Telescope (LSST), has built a comprehensive software tool that satisfies the needs of the next generation of cosmological analysis: the Core Cosmology Library\footnote{publicly available at \url{https://github.com/LSSTDESC/CCL}} (\ccl). \ccl is a software library providing the infrastructure to make theoretical predictions that are validated to a well-documented high degree of {\it numerical} accuracy for the purpose of constraining cosmology. In the context of this work, we establish the accuracy of \ccl predictions by comparing them to predictions from external packages. Thus, what we quantify is the {\it level of agreement} between independent pipelines.

\ccl computes standard cosmological functions including the Hubble parameter, cosmological distances, density parameters, the halo mass function, halo bias and linear growth functions. It calculates the matter power spectrum using various methods including common approximations, by calling external software such as \class \citep{class}, or emulators, such as the ``Cosmic Emulator'' of \citet{Lawrence17}. It computes 2-point angular power spectra and correlation functions from various probes, going beyond the Limber approximation. While \ccl incorporates state-of-the-art models available in the literature, this manuscript is mainly concerned with documenting their implementation and {\it numerical} accuracy, but does not address the {\it physical} accuracy of each model, for which we point the reader to the relevant references in the following sections. To our knowledge, no other adaptable, up-to-date and publicly available software tool for state-of-the-art cosmological predictions has undergone such a rigorous validation process as described in this manuscript.  

\begin{figure*}
\centering
\includegraphics[width=0.9\textwidth]{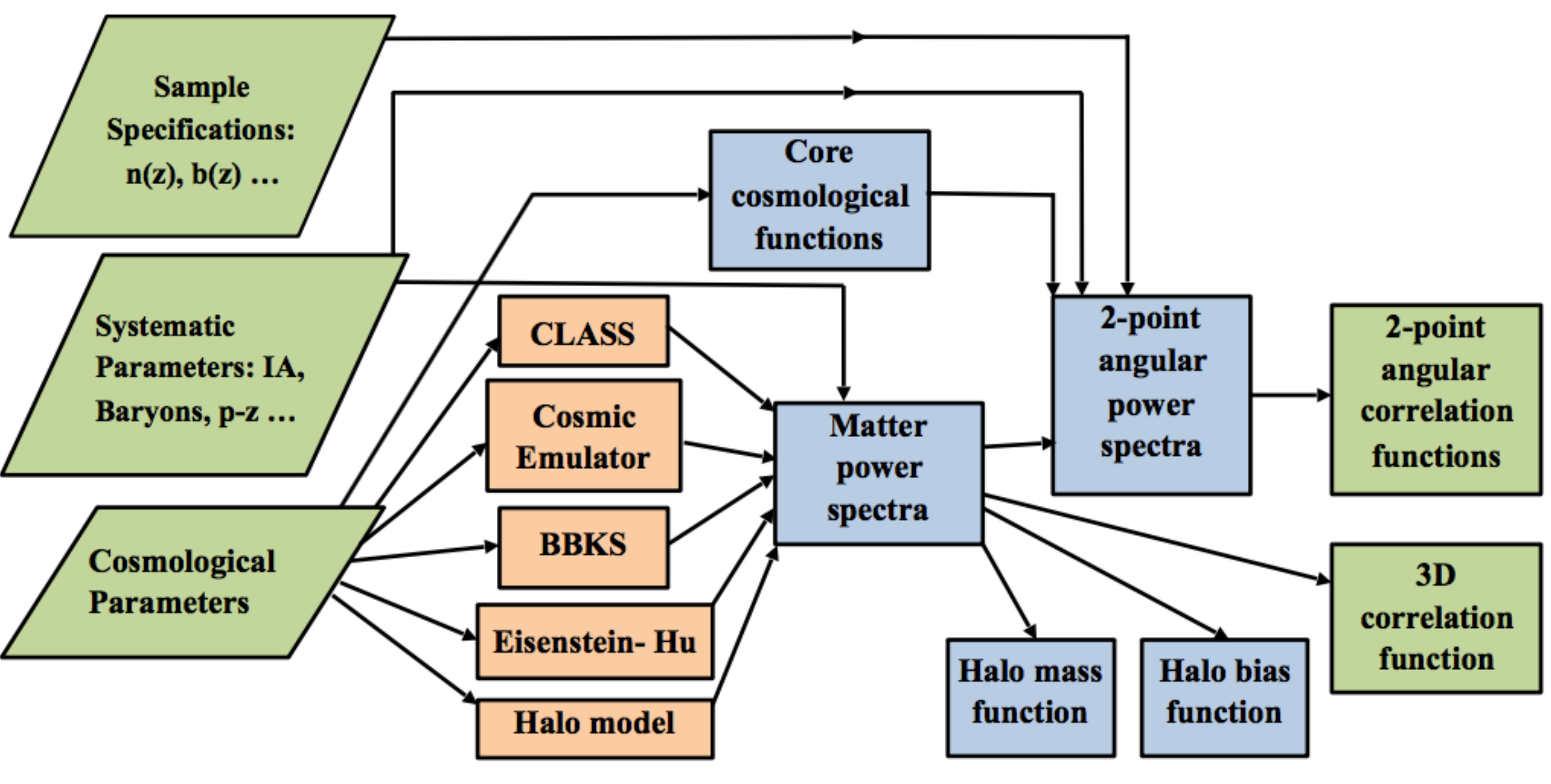}
\caption{\ccl structure flowchart. \ccl is written in C with a {\tt python} interface. \ccl routines calculate basic cosmological functions such as the Hubble function, density parameters, distances and growth function. The library uses various methods to compute the matter-power spectrum, including \class, the ``Cosmic Emulator'' developed by \citet{Lawrence17}, and other common approximations. \ccl computes 2-point angular power spectra and correlation functions from various probes, including typical astrophysical systematics and accounting for user-provided or pre-coded survey specifications.}
\label{fig:CCL_structure}
\end{figure*}

\ccl 's overall structure is illustrated in Figure \ref{fig:CCL_structure}. Our implementation has support for spatially flat and curved $\Lambda$-Cold Dark Matter ($\Lambda$CDM) cosmologies, and $w$CDM cosmologies with the option of using a time-dependent equation of state. It also allows for cosmologies with multiple massive neutrino species and can be linked to external software for modified gravity predictions \citep[\hiclass,][]{Zumalacarregui17}. While \ccl was built with LSST in mind, the goal is to produce a publicly available, user-friendly, well-documented, adaptable software that can be used in any theoretical modeling work in cosmology. This manuscript describes version {\tt 1.0.0} of the library.

The validation procedure to assess the {\it numerical} accuracy of each \ccl feature is key in this work. We compare the \ccl evaluation of each observable or function to an independent implementation from a stand-alone software package. For each prediction, we define an accuracy metric which surpasses our expected needs for accurate cosmological constraints and document the results obtained in this manuscript. Ultimately, the numerical uncertainties in the different \ccl functions propagate to our predictions for correlation functions, which we expect to be one of the main summary statistics used in the LSST cosmology analyses (similarly to current DES and KiDS efforts). Hence, our overall goal in this work is to demonstrate that correlation functions obtained by \ccl are accurate to within a fraction of the expected observational uncertainties for the models and in the range of scales of interest to LSST. In addition, we ensure that any prediction of the two-point statistics of the distribution of matter, necessary for predicting cross-correlations between probes, has a well-established accuracy.

This paper is organized as follows. Section \ref{sec:models} describes the cosmological models and observables supported by \ccl. In Section \ref{sec:implement}, we describe the details of the implementation of the quantities introduced in Section \ref{sec:models}. Section \ref{sec:validation} provides details of the validation procedure, the tests performed and the accuracy achieved. Section \ref{sec:usage} gives brief guidelines for the usage of \ccl, although we direct the reader to the software online repository, documentation and user manual for further information. We conclude in Section \ref{sec:conclusion} with an outlook towards the integration of \ccl in the LSST DESC pipelines, and we outline future additions to the software.

\section{Cosmological models and observables}
\label{sec:models}

The overarching goal of \ccl is to allow seamless integration of different cosmological models of interest to LSST. The cosmological models assume a homogeneous and isotropic space-time metric, and an inflationary model for the primordial universe described by a power-law with spectral index $n_s$, and amplitude $A_s$ defined at a pivot scale of $k_0=0.05\,{\rm Mpc}^{-1}$. The cosmological components include the matter density parameter $\Omega_m$, which is the sum of the baryonic component $\Omega_b$ and the cold dark matter component $\Omega_c$, the dark energy density\footnote{While we adopt $\Omega_\Lambda$ as notation, this quantity represents the dark energy density also in the case where dark energy is described by a dynamical field.} $\Omega_\Lambda$, the radiation density $\Omega_\gamma$ (excluding neutrinos), the curvature density $\Omega_K$, and the neutrino density of both mass-less and massive neutrinos, given by $\Omega_{\nu, {\rm rel}}$ and $\Omega_{\nu, {\rm m}}$ respectively. Unless otherwise specified, we refer to these densities at the present. The current expansion rate is given by the Hubble constant, $H_0 = 100~h~{\rm km}~{\rm s}^{-1}~{\rm Mpc}^{-1}$. The normalization of the density fluctuations is established either in terms of $A_s$ or in terms of the RMS variance in spheres of $8\,h\,\text{Mpc}^{-1}$ today, $\sigma_8$. 

The following set of models is supported in \ccl:
\begin{itemize}
 \item Flat $\Lambda$CDM cosmology governed by the parameters $\Omega_b$, $\Omega_m$, $H_0$, $n_s$, $A_s$ or $\sigma_8$, and a cosmological constant dark energy model with equation-of-state $w=-1$. 
\item The Chevallier-Polarski-Linder (CPL) model for dark energy, which adopts the following parametrization for $w$ as a function of the scale factor, $a$ (\citealt{Chevallier01} and \citealt{Linder03}),
  \begin{equation}
    w(a) = w_0+w_a(1-a).
  \end{equation}
  We note that models with constant $w$ are simply a subset of the above, where $w_a=0$.
 \item Non-zero curvature ($K$), so that the curvature density parameter $\Omega_K = 1 - \sum_i \Omega_i$, where $i$ refers to each of the density components. 
 \item Extra relativistic species, contributing to $N_{\rm eff}$ (the effective number of neutrinos).
  \item Massive neutrinos specified by either the sum of their masses $\Sigma m_\nu$ (which maps on to the density parameter $\Omega_{\nu, {\rm m}}$ above), or by the individual masses of each of the three neutrino species. This feature is allowed alongside non-zero curvature, extra relativistic species, and evolving dark energy.
 \item An arbitrary, user-defined modified growth function (see description in Section \ref{sec:growth}). This can be combined with a model that otherwise contains non-zero curvature and evolving dark energy.
\end{itemize}

In the particular case of cosmologies with massive neutrinos, if the user specifies a sum of masses, $\Sigma m_\nu$, \ccl will by default split $\Sigma m_\nu$ into three neutrino masses which are consistent with the normal hierarchy (see, e.g. \citealt{Gerbino2017} for a review). However, the user can alternatively ask for $\Sigma m_\nu$ to be split either into masses consistent with the inverted hierarchy, or into equal masses. Each neutrino species is then checked for whether it is non-relativistic (massive) at $z=0$, and this information is used in combination with the user-provided value of $N_{\rm eff}$ to set the number of relativistic neutrino species.

The following sub-sections describe the cosmological predictions implemented in \ccl. Not all \ccl features are available for the models described in this section. For a guide to which predictions are available for each model, see Table \ref{tab:cosmo}. Note that if users install their own version of the \class software (for example, \hiclass, \citealt{Zumalacarregui17}), \ccl can then make predictions for a more extended set of cosmologies. Users should take care to understand the validity of the \ccl assumptions for their own models.

\begin{table*}
  \begin{center}
    \caption{Cosmologies implemented in \ccl and observables supported for each of them. Note that the only reason why angular power spectra appear not to be supported in non-flat cosmologies is that the hyperspherical Bessel functions are currently not implemented, although their impact is fairly limited. Likewise, number counts power spectra are strictly not supported in the presence of massive neutrino cosmologies due to the scale-dependent growth rate that affects the redshift-space distortions term, even though the impact of this is also small for wide tomographic bins. The halo model can make matter power spectrum predictions for all cosmologies, but should not be used for massive neutrino models because the current version does not distinguish between the cold matter, relevant for clustering, and all matter. Finally, we note that \ccl can make predictions for the growth of perturbations for some modified gravity models through a user defined $\Delta f(a)$ as detailed in Section \ref{sec:growth}, and that other extensions are supported via integration of external modified gravity codes.\label{tab:cosmo}}
    \begin{tabular}{lcccccc}
      \hline\hline
      Observable/Model & flat $\Lambda$CDM & $\Lambda$CDM+$K$ & $\Lambda$CDM + $m_\nu$ & $w$CDM \\[3pt] 
      \hline
      Distances & \checkmark & \checkmark  & \checkmark & \checkmark \\
      Growth function  & \checkmark & \checkmark & $X$ & \checkmark    \\
      $P_m(k,z)$ & \checkmark & \checkmark & \checkmark & \checkmark \\
      Halo Mass Function & \checkmark & \checkmark & $X$ & \checkmark \\
      $C_l$, number counts & \checkmark & $X$ & $X$ & \checkmark \\
      $C_l$, weak/CMB lensing  & \checkmark & $X$ & \checkmark & \checkmark \\
      Correlation function & \checkmark & $X$ & \checkmark & \checkmark \\
      Halo model & \checkmark & \checkmark & $X$ & \checkmark \\
      \hline\hline
    \end{tabular}
  \end{center}
\end{table*}

\subsection{Background cosmology}

The models that are specified above map directly onto cosmological observables such as the expansion rate of the Universe, which is parameterized through the Hubble parameter as
\begin{align}\label{eq:Ha}
\frac{H(a)}{H_0} &= a^{-3/2}\Big(\Omega_{m}+\Omega_{\Lambda} a^{-3(w_0+w_a)}
\exp[3 w_a (a-1)]~+ \nonumber \\ &\Omega_{K} a + (\Omega_{\gamma} + \Omega_{\nu, {\rm rel}}) a^{-1} + \tilde\Omega_{\nu, {\rm m}}(a)a^3\Big)^{\frac{1}{2}} ,
\end{align}
and is a function of the scale factor and of the energy density in the different components today. In this expression, we have assumed the CPL parameterization described above for the dark energy equation of state and we have defined $\tilde\Omega_{\nu, {\rm m}}(a) \equiv \rho^{-1}_{\rm crit} \rho_{\nu, {\rm m}}(a)$ as the fractional energy density of massive neutrinos as a function of time, where $\rho_{\rm crit}$ is the critical density of the Universe today.

In general, the density parameter $\Omega_X(a)$ of a given species $X$ at a given time is defined in terms of the physical background densities $\bar\rho_X(a)$ via $\Omega_X(a) \equiv \rho^{-1}_{\rm crit}(a) \bar\rho_X(a)$, where the critical density of the Universe at a given time is
\begin{equation}
  \rho_{\rm crit}(a) = {( 8 \pi G)}^{-1} 3c^2H^2(a) = \rho_{\rm crit} H_0^{-2} H^2(a).
  \label{eq:rhocrit}
\end{equation}
As an example, the physical density of matter is given by
\begin{equation}
  \bar\rho_m(a) = \bar\rho_{m} a^{-3} = \rho_{\rm crit} \Omega_{m} a^{-3},
\end{equation}
and its density parameter is
\begin{equation}
  \Omega_m(a) = \Omega_{m} H_0^{2} a^{-3} H^{-2}(a).
\end{equation}
Moreover, \ccl allows for comoving physical densities $\bar\rho_{X, {\rm com}}(a) = \bar\rho_X(a) a^3$ to be extracted, which in the case of matter reduces to a time-independent $\bar\rho_{m, {\rm com}} = \rho_{\rm crit} \Omega_{m}$. We include bars for $\rho_X$ to distinguish from spatially-varying densities in later sections.

The specific case of $\tilde\Omega_{\nu, {\rm m}}(a)$ in Eq. (\ref{eq:Ha}) is calculated via
\begin{align}
\tilde\Omega_{\nu, {\rm m}}(a) &= \frac{7}{8}\sum_{i=1}^{N_\nu} \frac{4 \sigma_{B}}{c \rho_{{\rm crit}}} \left(\frac{T_{\nu}^{\rm eff}}{a}\right)^4  \int_0^{\infty} dx \, x^2 \frac{\sqrt{x^2 + \left(\tilde{m}^i\right)^2}}{\exp(x) + 1}.
\label{Omnu}
\end{align}
Here, $\sigma_B$ is the Stefan-Boltzmann constant, $c$ is the speed of mass-less particles, $\rho_{{\rm crit}}$ is the present critical density, and $T_{\nu}^{\rm eff}$ is the present effective temperature of the massive neutrinos. $T_{\nu}^{\rm eff}$ is related to the temperature of the CMB via $T_{\nu}^{\rm eff} = T_{\rm CMB} T_{\rm NCDM}$, where $T_{\rm NCDM}$ is a dimensionless factor ($\simeq1$) used by e.g. \class to set the ratio ${\sum m_\nu}/{\Omega_{\nu, {\rm m}}}$ to its experimentally measured value. Note that $T_{\rm NCDM}$ is used to modulate the effective temperature of massive neutrinos only; the temperature of relativistic neutrinos follows the usual relation in which $T_\nu = T_{\rm CMB} \left(\frac{4}{11}\right)^{1/3}$. Finally, $\tilde{m}^i$ is a per-species mass-dependent dimensionless constant, given by $\tilde{m}^i = m_{\nu}^{i}c^2 a / (k_B T_{\nu}^{\rm eff})$ where $k_B$ is the Boltzmann constant.

Fitting models to cosmological observables requires predicting cosmological distances for a given model. We consider the comoving radial distance, which is calculated via a numerical integral as
\begin{equation}
 \chi(a)= c \int_a^1 \frac{da'}{a'^2 H(a')}.
 \label{eq:comrdist}
\end{equation}
The comoving angular diameter distance is then computed in terms of the comoving radial distance,
\begin{equation}\label{eq:angdist}
 r(\chi)=\left\{\begin{array}{cc}
                 K^{-1/2}\sin(K^{1/2}\chi) & K>0\\
                 \chi & K=0\\
                 |K|^{-1/2}\sinh(|K|^{1/2}\chi) & K<0\\
                \end{array}\right.
\end{equation}
where $K \equiv \Omega_K H_0^2 c^{-2}$ is the curvature.
The angular diameter distance is given by $d_A=a\,r(\chi(a))$, and the luminosity distance is
$d_L=r(\chi(a))/a,$ leading to the familiar relation $d_A = a^2d_L$.
The \ccl suite also has the functionality to compute the distance modulus, defined as
\begin{equation}\label{eq:distmod}
    \mu = 5 \log_{10}(d_L / {\rm pc})-5,
\end{equation}
along with $a(\chi)$, the inverse function of $\chi(a)$.

\subsection{Growth of perturbations}
\label{sec:growth}

In conjunction with the expansion rate, the growth history of the Universe can allow us to distinguish between cosmological models. To compute the linear growth factor of matter perturbations, $D(a)$, \ccl solves the following differential equation:
\begin{equation}
  \frac{d}{da}\left(a^3H(a)\frac{dD}{da}\right)=\frac{3}{2}\Omega_m(a)aH(a)D,
  \label{eq:growth}
\end{equation}
using a Runge-Kutta Cash-Karp algorithm. We define $g(a)\equiv D/a$ and adopt as initial conditions $g(a)=1$ and $g'(a)=0$ at sufficiently high redshift, during the matter-dominated era. \ccl simultaneously computes the logarithmic growth rate $f(a)$, defined as:
\begin{equation}
  f(a)\equiv \frac{d\ln D}{d\ln a}.
  \label{eq:lingrowthf}
\end{equation}

\ccl provides functions that return the growth normalized either to $D(a=1)=1$ or to $D(a\ll1)\rightarrow a$. It employs an accelerated spline that is linearly spaced in the scale factor to interpolate the growth functions (for more details, see Section \ref{sec:implement}). The growth calculations cover flat and curved $\Lambda$CDM and $w$CDM cosmologies. However, it should be noted that the above treatment is ill-defined in the presence of massive neutrinos, and attempts to compute the growth rate in cosmologies with massive neutrinos will produce an inconsistency between growth predictions and the matter power spectrum (Section \ref{sec:matterps}), for example.

Finally, \ccl allows for growth predictions with an alternative `modified gravity' cosmological model defined by a regular curved $w$CDM background as well as a user-defined $\Delta f(a)$, such that the true growth rate in this model is given by
\begin{equation}
  f(a)=f_0(a)+\Delta f(a),
\label{eq:mgrowth}
\end{equation}
where $f_0(a)$ is the growth rate in the background model. Note that this model is only consistently implemented with regards to the computation of the linear growth factor and growth rates and does not feed into other observables. This model, and the interpretation of the predictions given by \ccl, should therefore be used with care. 

\subsection{Matter power spectrum}
\label{sec:matterps}

Theoretical predictions for cosmological observables such as galaxy clustering, gravitational lensing and cluster mass functions rely on knowledge of the distribution of matter from small to large scales in the Universe. To second order, the distribution of matter density fluctuations at a given wavenumber ${\bf k}$ and redshift is described by the matter power spectrum, $P(k,z)$, defined as
\begin{equation}
  \langle \tilde\delta({\bf k},z)\tilde\delta({\bf k}^\prime,z)\rangle = (2\pi)^3P(k,z)\
\delta_D^3({\bf k}+{\bf k}')
\end{equation}
where $\tilde\delta({\bf k})$ is the Fourier component of the density field at a given wavenumber and $\delta_D^3$ is the Dirac delta function. $P(k,z)$ has units of volume and a dimensionless analogue is often defined as
\begin{equation}
  \Delta^2(k,z) \equiv \frac{k^3}{2\pi^2}P(k,z).
\end{equation}
At sufficiently large scales (small $k$), $P(k,z)$ can be obtained from solving linear perturbation theory equations. In this case, $P(k,z)$ is referred to as the ``linear'' matter power spectrum. At small scales, where perturbation theory breaks down, other approaches based on numerical simulations are needed. In this more general case, $P(k,z)$ is referred to as the ``non-linear'' matter power spectrum.

\ccl implements several different methods for making predictions for the matter power spectrum. Two of those methods, the BBKS \citep{BBKS} and \citet{1998ApJ...496..605E} approximations, are only accurate to within a few per cent and are implemented for validation purposes mainly. These approximations provide analytical expressions for the transfer function, $T(k)$, which is related to the matter power spectrum by $\Delta^2(k) \propto T^2(k) k^{3+n_s}$. There are two alternative ways to normalize the power spectrum. One option, which establishes a normalization at $z=0$, is to provide a value for $\sigma_8$. The second option is to set the normalization at high redshift by giving a value for the amplitude of primordial fluctuations, $A_s$. From the point of view of the \ccl implementation, if the user provides $\sigma_8$, \ccl also calculates the corresponding $A_s$ for the specified cosmology.

The default \ccl implementation uses the \class algorithm \citet{class} to obtain predictions for $P(k,z)$. \class uses a Boltzmann solver to compute the linear power spectrum and also includes the \halofit \citep{Smith2003,CLASS_halofit} fitting function for the non-linear spectrum. In addition, \ccl can also generate $P(k,z)$ predictions by emulation of cosmological numerical simulations using the ``Cosmic Emulator'' developed by \citet{Lawrence17}.

We also provide a basic halo model calculation of the non-linear matter power spectrum which uses the included halo bias, halo mass function and halo density profiles (see Sections \ref{sec:hmfdef}, \ref{sec:hbdef} and \ref{sec:halo_model}). The power spectrum calculated via this method is not accurate enough for precision cosmology, with deviations of as great as 50 per cent compared to numerical simulations, but is pedagogically useful and we envisage to expand its functionalities to make it more realistic in the future.

None of the above methods account for the impact of baryonic physics on the distribution of matter, which is known to exceed the per cent level at scales $k \gtrsim 1\,\text{Mpc}^{-1}$ \citep{vanDaalen11,Illustris,Hellwing16,Springel17,Chisari18} and can affect the extraction of cosmological parameters \citep{Semboloni11,Semboloni13,Mohammed14,Eifler15,Mohammed17}. To account for this effect, we incorporate in \ccl an effective parametrization \citep{Schneider15} of the redistribution of matter as a consequence of feedback from Active Galactic Nuclei and adiabatic cooling. We give an overview of each method to predict the matter power spectrum in what follows.

\paragraph{\bf BBKS approximation.} \ccl implements the analytical BBKS approximation to the transfer function \citep{BBKS}, given by
\begin{eqnarray}
  \label{eq:bbks}
  T(q) &=& \frac{\ln[1+2.34q]}{2.34q}\times\\
  &&[1+3.89q+(16.2q)^2+(5.47q)^3+(6.71q)^4]^{-0.25}\nonumber
\end{eqnarray}
where $q$ is defined as follows \citep{Sugiyama95}
\begin{equation}
  q \equiv k/\{\Omega_m h^2 e^{-\Omega_b[1+\sqrt{2h}/\Omega_m]}{\rm Mpc}^{-1}\},
\end{equation}
where $k$ has units of ${\rm Mpc}^{-1}$. The BBKS power spectrum option is primarily used as a precisely-defined input for testing the numerical accuracy of \ccl routines (as described in Section \ref{sec:implement}), and it is not recommended for other uses.

\paragraph{\bf Eisenstein \& Hu approximation.} \ccl also provides an approximation to the matter power spectrum as implemented by \citet{1998ApJ...496..605E} (we refer the reader to this paper for a detailed discussion of the fitting formulae).\footnote{Note that the implementation in \ccl modifies Eq. 5 of \citet{1998ApJ...496..605E} using $a^{-1}=1+z$ instead of the approximation $a^{-1}\sim z$. The difference in the resulting power spectra is negligible, but larger than 1 part in $10^4$ for $k<10\,h\,{\rm Mpc}^{-1}$.}

\paragraph{\bf \tt CLASS.} The default configuration of \ccl adopts predictions for the linear and non-linear matter power spectrum from the publicly available software \citep{class}. \class uses a Boltzmann solver to compute the linear power spectrum and makes predictions for the non-linear power spectrum using the \halofit prescription of \cite{CLASS_halofit}.

\paragraph{\bf Cosmic emulator.} An emulator method trained on numerical simulations \citep{Lawrence17} provides accurate predictions for the non-linear matter power spectrum for $z\leq 2$ and in the wavenumber range $k=[10^{-3},5]$ Mpc$^{-1}$. The allowed range of cosmological parameters that can be passed to the emulator is as follows\footnote{$w_a$ and $w_0$ are constrained jointly to be $0.3\leq (-w_0-w_a)^{1/4}$.}:
 \begin{eqnarray}
 0.12&\leq& \Omega_{m} h^2 \leq 0.155,\nonumber\\
 0.0215&\leq& \Omega_{b} h^2 \leq 0.0235,\nonumber\\
 0.7&\leq& \sigma_8 \leq 0.9,\nonumber\\
 0.55&\leq& h \leq 0.85,\nonumber\\
 0.85&\leq& n_s\leq 1.05,\nonumber\\
 -1.3&\leq& w_0\leq-0.7,\nonumber\\
 -1.73&\leq& w_a\leq -0.7,\nonumber\\
 0.0&\leq& \Omega_{\nu, {\rm m}} h^2 \leq 0.01.
 \end{eqnarray}
In the case of the emulator, the effective number of relativistic neutrino species is set to $N_{\rm eff}=3.04$ and $\Omega_\gamma=0$. In \citet{Lawrence17}, the neutrino component of the power spectrum is not simulated, but either linearly evolved and added to the simulated power spectra at low redshift, or accounted for by a scale-dependent correction to the growth function. The typical accuracy of the emulator with respect to simulated power spectra is at the $<3\%$ level and depends on the cosmological model. More details on this method and its accuracy can be found in \citet{Upadhye14,Castorina15,Heitmann16}.
 
\paragraph{\bf Baryonic correction model (BCM).} \ccl incorporates the impact of baryons on the total matter power spectrum via the BCM of \citet{Schneider15}. The main consequences of baryonic processes are: to suppress the power spectrum at intermediate scales ($k\sim$ a few $\text{Mpc}^{-1}$) due to the heating and ejection of gas by Active Galactic Nuclei feedback, and to enhance it at smaller scales due to adiabatic cooling. To account for these effects, BCM uses an effective decomposition for the impact of gas ejection ($G$) and the enhancement of the small scale profile due to star formation ($S$) to estimate the fractional effect of baryonic processes on the dark matter-only power spectrum ($P_{\rm DMO}$):
\begin{equation}
  P_{\rm BCM}(k,z)=P_{\rm DMO}(k,z) G(k|M_c,\eta_b,z)S(k|k_s)
  \label{eq:bcm}
\end{equation}
Three effective parameters govern the contribution of baryonic processes to modifying the total matter power spectrum:
 \begin{itemize}
   \item $\log_{10} [M_c/($M$_\odot/h)]$: the mass of the clusters responsible for feedback, which regulates the amount of suppression of the matter power spectrum at intermediate scales;
   \item $\eta_b$: a dimensionless parameter which determines the scale at which suppression peaks;
   \item and $k_s$ [$h\,\text{Mpc}^{-1}$]: the wavenumber that determines the scale of the stellar distribution of matter in the center of halos.
 \end{itemize}
 If these parameters are not specified by the user, \ccl assumes the default parameters of \citet{Schneider15}, calibrated through different comparisons with observations and simulations in that work.

\subsection{Two-point correlators}\label{ss:2point}

The matter power spectrum is one of the necessary components to produce theoretical expectations for the two-point correlators of pairs of quantities (fields) that trace the matter density field in the Universe. In this section, we will define these fields on the sky, such as galaxy positions or galaxy shapes. These fields can be classified in terms of their spin $s$ under rotations on the plane tangent to the sphere. In general a spin-$s$ field is defined by two real-valued functions of the spherical coordinates $a_1(\nv)$ and $a_2(\nv)$ (e.g. $\gamma_1$ and $\gamma_2$ for weak lensing or the Stokes parameters $Q$ and $U$ in the case of polarized intensity), from which one can form the complex field $a=a_1+ia_2$. For galaxy clustering, the galaxy density is a spin-$0$ field described by a scalar.

Spin-$s$ quantities can be decomposed into their harmonic coefficients $\,_sa_{\ell m}$ through a spherical harmonic transform \citep{1997PhRvD..55.1830Z,2011A&A...526A.108R}:
\begin{equation}\nonumber
  \,_sa_{\ell m}=\int d\nv\,a(\nv)\,_sY^*_{\ell m}(\nv),\hspace{12pt}
  a(\nv)=\sum_{\ell m}\,_sa_{\ell m}\,_sY_{\ell m}(\nv)
\end{equation}
where $_sY_{\ell m}$ are the spin-weighed spherical harmonics. The harmonic coefficients can then be associated with parity-even and parity-odd components ($E$-modes and $B$-modes respectively) as\footnote{We note that for spin-$0$ quantities the minus sign preceding these equations is usually omitted, and we do so in what follows. Also, all scalar fields discussed here are real-valued, and therefore have zero $B$-modes.}
\begin{align}\nonumber
   E_{\ell m}&=-\frac{1}{2}\left[\,_{s}a_{\ell m}+(-1)^s\,_{-s}a_{\ell m}\right]\\
  iB_{\ell m}&=-\frac{1}{2}\left[\,_{s}a_{\ell m}-(-1)^s\,_{-s}a_{\ell m}\right],
\end{align}
where $_{-s}a_{\ell m}$ is defined as
\begin{equation}\nonumber
  \,_{-s}a_{\ell m}=\int d\nv\,a^*(\nv)\,_{-s}Y^*_{\ell m}(\nv).
\end{equation}

In what follows we will focus on scalar ($s=0$) quantities such as the overdensity of source number counts or the CMB lensing convergence, and on spin-2 fields such as the lensing shear. We will also distinguish between {\sl tracers} (fields observed on the sky, such as number counts in a redshift bin, shear, or CMB temperature fluctuations) and {\sl contributions} to the total observed fluctuations of these tracers (such as the biased matter density term in number counts, redshift-space distortions, magnification, etc.).

\subsubsection{Angular power spectra}\label{sssec:2pt.pspec}

The angular power spectrum $C^{ab}_\ell$ between two tracers $a$ and $b$ is defined as
\begin{equation}
  \langle a_{\ell m}b^*_{\ell m}\rangle\equiv C^{ab}\delta_{\ell\ell'}\delta_{mm'},
  \label{eq:clgen}
\end{equation}
where $a_{\ell m}$ and $b_{\ell m}$ can be either the $E$-mode or $B$-mode component of the corresponding field. In what follows we will only work with fields for which the $B$-modes are exactly or nearly $0$, and which we will take to be identically $0$. Therefore all equations refer to the $E$-$E$ power spectrum. In general, this power spectrum can be written as:
\begin{equation}
  C^{ab}_\ell=4\pi\int_0^\infty \frac{dk}{k}\,\mathcal{P}_\Phi(k)\Delta^a_\ell(k)\Delta^b_\ell(k),
  \label{eq:cls}
\end{equation}
where $\mathcal{P}_\Phi(k)$ is the dimensionless power spectrum of the primordial curvature perturbations, and $\Delta^a$ and $\Delta^b$ are the transfer functions corresponding to these tracers. Each transfer function will receive contributions from different terms. \ccl supports three types of tracers: number counts, galaxy shape distortions and CMB lensing convergence, with the following contributions\footnote{Note that we use units where the speed of light is $c=1$ throughout.}:

\paragraph{\bf Number counts.} The transfer function for number counts can be decomposed into three contributions: $\Delta^{\rm NC}=\Delta^{\rm D}+\Delta^{\rm RSD}+\Delta^{\rm M}$, where
\begin{itemize}
  \item $\Delta^{\rm D}$ is the standard density term proportional to the matter density:
        \begin{equation}\label{eq:transfer_nc}
          \Delta^{\rm D}_\ell(k)=\int dz\,p_z(z)\,b(z)\,T_\delta(k,z)\,j_\ell(k\chi(z)),
        \end{equation}
        where $j_\ell(x)$ is $\ell$-th order spherical Bessel function, $T_\delta$ is the matter overdensity transfer function, $b(z)$ is the linear clustering bias for this tracer and $p_z(z)$ is the normalized distribution of sources in redshift. The fluctuations in the number density of sources in different redshift bins are therefore treated by \ccl as different tracers. Note that \ccl does not currently support non-linear or scale-dependent bias, but future releases will do so under a number of schemes, including perturbative approaches as implemented in, e.g., \citet{FASTPT}.
        
        It is also worth noting that the matter overdensity transfer function $T_\delta$ in Eq. \ref{eq:transfer_nc} is not the same as the transfer function used in Section \ref{sec:matterps}. While $T(k)$ is defined as \citep{1998ApJ...496..605E} 
        \begin{equation}
          T(k)=\frac{\delta(k,z=0)}{\delta(k,z=\infty)}\frac{\delta(k=0,z=\infty)}{\delta(k=0,z=0)},
        \end{equation}
        all subscripted transfer functions $T_X$ used here are defined as the ratio between the subscript quantity $X$ and the primordial curvature perturbations:
        \begin{equation}
          X({\bf k},z)=T_X(k,z)\,\Phi({\bf k}).
        \end{equation}        
  \item $\Delta^{\rm RSD}$ is the linear contribution from redshift-space distortions (RSDs):
        \begin{equation}\label{eq:transfer_rsd}
          \Delta^{\rm RSD}_\ell(k)=\int dz\,\frac{(1+z) p_z(z)}{H(z)}T_\theta(k,z) j_\ell''(k\chi(z)),
        \end{equation}
        where $T_\theta(k,z)$ is the transfer function of $\theta$, the divergence of the comoving velocity field, and $j''_\ell$ is the second order derivative of the spherical Bessel function, $j_\ell$. Note that the RSD contribution to number counts is computed by \ccl assuming a linear-theory relation between the matter overdensity and peculiar velocity fields, mediated by the scale-independent growth rate $f$ (Eq. \ref{eq:lingrowthf}). While this should not be problematic for wide photometric redshift bins and standard cosmological models, users should exercise caution when interpreting results for narrow window functions or exotic cosmologies. Additionally, number count tracers with RSD in cosmologies with massive neutrinos are not currently supported.
  \item $\Delta^{\rm M}$ is the contribution from lensing magnification:
        \begin{align}\nonumber
          \Delta_\ell^{\rm M}(k)=-\ell(\ell+1)\int & \frac{dz}{H(z)} W^{\rm M}(z) \\\label{eq:deltaM}
          &T_{\phi+\psi}(k,z) j_\ell(k\chi(z)),
        \end{align}
        where $T_{\phi+\psi}$ is the transfer function for the Newtonian-gauge scalar metric perturbations, and $W^{\rm M}$ is the magnification window function:
        \begin{equation}\label{eq:window_mag}
          W^{\rm M}(z)\equiv\int_z^\infty dz'\,p_z(z')\frac{2-5s(z')}{2}\frac{r(\chi'-\chi)}{r(\chi')r(\chi)}.
        \end{equation}
        Here, $s(z)$ is the logarithmic derivative of the number of sources with magnitude limit, and $r(\chi)$ is the angular comoving distance (see Eq. \ref{eq:angdist}).
\end{itemize}
Note that \ccl does not currently compute relativistic corrections to number counts other than magnification bias \citep{2011PhRvD..84d3516C,2011PhRvD..84f3505B}. Although these will be included in the future, their contribution to the total fluctuation is largely sub-dominant (see \citealt{GReffects} and the two references above), and therefore it is safe to ignore them for our purposes.

\paragraph{\bf Correlated galaxy shapes.} The transfer function for correlated galaxy shapes (intrinsic and lensed) is decomposed into two terms: $\Delta^{\rm SH}=\Delta^{\rm WL}+\Delta^{\rm IA}$, where
\begin{itemize}
  \item $\Delta^{\rm L}$ is the standard lensing (``cosmic shear'') contribution:
    \begin{align} \label{eq:transfer_lensing}
      \nonumber
      \Delta_\ell^{\rm L}(k)=-\frac{1}{2}\sqrt{\frac{(\ell+2)!}{(\ell-2)!}}\int &\frac{dz}{H(z)} W^{\rm L}(z)T_{\phi+\psi}(k,z)\\
      &j_\ell(k\chi(z)),
    \end{align}
    where $W^{\rm L}$ is the lensing kernel, given by
    \begin{equation}\label{eq:window_shear}
      W^L(z)\equiv\int_z^\infty dz' p_z(z')\frac{r(\chi'-\chi)}{r(\chi')r(\chi)}.
    \end{equation}
  \item $\Delta^{\rm IA}$ is the transfer function for intrinsic galaxy alignments. \ccl supports the so-called ``non-linear alignment model'' (NLA), in which the intrinsic galaxy inertia tensor is proportional the local tidal tensor \citep{Catelan01,2004PhRvD..70f3526H,2007MNRAS.381.1197H}:
    \begin{align}\nonumber
      \Delta_\ell^{\rm IA}(k)=\sqrt{\frac{(\ell+2)!}{(\ell-2)!}}\int &dz\,p_z(z)\,b_{\rm IA}(z)\,f_{\rm IA}(z)\\
      &T_\delta(k,z)\,\frac{j_\ell(k\chi(z))}{[k\chi(z)]^2}.
      \label{eq:transfer_ia}
    \end{align}
    Here, $b_{\rm IA}$ is the so-called alignment bias, and $f_{\rm IA}$ is the fraction of aligned galaxies in the sample. Notice that $b_{\rm IA}(z)$ absorbs the typical normalization factors used in the literature for intrinsic alignment amplitude and redshift evolution. It is thus not to be confused with $C_1$ or $A_{\rm IA}$, typical parameters for adopted in works such as \citet{vanUitert18,Joudaki18,Hildebrandt17}. In particular, the product $b_{\rm IA}\,f_{\rm IA}$ is equivalent to the quantity defined in Equation 8 of \cite{Hildebrandt17}. The NLA model has limitations in the modelling of small-scale correlations \citep{Singh15} and does not predict any $B$-mode contributions that can arise from non-linearities at such scales. However, it is a commonly adopted approximation in the current literature and going beyond it is outside of the scope of this work, though future versions of LSST DESC software will provide alternative modeling options \citep{Blazek17}.
    
\end{itemize}

\paragraph{\bf CMB lensing.} The transfer function for the lensing convergence, $\kappa$, of a given source plane at redshift $z_*$ receives only one contribution, given by
\begin{equation}
  \Delta_\ell^\kappa(k)=-\frac{\ell(\ell+1)}{2}\int_0^{\chi_*}\frac{dz}{H(z)}\,\frac{r(\chi_*-\chi)}{r(\chi)r(\chi_*)}T_{\phi+\psi}(k,z),
  \label{eq:cmblens}
\end{equation}
where $\chi_*\equiv\chi(z_*)$.

\noindent
It is worth noting that the equations above should be modified for non-flat cosmologies by replacing the spherical Bessel functions $j_\ell$ with their hyperspherical counterparts \citep{1994ApJ...432....7K}. These are currently not supported by \ccl, and their impact is mostly relevant on low multipoles. The library also assumes a factorizable matter power spectrum at unequal times $P_\delta(k,z_1,z_2)=T_\delta(k,z_1)T_\delta(k,z_2)\,2\pi^2\mathcal{P}_\Phi(k)$. This approximation is widely used in the literature, but further work is needed to assess its impact on LSST observables \citep{2017PhRvD..95f3522K}. Furthermore, \ccl assumes a relation between transfer functions $T_\delta$, $T_\theta$ and $T_{\phi+\psi}$ that is strictly only valid in vanilla $\Lambda$CDM\footnote{Note that the transfer functions are defined here for the full non-linear density field, as opposed to the more common linear transfer functions.}:
\begin{equation}
  T_\delta=-\frac{1+z}{H(z)f(z)}T_\theta=-\frac{k^2}{3H_0^2\Omega_m}\frac{T_{\phi+\psi}}{1+z}.
\end{equation}
These approximations will be revisited in future versions of the library.

\subsubsection{Correlation functions}

Fields are correlated in configuration space, and the corresponding correlators are called correlation functions. Let $a$ and $b$ be two fields with spins $s_a$ and $s_b$. We start by defining $\tilde{a}(\nv_1)$ and $\tilde{b}(\nv_2)$ as the fields $a$ and $b$ rotated such that the $x$-axis of the tangential coordinate systems at directions $\nv_1$ and $\nv_2$ become aligned with the vector connecting both points. We can then define two correlation functions:
\begin{equation}\nonumber
  \xi^{ab}_+(\theta)\equiv\left\langle\tilde{a}(\nv_1)\tilde{b}^*(\nv_2)\right\rangle,\hspace{12pt}
  \xi^{ab}_-(\theta)\equiv\left\langle\tilde{a}(\nv_1)\tilde{b}(\nv_2)\right\rangle,
  \label{eq:xipm}
\end{equation}
where $\nv_1\cdot\nv_2\equiv\cos\theta$. 

$\xi_{\pm}$ can be related to the angular power spectra (Eq. \ref{eq:clgen}) as
\begin{equation}\label{eq:cl_xi}
 \xi^{ab}_\pm = \sum_\ell\frac{2\ell+1}{4\pi}\,(\pm1)^{s_b}\,C^{ab\pm}_\ell\,d^\ell_{s_a,\pm s_b}(\theta),
\end{equation}
where $d^\ell_{mm'}$ are the Wigner-$d$ matrices \citep{Ng1999,2004MNRAS.350..914C} and we have defined the power spectra
\begin{equation}
  C^{ab\pm}_\ell\equiv\left(C^{a_Eb_E}_\ell\pm C^{a_Bb_B}_\ell\right)+i\left(C^{a_Bb_E}_\ell\mp C^{a_Eb_B}_\ell\right),
\end{equation}
which reduces to the $EE$ power spectrum when all $B$-modes are 0.

Note that, as scalar quantities are real, any correlation involving at least one spin-$0$ field only has one unique correlation function. In these cases, the Wigner-$d$ matrices can also be expressed in terms of associated Legendre polynomials $P^m_\ell$, and therefore Eq. (\ref{eq:cl_xi}) becomes
\begin{align}
  \xi^{ab}(\theta)&=\sum_\ell\frac{2\ell+1}{4\pi}\,C^{ab}_\ell\sqrt{\frac{(\ell-s_a)!}{(\ell+s_a)!}}P^{s_a}_\ell(\cos\theta)\label{eq:xigg},
\end{align}
where we have assumed $s_b=0$.

In the flat-sky approximation we can take the small-scale limit $\ell\gg s_a,s_b$ and approximate 
\begin{equation}
  d_{s_as_b}^\ell(\theta)\longrightarrow J_{s_a-s_b}(\ell\theta),
\end{equation}
where $J_\alpha(x)$ is the Bessel function of order $\alpha$. Eq. (\ref{eq:cl_xi}) then becomes\footnote{See the weak lensing review by \citet{Bartelmann01} and \citet{Joachimi10}.}
\begin{equation}
  \xi^{ab}_{\pm}(\theta)=\left(\pm1\right)^{s_b}\int\frac{d\ell\,\ell}{2\pi}\,C^{ab\pm}_\ell J_{s_a\mp s_b}(\ell\theta).
\end{equation}

In summary, for spins 0 and 2, the three relevant cases for the cosmological observables supported by \ccl are:
\begin{itemize}
  \item $s_a=s_b=0$ (e.g. galaxy-galaxy, galaxy-$\kappa$ and $\kappa$-$\kappa$):
    \begin{align}\label{eq:xi00full}
      \xi^{ab}(\theta)&=\sum_\ell\frac{2\ell+1}{4\pi}\,C^{ab}_\ell P_\ell(\cos\theta)\hspace{5pt}\text{(full-sky)}\\\label{eq:xi00flat}
                      &=\int_0^\infty\frac{d\ell\,\ell}{2\pi}\,C^{ab}_\ell J_0(\ell\theta)\hspace{25pt}\text{(flat-sky)}
    \end{align}
  \item $s_a=2$, $s_b=0$ (e.g. galaxy-shear, $\kappa$-shear):
    \begin{align}\label{eq:xi02full}
      \xi^{ab}(\theta)&=\sum_\ell\frac{2\ell+1}{4\pi}\,C^{ab}_\ell d^\ell_{2,0}(\theta)\hspace{15pt}\text{(full-sky)}\\\label{eq:xi02flat}
                      &=\int_0^\infty\frac{d\ell\,\ell}{2\pi}\,C^{ab}_\ell J_2(\ell\theta)\hspace{25pt}\text{(flat-sky)}
    \end{align}
  \item $s_a=s_b=2$ (e.g. shear-shear):
    \begin{align}\label{eq:xi22full}
      \xi^{ab}_\pm(\theta)&=\sum_\ell\frac{2\ell+1}{4\pi}\,C^{ab}_\ell d^\ell_{2,\pm2}(\theta)\hspace{12pt}\text{(full-sky)}\\\label{eq:xi22flat}
                      &=\int_0^\infty\frac{d\ell\,\ell}{2\pi}\,C^{ab}_\ell J_{2\mp2}(\ell\theta)\hspace{16pt}\text{(flat-sky)}
    \end{align}
\end{itemize}
In the following sections, we will specifically refer to the clustering correlation function in Eq. (\ref{eq:xi00flat}) as $\xi_{gg}$.

\subsubsection{Three-dimensional spatial correlation function}
In addition to the angular correlation functions, \ccl can also compute the three-dimensional spatial correlation function, $\xi(r)$, from the following transformation of the matter power spectrum:
\begin{equation}
\xi(r) = \frac{1}{2 \pi^2} \int_0^\infty dk \; k^2 P(k) \frac{\sin(kr)}{kr}.
\label{eq:xi3d}
\end{equation}
In the future \ccl will be expanded to incorporate the calculation of the higher-order multipoles needed to characterize the redshift-space three-dimensional correlation function in the presence of RSDs.

\subsection{Halo mass function}
\label{sec:hmfdef}

Being able to calculate the halo abundance as a function of mass is a necessary step to constrain cosmology with probes such as galaxy clusters \citep{Paranjape2014}. Modern cosmology makes extensive use of fitting functions in order to predict the evolution of halo abundances, which necessarily require derivation from cosmological simulations \citep{Tinker2008, Tinker2010, Angulo2012}. We implement halo mass functions with parameters fit to these simulations. The calculation of the halo mass function focuses around the spherical overdensity method of halo finding, in which a halo can be defined as a region of average density
\begin{equation}
  \bar{\rho}(r_{\Delta}) = \Delta_\mathrm{v} \times \bar{\rho}_{\mathrm{m}},
  \label{eq:halodef}
\end{equation}
equal to the overdensity parameter $\Delta_\mathrm{v}$ times the mean background density of the universe at a given redshift, $\bar\rho_{\mathrm{m}}(z)$, and with radius $r_{\Delta}$. Within the literature, the choice of $\Delta_\mathrm{v}$ can vary considerably, as observations focusing on the compact cores of halos often take much larger values of $\Delta_\mathrm{v}$ than the fiducial definition in most halo clustering studies, $\Delta_\mathrm{v} = 200$. We note that an alternative definition exists which utilizes the critical density of the universe, $\rho_{\mathrm{crit}}$, instead of the mean in Eq. (\ref{eq:halodef}); this introduces a simple conversion factor between the two definitions that must be accounted for. \ccl only accepts overdensity parameters with respect to the mean matter density, but we plan to allow for self-consistent handling of critical density based definitions in the future. 

The halo mass function is defined as
\begin{equation}
\frac{\mathrm{d}n}{\mathrm{d}M}=f(\sigma)\frac{\bar{\rho}_\mathrm{m}}{M}\frac{\mathrm{d}\ln{\sigma^{-1}}}{\mathrm{d}M},
\label{eq:halo_mass_function}
\end{equation}
where $n$ is the number density of halos of a given mass $M$ associated with the RMS variance of the matter density field, $\sigma^2$, at a given redshift and $f$ is a fitting function\footnote{Not to be confused with the linear growth rate of structure defined in Eq. (\ref{eq:lingrowthf}).}. \ccl makes predictions for the mass function in logarithmic mass bins, $dn/d\log_{10}{M}$, where the input is the halo mass $M$ and scale factor $a$.

The halo mass $M$ is related to $\sigma$ by first computing the radius $R$ that would enclose a mass $M$ in a homogeneous Universe at $z=0$:
\begin{equation}
  M=\frac{H_0^2}{2G}R^3\,\rightarrow \frac{M}{M_\odot}=1.162\times10^{12}\Omega_mh^2\,\left(\frac{R}{1\,{\rm Mpc}}\right)^3.
\end{equation}
The RMS density contrast in spheres of radius $R$ can then be computed as
\begin{equation}
  \sigma_R^2 = \frac{1}{2\pi^2}\int dk\,k^2\,P_{\rm lin}(k)\,\tilde{W}_R^2(k)
  \label{eq:sigR}
\end{equation}
where $P_{\rm lin}(k)$ is the linear matter power spectrum at $z=0$ and $\tilde{W}(kR)$ is the Fourier transform of a spherical top hat window function,
\begin{equation}
\tilde{W}_R(k) = \frac{3}{(kR)^3}[\sin(kR)-kR\cos(kR)].
\end{equation}

This is commonly related in terms of the mass inside of the Lagrangian scale of the halo, using the following transformation:
\begin{equation}
R = (3M/4\pi{\bar\rho_{\mathrm{m}}})^{1/3}.
\label{eq:lagrangemass}
\end{equation}
As a consequence, one can also define $\sigma_M$ as the RMS variance of the density field smoothed on some scale $M$, analogously to Eq. (\ref{eq:sigR}).

One commonly used halo mass function definition within the literature is the \citet{Tinker2010} fitting function. This fitting function has been developed using collisionless $N$-body simulation data, using halos identified by spherical overdensities. This is an extension of the \citet{Tinker2008} halo mass function, which is also included within \ccl as a comparative option. This fitting function assumes no change with respect to cosmological parameters except for changes in $\sigma_M(z)$\footnote{\citet{Tinker2008} stated that the difference in the mass function between adopting {\it WMAP}1 and {\it WMAP}3 cosmologies was within 5\%.}. Further, it includes a redshift scaling which is assumed to sharply end at a redshift of $z = 3$. This halo mass function is calibrated within the range of $10^{10.5} h\,\mathrm{M}_\odot \leq M \leq 10^{15.5} h\,\mathrm{M}_\odot$ at a redshift of $z = 0$.

For comparison purposes, we also have included the results of \citet{Angulo2012}, which uses the Millennium XXL simulation in order to study galaxy cluster scaling relations. As part of their study, they calculated their own best fit parameters for the \citet{Tinker2010} fitting function. While this additional halo mass function is available, it has not been extended to a broad range of overdensity parameter $\Delta_\mathrm{v}$, nor has it been extended beyond a redshift of $z = 0$.

The \citet{Tinker2008} fitting function uses the following parameterisation:
\begin{equation}
f(\sigma)=A\Big[\Big(\frac{\sigma}{b}\Big)^{-a}+1\Big]e^{-c/{\sigma}^2},
\end{equation}
where $A$, $a$, $b$, and $c$ are fitting parameters that have additional redshift scaling. This basic form is modified for the \citet{Angulo2012} formulation. The resulting form is
\begin{equation}
f(\sigma)=A\Big[\Big(\frac{b}{\sigma}+1\Big)^{-a}\Big]e^{-c/{\sigma_M}^2},
\end{equation}
where the only change is in the formulation of the second term. Note that the fitting parameters in the \citet{Angulo2012} formulation do not contain any redshift dependence and the use of it is primarily for testing and benchmarking purposes.

The \citet{Tinker2010} model parameterizes the halo mass function in terms of the peak height, $\nu \equiv \delta_c/\sigma_M$, where $\delta_c=1.686$ is the critical density for collapse (taken to be independent of cosmological model). The function is then re-expressed as
\begin{equation}
  f(\nu) = \alpha[1+(\beta\nu)^{-2\phi}]\nu^{2\eta}e^{-\gamma\nu^2/2}.
  \label{eq:tinkerf}
\end{equation}

\citet{Tinker2008,Tinker2010} quote 5\% accuracy of their parametrised mass functions, compared to the simulations used to calibrate them. This result is consistent with the work of \citet{Watson2013}, which also finds a $5\%$ level difference in comparison to the \citet{Tinker2008} fitting function. Further study will be required in the future in order to gain per cent level accuracy in determining the halo mass function.

We note that these halo mass functions, while implemented to high {\em numerical} accuracy in \ccl, carry their own uncertainties. It has not been significantly studied whether the halo mass function is universal with respect to changes in dark energy parameterisation or, in general, any other changes in cosmological parameters.

We also include the mass function from \cite{Sheth1999}:
\begin{equation}
f(\nu)=A\left[1+\frac{1}{(q\nu^2)^p}\right]\mathrm{e}^{-q\nu^2/2}\ ,
\label{eq:st_mf}
\end{equation}
with $p=0.3$, $q=0.707$ and $A\simeq 0.21616$, where $A$ is fixed such that the mass function is normalized.

This mass function was fitted to halos measured in $N$-body simulations where halos were identified with a cosmology-dependent overdensity criterion from the spherical-collapse model ($\Delta_\mathrm{v}\sim 300$ for $\Omega_\mathrm{m}\sim 0.3$ $\Lambda$CDM; $\Delta_\mathrm{v}\sim 178$ for $\Omega_\mathrm{m}\sim 1$). For the cosmology dependence, we use the fitting-formula of \cite{Bryan1998}
\begin{equation}
\Delta_\mathrm{v}(z)=\frac{1}{\Omega_\mathrm{m}(z)}\left(18\pi^2-82x-39x^2\right)\ ,
\label{eq:Deltav_Bryan}
\end{equation}
where $x=1-\Omega_\mathrm{m}(z)$. In addition, in \cite{Sheth1999} the relation between $M$ and $\nu$ was taken to include the cosmology-dependence of $\delta_\mathrm{c}(z)$, which derives from the spherical-collapse model. For this we use the fitting formula of \cite{Nakamura1997}:
\begin{equation}
\delta_\mathrm{c}(z)=\frac{3(12\pi)^{2/3}}{20}\left\{1+0.012299\log_{10}[\Omega_\mathrm{m}(z)]\right\}\ .
\label{eq:deltac_Nakamura}
\end{equation}

\subsection{Halo bias}
\label{sec:hbdef}

An important step in many interpretations of the halo model is to have a measure of the bias of dark matter halos, defined as the ratio of the halo power spectrum, $P_h(k)$, to the linear dark matter power spectrum,
\begin{equation}
  b^2(k) = \frac{P_h(k)}{P_{\mathrm{lin}}(k)}.
  \label{eq:halo_bias}
\end{equation}
This is implemented as a stand-alone function in \ccl and does not currently feed into predictions for galaxy or halo clustering described in Section \ref{ss:2point}.

As with measures of the halo mass function, high accuracy cosmological constraints require the use of numerical simulations to develop fitting functions and emulators. Here, we define halos as in the above sub-section. {\tt CCL} implements the halo bias fitting function results from \citet{Tinker2010}, though future improvements will likely require the use of emulator methods.

The \citet{Tinker2010} model parameterizes the halo bias in terms of the peak height and the critical density for collapse (similarly to Eq.~\ref{eq:tinkerf}) as
\begin{eqnarray}
  b(\nu) &=& 1 - A\frac{\nu^a}{\nu^a + {\delta_c}^a} + B\nu^b+C\nu^c,
\end{eqnarray}

\citet{Tinker2010} found a $\sim6\%$ scatter when determining the halo bias due to differences in simulations alone. There is remaining uncertainty to the {\em physical accuracy} of this model, as this parameterization does not consider any impact due to changes in the dark energy equation of state. As with the halo mass function, studies will be required to reach accuracy at the per cent level for any cosmological predictions \citep[e.g.][]{Gao2005, Schulz2006, Smith2007, Croton2007, Parfrey2011, Sunayama2016, Villarreal2017, Mao2018}.

\ccl can also make predictions for the halo bias from \cite{Sheth1999},
\begin{equation}
b(\nu)=1+\frac{1}{\delta_\mathrm{c}(z)}\left[q\nu^2-1+\frac{2p}{1+(q\nu^2)^p}\right]\ ,
\label{eq:st_bias}
\end{equation}
which can be derived using the peak-background split applied to Eq.~(\ref{eq:st_mf}). Similar to that equation, $p=0.3$, $q=0.707$ and $\delta_\mathrm{c}(z)$ is defined in Eq.~(\ref{eq:deltac_Nakamura}).


\subsection{Halo model}
\label{sec:halo_model}

In this section we review a basic halo model computation \citep{Seljak2000,Peacock2000,Cooray2002} of the cross-correlation between any two cosmological scalar fields. The calculation only requires knowledge of the halo profiles of the field in question. For example, in the case of the matter-density auto spectrum we need only know the halo density profiles. For the galaxy spectrum we would require knowledge of the number of, and distribution of, galaxies as a function of halo mass (the so-called halo-occupation distribution). In this simple form the halo model is approximate and makes the assumption that halos are \emph{linearly} biased with respect to the \emph{linear} matter field and also assumes that halos are spherical with properties that are determined solely by their mass. For the matter power spectrum, these assumptions mean that the matter power spectrum is only accurate to within a factor of two compared to that measured from numerical simulations \citep{Mead2015}. It is possible to go beyond these simplistic assumptions, and we direct the interested reader to \cite{Cooray2002,Smith2007,Giocoli2010,Smith2011} for this.

The eventual aim for \ccl is to have a halo model that can calculate the auto- and cross-spectra for any cosmological field combinations with parameters that can be taken either from numerical simulations or observational data. So far, we have only implemented the halo model calculation of the density power spectrum, but we keep the notation as general as possible in the following.

Consider two three-dimensional cosmological scalar fields $\rho_i$ and $\rho_j$, the cross power spectrum at a given redshift can be written as a sum of a two- and a one-halo term. The two-halo term accounts for power that arises due to the distribution of halos with respect to one another, while the one-halo term accounts for power that arises due to the internal structure of individual halos. These terms are given by
\begin{equation}
P_{2\mathrm{H},ij}(k)=P_{\mathrm{lin}}(k)
\prod_{n=i,j}\left[\int_0^\infty b(M)\frac{\mathrm{d}n}{\mathrm{d}M}W_n(M,k)\;\mathrm{d}M\right]\ ,
\label{eq:two_halo}
\end{equation}
and
\begin{equation}
P_{1\mathrm{H},ij}(k)=\int_0^\infty \frac{\mathrm{d}n}{\mathrm{d}M}W_i(M,k)W_j(M,k)\;\mathrm{d}M\ ,
\label{eq:one_halo}
\end{equation}
where $M$ is the halo mass, $\mathrm{d}n/\mathrm{d}M$ is the halo mass function defined in Eq.~(\ref{eq:halo_mass_function}) and $b(M)$ is the linear halo bias with respect to the linear matter density field, defined as the large-scale limit of Eq.~(\ref{eq:halo_bias}). The full halo model power is then simply the sum
\begin{equation}
P_{\mathrm{HM},ij}=P_{2\mathrm{H},ij}+P_{1\mathrm{H},ij}\ .
\label{eq:halo_model_power}
\end{equation}

Eqs.~(\ref{eq:two_halo}) and (\ref{eq:one_halo}) contain the (spherical) Fourier transform of the halo profile, or halo ``window function'':
\begin{equation} 
W_i(M,k)=\int_0^\infty4\pi r^2\frac{\sin(kr)}{kr}\rho_{\mathrm{H},i}(M,r)\;\mathrm{d}r\ ,
\label{eq:window_function}
\end{equation}
where $\rho_{\mathrm{H},i}(M,r)$ is the radial profile for the field $i$ in a host halo of mass $M$. For example, if one is interested in calculating the matter power spectrum then $\rho_{\mathrm{H},i}(M,r)$ would be the density \emph{contrast} profile of a halo of mass $M$.

By default, in the halo model calculation in \ccl we use the mass function and bias from \cite{Sheth1999}. Note that the halo mass function and bias \emph{must} satisfy the following properties for the total power spectrum to have the correct large-scale limit, which is that the power should revert to the linear power spectrum
\begin{equation}
\frac{1}{\bar\rho_\mathrm{m}}\int_0^\infty M\frac{\mathrm{d}n}{\mathrm{d}M}\;\mathrm{d}M=1\ ,
\label{eq:mf_normalisation}
\end{equation}
and
\begin{equation}
\frac{1}{\bar\rho_\mathrm{m}}\int_0^\infty Mb(M)\frac{\mathrm{d}n}{\mathrm{d}M}\;\mathrm{d}M=1\ .
\label{eq:bias_normalisation}
\end{equation}
If one uses a mass function and bias pair that are related via the peak-background split formalism \citep{Mo1996,Sheth2001} then these conditions are automatically satisfied. In words, these equations enforce that all matter is associated with a halo. In the convention used in \ccl the units of $P_{\mathrm{HM},ij}(k)$ will be exactly the units of $\rho_i\rho_j / \mathrm{Mpc}^3$. The units of the $W_i$ are those of the field $\rho_i$ multiplied by volume. 

For the matter power spectrum, we use the halo profiles of \citeauthor*{Navarro1997} (NFW; \citeyear{Navarro1997}):
\begin{equation}
\rho_\mathrm{H}(M,r)\propto\frac{1}{r/r_\mathrm{s}(1+r/r_\mathrm{s})^2}\ .
\label{eq:NFW_profile}
\end{equation}
The NFW profile is written in terms of a scale radius $r_\mathrm{s}$. The constant of proportionality is fixed by the condition that the halo has total mass $M$ integrated within the virial radius $r_\mathrm{v}$. This radius is in turn set such that the halo has a fixed density $\Delta_\mathrm{v}$ with respect to the mean. Hence, the following relation holds between mass, density and radius:
\begin{equation}
M=4\pi r_\mathrm{v}^3\Delta_\mathrm{v}\bar\rho_\mathrm{m}\ .
\label{eq:virial_radius}
\end{equation}
Finally, the scale radius is usually expressed in terms of the mass-dependent halo concentration parameter $c(M)=r_\mathrm{v}/r_\mathrm{s}$.

We use the mass-concentration relation from \cite{Duffy2008} appropriate for the full sample of halos defined using a virial $\Delta_\mathrm{v}$ criterion
\begin{equation}
  c(M,z)=7.85\left(\frac{M}{M_\mathrm{p}}\right)^{-0.081}(1+z)^{-0.71}\ ,
  \label{eq:cDuffy}
\end{equation}
with $M_\mathrm{p}=2\times10^{12}\,h^{-1}\mathrm{M}_\odot$.
In order to be consistent one \emph{must} use values of $\Delta_\mathrm{v}$ and $c(M)$ that are consistent with the halo definition used for the halo mass function and bias. This consistency check is enforced by \ccl and we do not allow mixing of halo properties defined with different overdensity criteria.

\subsection{Photometric redshifts}

Redshifts of LSST galaxies will be obtained via photometry. Therefore, performing any cosmological analysis which incorporates redshift information requires a model for the probability of measuring a photometric redshift $z_{\rm ph}$ for an object with true redshift $z_{\rm t}$.  In order to maintain agnosticism towards the optimal model, and hence to allow for the future inclusion of advancements from ongoing research, \ccl allows the user to flexibly input a photometric redshift model. In addition, for ease of use, \ccl provides the option of using a built-in function for a simple Gaussian photometric redshift probability distribution.

We define $dN/dz$ as the true redshift distribution of a sample of galaxies, and $dN^i/dz$ as the true redshift distribution of those galaxies that belong to photometric redshift bin $i$. The photometric redshift model can then be used, for example, when computing $dN^i/dz$ as given by:
\begin{equation}
\frac{dN}{dz}^i = \frac{\frac{dN}{dz}\int_{z_i}^{z_{i+1}} dz' p(z,z')}{\int_{z_{\rm min}}^{z_{\rm max}}dz \frac{dN}{dz} \int_{z_i}^{z_{i+1}}dz' p(z, z')}
\label{photoz}
\end{equation}
where $p(z,z')$ is the photometric redshift probability distribution, and $z_{i}$ and $z_{i+1}$ are the photo-$z$ edges of the bin in question. In the case of the simple Gaussian photometric redshift model for which native support is included in \ccl , $p(z, z')$ is given by
\begin{equation}
p(z,z') = \frac{1}{\sqrt{2 \pi}\sigma_z} \exp\left(-\frac{(z-z')^2}{2\sigma_z^2}\right),
\label{pz_gauss}
\end{equation}
where the user can set the value of $\sigma_z$, or indeed any arbitrary function may be provided for $p(z,z')$.

\section{Implementation of high-accuracy cosmological functions}
\label{sec:implement}

In this section, we note some of the assumptions and implementation details that are relevant when making accurate cosmological predictions. In general, we use the publicly available {\tt GSL} library\footnote{\url{https://www.gnu.org/software/gsl/}} to perform all of the integrations and interpolations. Most interpolations use the {\tt gsl\_interp\_akima} method, and the power spectra interpolation use a bicubic spline provided by {\tt gsl\_interp2d\_bicubic}. We work with double precision quantities throughout. The validation tests performed for {\tt CCL} are described in detail in Section \ref{sec:validation}.

\subsection{Background functions \& growth of perturbations}
\label{sec:distances}

Cosmological predictions require making assumptions on the values of several physical constants, as defined in the previous sections. \ccl adopts physical constant values from CODATA 2014 \citep{CODATA14} with the exception of the solar mass, which is not provided by this source and which we take from IAU 2015 \citep{IAU15}.

We have performed a comparison of the physical constants used in \ccl to those used in {\tt GSL} and \class as well as published sources such as the NIST\footnote{\url{https://www.nist.gov}} Handbook and Particle Data Group (PDG) Review of Particle Physics \citep{Beringer:1900zz}. In general, we have found better than $10^{-4}$ agreement except for the gravitational constant and the value of the solar mass, where the discrepancies are nevertheless $<10^{-3}$. Notice that the value of these constants enters into the definition of the critical density (Eq. \ref{eq:rhocrit}).

\subsection{Matter power spectrum}\label{ssec:mpspec}

For speed, the initialization of a cosmological model within \ccl performs initial computations of the linear and non-linear matter power spectra, which are then interpolated to be used whenever required. A bicubic spline is performed in two variables. The first one is the logarithmically-spaced wavenumber. For the scale factor, we adopt a hybrid spacing scheme where this quantity is linearly-spaced for $a>0.1$ and logarithmically-spaced otherwise. The goal of this hybrid scheme is to allow sufficiently fine sampling at low redshift for LSST observables, while at the same time allowing for predictions for CMB lensing without significantly slowing down the computations, as would result from a linear-spacing throughout. The spline interpolation causes some precision loss in the power spectra output (compared to, for example, direct outputs from \class or the Cosmic Emulator) which is quantified in Section \ref{sec:validation}.

We introduce a maximum value $k$ (in units of $\text{Mpc}^{-1}$) up to which we evaluate the power spectra for interpolation; we call this parameter {\tt K$\_$MAX$\_$SPLINE}. A separate {\tt K$\_$MAX} parameter sets the limit of evaluation of the matter power spectrum. The range between {\tt K$\_$MAX$\_$SPLINE}~$<k<$~{\tt K$\_$MAX} is evaluated by performing a second order Taylor expansion in $\ln k$.

The Taylor expansion is implemented as follows: first, we compute the first and second derivative of $\ln P(k,z)$ at $k_0={\rm \tt K\_MAX\_SPLINE}-2\Delta\ln k$ via finite difference derivatives using {\tt GSL}. The fiducial choice for $\Delta\ln k$ is $10^{-2}$. We then apply a second order Taylor expansion to extrapolate the matter power spectrum to $k>$~{\tt K$\_$MAX$\_$SPLINE}. The Taylor expansion gives
\begin{eqnarray}
  \ln P(k,z) &\simeq& \ln P(k_0,z) + \frac{d\ln P}{d\ln k}(k_0,z) (\ln k-\ln k_0)  \nonumber\\
  &+& \frac{1}{2}  \frac{d^2\ln P}{d\ln k^2}(k_0,z) (\ln k-\ln k_0)^2.
  \label{eq:NLPSTaylor}
\end{eqnarray}

We also extrapolate the power spectrum at small wavenumbers. In this case, we introduce the parameter {\tt K$\_$MIN$\_$SPLINE}, the wavenumber below which the power spectra are obtained by a power-law extrapolation with index $n_s$:
\begin{eqnarray}
  \log P(k<{\tt K\_MIN\_SPLINE},z) = \nonumber\\
  \log P({\tt K\_MIN\_SPLINE},z) + \nonumber\\
  n_s (\log k-\log{\tt K\_MIN\_SPLINE})
\end{eqnarray}
Note that an additional parameter, {\tt K\_MIN}, sets the minimum $k$ for integrations. This is set to {\tt K\_MIN}$=5\times 10^{-5}\,\text{Mpc}^{-1}$.

The value adopted for {\tt K\_MIN\_SPLINE} depends on the choice of power spectrum method is not accessible by the user. For {\tt CLASS} and the nonlinear power spectrum, we adopt {\tt K\_MIN\_SPLINE} that coincides with the smallest wavenumber output by {\tt CLASS}, {\tt K\_MIN\_SPLINE}$=7\times 10^{-6}$ Mpc$^{-1}$. Hence, in practice, no extrapolation is occurring in this case. For BBKS, the power spectrum is computed analytically at all $k$, there is no extrapolation. For the \citet{1998ApJ...496..605E} implementation, the splines of the power spectrum span {\tt K\_MIN}$<k<${\tt K\_MAX\_SPLINE}, so there is only extrapolation at high $k$. For the nonlinear matter power spectrum from the emulator, {\tt K\_MIN\_SPLINE} and {\tt K\_MAX\_SPLINE} are set to fixed values that are determined from the range of validity of the emulator:  {\tt K\_MIN\_SPLINE}$=10^{-3}$ Mpc$^{-1}$ and {\tt K\_MAX\_SPLINE}$=5$ Mpc$^{-1}$.

\subsection{Angular power spectra}

Different numerical approaches have been implemented in the library in order to expedite the computation of angular power spectra. We describe these here. In all cases, to avoid calculating power spectra at all integer values of $\ell$, by default \ccl samples the power spectra at particular values of $\ell$ and interpolates between them to obtain the result at the $\ell$ values requested by the user. The sampling scheme is based on a combination of logarithmic samples at low-$\ell$ and linear samples at high-$\ell$, although the particulars of the sampling scheme can be configured by the user. A cubic-spline method is used to do the interpolation.

\subsubsection{Limber approximation}

As shown in Section \ref{sssec:2pt.pspec}, computing each transfer function contributing to a given power spectrum involves a radial projection (i.e. an integral over redshift or $z$ or $\chi$), and thus computing full power spectra consists of a triple integral for each $\ell$. This can be computationally intensive, but can be significantly simplified in certain regimes by using the Limber approximation \citep{1954ApJ...119..655L,2004PhRvD..69h3524A}, given by:
\begin{equation}
 j_\ell(x)\simeq\sqrt{\frac{\pi}{2\ell+1}}\,\delta\left(\ell+\frac{1}{2}-x\right).
\end{equation}
This eliminates the integrals associated with each of the two transfer functions, accelerating the calculation significantly.

Thus, for each $k$ and $\ell$ we define a radial distance $\chi_\ell\equiv(\ell+1/2)/k$, with corresponding redshift $z_\ell$. Substituting this in the expressions presented in Section \ref{sssec:2pt.pspec}, the power spectrum can be computed as a single integral:
\begin{equation}\label{eq:limber}
 C^{ab}_\ell=\frac{2}{2\ell+1}\int_0^\infty dk\,P_\delta\left(k,z_\ell\right)
 \tilde{\Delta}^a_\ell(k)\tilde{\Delta}^b_\ell(k)
\end{equation}
where
\begin{align}
 &\tilde{\Delta}_\ell^{\rm D}(k)=p_z(z_\ell)\,b(z_\ell)\,H(z_\ell)\\
 &\tilde{\Delta}_\ell^{\rm RSD}(k)=
 \frac{1+8\ell}{(2\ell+1)^2}\,p_z(z_\ell)\,f(z_\ell)\,H(z_\ell)-\\\nonumber
 &\hspace{48pt}\frac{4}{2\ell+3}\sqrt{\frac{2\ell+1}{2\ell+3}}p_z(z_{\ell+1})\,f(z_{\ell+1})\,H(z_{\ell+1})\\
 &\tilde{\Delta}_\ell^{\rm M}(k)=3\Omega_{M,0}H_0^2\frac{\ell(\ell+1)}{k^2}\,
 \frac{(1+z_\ell)}{\chi_\ell}W^{\rm M}(z_\ell)\\
 &\tilde{\Delta}_\ell^{\rm L}(k)=\frac{3}{2}\Omega_{M,0}H_0^2\sqrt{\frac{(\ell+2)!}{(\ell-2)!}}\frac{1}{k^2}\,
 \frac{1+z_\ell}{\chi_\ell}W^{\rm L}(z_\ell)\\
 &\tilde{\Delta}_\ell^{\rm IA}(k)=\sqrt{\frac{(\ell+2)!}{(\ell-2)!}}\frac{p_z(z_\ell)\,b_{\rm IA}(z_\ell)f_{\rm red}(z_\ell)H(z_\ell)}{(\ell+1/2)^2}.
\end{align}
The Limber approximation works best for wide radial kernels and high $\ell$. The integration in Eq. (\ref{eq:limber}) is performed via Gauss-Kronrod quadrature, as are the integrals needed to estimate the lensing and magnification window functions (Eqs. \ref{eq:window_mag} and \ref{eq:window_shear}). The integration limits for Eq. \ref{eq:limber} are adapted to the shape of the window functions entering $\tilde{\Delta}^{a,b}$, with absolute limits given by the {\tt K\_MIN} and {\tt K\_MAX} parameters described in Section \ref{ssec:mpspec}. 

\subsubsection{Beyond Limber: \texttt{Angpow}}
\label{sec:angpow}

The computation of the $C^{ab}_\ell$ without the Limber approximation is extremely costly in terms of computing time using this method, particularly if one wants to extensively explore a full cosmological parameter space. To overcome this issue, \ccl provides fast non-Limber predictions by calling the \texttt{Angpow} software \citep{2017A&A...602A..72C}. 

The angular power spectrum for two tracers $C_{\ell}^{ab}$ is computed in \texttt{Angpow} according to the following expression
\begin{equation}
  C_{\ell}^{ab} = \iint_0^\infty \mathrm{d} z \mathrm{d} z^\prime  p_{z_1}(z_1) p_{z_2}(z^\prime) \int_0^\infty \mathrm{d} k\ f_{\ell}(z, k) f_{\ell}(z^\prime, k).
  \label{eq-clz1z2-obs}
\end{equation}
The auxiliary function $f_\ell(z,k)$ is defined as
\begin{equation}
f_\ell(z,k) \equiv  \sqrt{\frac{2}{\pi}}\  k \sqrt{P_\delta(k,z)}\ \widetilde{\Delta}_\ell(z,k)\label{eq-fell-func}
\end{equation}
with $\widetilde{\Delta}_\ell(z,k)$ the function describing the physical processes such as matter density fluctuations and redshift-space distortions as described for instance in \citet{2008cmb..book.....D,2009PhRvD..80h3514Y,2010PhRvD..82h3508Y, 2011PhRvD..84d3516C,2011PhRvD..84f3505B}.

The \texttt{Angpow} version delivered with \ccl can only model galaxy clustering tracers (no gravitational lensing), and this without the magnification lensing term (Eq.~\ref{eq:deltaM}). The incorporation of those transfer functions is left for future work, but in principle this is a straightforward extension of \texttt{Angpow}. For galaxy clustering tracers we define $\widetilde{\Delta}_\ell(z,k)$ as 
\begin{equation}
 \widetilde{\Delta}_\ell(z,k) \equiv b(z) j_\ell(k \chi(z)) - f(z) j_\ell^{\prime\prime}(k \chi(z)) 
\end{equation}
with $j_\ell(x)$ and $j_\ell^{\prime\prime}(x)$ the spherical Bessel function of order $\ell$ and its second derivative, and $f(z)$ the growth rate of structure.

In \texttt{Angpow}, the inner integral in $k$ is computed first.
To conduct such computation where the integrand is a highly oscillating function, the 3C-algorithm described in details in \citet{2017A&A...602A..72C} is used. In brief, it relies on the projection of the oscillating $f_\ell(z_,k)$ onto a Chebyshev series of order $2^N$. The product of the two Chebyshev series is performed with a $2^{2N}$ Chebyshev series; then the integral is computed using Clenshaw-Curtis quadrature. Finally, the integrals over $z$ are performed once again via an optimised Clenshaw-Curtis quadrature. All the Chebyshev expansions and the Clenshaw-Curtis quadrature are performed via the \textit{Discrete Cosine Transform} of type I from the DCT-I fast transform of the FFTW library \citep{FFTW}.

As in the general case the Limber approximation is valid at high $\ell$ values, the \ccl user can define an $\ell$ threshold to switch from the non-Limber computation to the faster Limber approximation.

\subsection{Correlation functions}

Computing the angular correlation functions essentially involves performing a linear transformation on the power spectra to go from harmonic to real space. The exact equations (\ref{eq:xi00full}, \ref{eq:xi02full}, \ref{eq:xi22full}) relating both quantities involve carrying out $N_\theta\times\ell_{\rm max}$ operations, where $\ell_{\rm max}\sim 10^{4-5}$ is the maximum multipole needed to achieve convergence and $N_\theta$ is the number of angular scales $\theta$ at which the angular correlation function needs to be computed. Thus, evaluating these expressions directly can become prohibitively slow and should be avoided except in regimes where other approximations are not valid. In particular \ccl only supports the brute-force evaluation of these equations for correlations involving at least one spin-0 field. The default method in \ccl is to use the flat-sky approximation and evaluate the Hankel transforms (Eqs. \ref{eq:xi00flat}, \ref{eq:xi02flat}, \ref{eq:xi22flat}).

\ccl provides two methods to compute Hankel transforms:

\paragraph{\bf Brute-force integration.} \ccl allows users to compute Hankel transforms by brute-force integration over the Bessel functions using an adaptive Gauss-Kronrod algorithm. The oscillating nature of these functions makes this method slow and not appropriate for likelihood-sampling.

Thus, despite the higher precision of the brute-force integration approach, the preferred method to compute correlation functions is through the use of {\tt FFTlog} (see below), and we support the brute-force method primarily for testing and validation.

\paragraph{\bf FFTlog.} The public code {\tt FFTlog}\footnote{\url{http://casa.colorado.edu/~ajsh/FFTLog/}} is able to compute fast Hankel transforms through the assumption that the kernels of these transforms are periodic functions in logarithmic space. The Hankel transform can then be solved using Fast Fourier Transforms at a much lower computational expense than brute-force integration \citep{Hamilton2000,Talman2009}. \ccl incorporates a version of the {\tt FFTlog} method with only minor modifications from the original. The only potential drawback of this method is the need to sample the kernels (i.e., the $C_\ell$) on very small scales to ensure the convergence of the method. To do this, \ccl extrapolates the power spectrum as a power law, assuming $C_\ell\propto\ell^\beta$, with a tilt $\beta$ estimated from the logarithmic slope of the two last values of the $C_\ell$ provided as input. We have verified that this method agrees with the brute-force integration to well within cosmic-variance uncertainties.

We should also note that other approaches relating the correlation functions directly with the three-dimensional matter power spectrum (e.g. \citealt{2017ApJ...845...28C}) could be useful in accelerating this computation, and we will explore these in the future.

\subsection{Halo mass function, halo bias \& halo model}

The computation of the halo mass function requires obtaining the derivative of $\sigma^{-1}$ with respect to mass, Eq. (\ref{eq:halo_mass_function}). These derivatives are calculated utilizing a spline interpolation of $\sigma(M)$. These splines cover the range from $10^6$ to $10^{17} M_\odot$. For each value of $\log(M)$ in our spline evaluation, we calculate the value of $\sigma(M)$ half a step in either direction. We use the difference compared to the mass spacing to calculate an approximate derivative, which is then used in the spline interpolation. The precision of this method was established for the halo mass function within the mass range explored by \citet{Tinker2010} and we give details on this in the next section. We note that the accuracy is reduced at the edges of these splines and exploring extreme mass ranges may require changes in the parameters to initialize these splines.

In order to accommodate a wide range of values of the overdensity parameter $\Delta_\mathrm{v}$, we have generated a spline interpolation between best fit values as defined by \citet{Tinker2008} and \citet{Tinker2010}. This covers a dynamic range from $\Delta_\mathrm{v}=200$ to $3200$, with respect to the mean density. Within this range, we interpolate in the space of the fit parameter and $\log\Delta_\mathrm{v}$ using Akima interpolation built from piecewise third order polynomials. We have chosen this rather than the fitting formulas utilized in \citet{Tinker2010} in order to assure high precision match to the Tinker halo mass function when choosing a value of $\Delta_\mathrm{v}$ directly from the paper. 

Calculations required to make predictions for halo bias are analytical and are thus implemented in \ccl. In the case of the halo model, this phenomenological approach to modeling the matter power spectrum requires us to perform the integrations of the two-halo term (Eq. \ref{eq:two_halo}), the one-halo term (Eq. \ref{eq:one_halo}) and the window function (Eq. \ref{eq:window_function}). For both Eq.~(\ref{eq:two_halo}) and Eq.~(\ref{eq:one_halo}) we use {\tt GSL$\_$INTEG$\_$GAUSS41} to perform the integration between the limits of $10^{7}$ and $10^{17}$ solar masses with a relative error tolerance of $10^{-4}$. Achieving the correct $k\to0$ limit for the two-halo term, which should be exactly the linear power spectrum, is difficult numerically because of the large amount of mass contained in low-mass halos according to most popular mass functions. We deal with this for an arbitrary lower mass limit by enforcing the large-scale limiting behavior of the halo mass function (Eq.~\ref{eq:mf_normalisation}) by adding the mass missing from the integral as a delta function in mass at the lower limit in the two-halo integral in Eq.~(\ref{eq:two_halo}). For NFW haloes (Eq.~\ref{eq:NFW_profile}) the integral required for the window function (Eq.~\ref{eq:window_function}) is analytical:
\begin{equation}
\begin{split}
W_\delta(M,k) = &\, 4\pi r_\mathrm{s}^3 \times \\
& \left\{\sin(kr_\mathrm{s})\left[\mathrm{Si}(\{1+c\}kr_\mathrm{s})-\mathrm{Si}(kr_\mathrm{s})\right]\right. \\
+ & \cos(kr_\mathrm{s})\left[\mathrm{Ci}(\{1+c\}kr_\mathrm{s})-\mathrm{Ci}(kr_\mathrm{s})\right] \\
- & \left.\frac{\sin(kr_\mathrm{v})}{(1+c)kr_\mathrm{s}}\right\}\ ,
\label{eq:NFW_Fourier}
\end{split}
\end{equation}
where $\mathrm{Si}(x)$ and $\mathrm{Ci}(x)$ are the sine and cosine integral functions and $c$ is the concentration parameter defined in Section \ref{sec:halo_model}.

\subsection{Massive neutrinos}

When initializing a cosmology with massive neutrinos within \ccl , the user can provide either a single value for $m_\nu$, corresponding to a sum of the masses of three neutrinos, or a set of three values, corresponding directly to the three masses. In the former case, one can also specify how the sum of masses should be split for calculations. The default behavior of \ccl is to split the sum into three masses which are consistent with the normal neutrino mass hierarchy, but an inverted hierarchy or equal splitting can also be requested. (For a review of the neutrino mass hierarchies and relevant particle physics results, see for example \citealt{Gerbino2017, Lesgourgues2012}.)

For equal splitting, it is clearly trivial to compute the three neutrino masses. If splitting with respect to the normal or inverted hierarchy is desired, the mass calculation of the three masses is only marginally more complicated. The relevant known quantity which has been determined via particle physics experiments is the square of the difference of neutrino masses (up to a sign for one of the differences, hence the two possible hierarchies, see \citealt{Lesgourgues2012, Gerbino2017}). Because we know the square of the differences rather than the differences themselves, we must solve a set of quadratic equations for the neutrino masses. This is accomplished via a simple implementation of Newton's method, which converges to within machine precision in a few iterations.

Having then a set of three neutrino masses, we check which of the corresponding neutrino species is non-relativistic today ($m_\nu>0.00017$, \citealt{Lesgourgues2012}), and obtain the number of massive neutrinos in the cosmology. We use this, along with $N_{\rm eff}$, to set the number of relativistic neutrinos species, which is required in computing $\Omega_\gamma$ and $\Omega_{\nu, {\rm rel}}$. We must be careful in doing so, as only for massive neutrinos do we modify the relationship between the temperature of the CMB and the neutrino temperature as described following Eq.~(\ref{Omnu}) above. The value of $N_{\nu, {\rm rel}}$ consistent with the user-provided $N_{\rm eff}$ is given by:
\begin{equation}
N_{\nu, {\rm rel}} = N_{\rm eff} - \left(T_{\rm NCDM}\right)^{4} \left(\frac{4}{11}\right)^{-\frac{4}{3}} N_{\nu, {\rm m}}.
\label{Nnurel}
\end{equation}

In Eq.~(\ref{Omnu}) above, we specify how $\Omega_{\nu, {\rm m}}$ is computed for a given cosmology with massive neutrinos. Within this expression is a phase-space integral:
\begin{equation}
\int_0^{\infty} dx \, x^2 \frac{\sqrt{x^2 + \tilde{m}^2}}{\exp(x) + 1}.
\label{phasespacenu}
\end{equation}
At high and low $\tilde{m}$, corresponding to high and low mass neutrinos, this integral need not be evaluated numerically. At high $\tilde{m}$, we set the integral equal to $5\zeta(3)\tilde{m}/(18\pi^4)$ (where $\zeta$ is the Riemann zeta function), while at low $\tilde{m}$, it goes to $\frac{7}{8}$. The $\tilde{m}$ values at which these approximations are taken can be set by the user. Outside of the regime in which these approximations are valid, the integral is computed numerically using {\tt GSL}, splined, and stored such that for a single cosmology it must only be computed once.

It may sometimes be preferable or necessary to specify a cosmology in terms of $\Omega_{\nu, {\rm m}}$ instead of $m_\nu$. To facilitate this, \ccl includes a convenience function which returns $m_\nu$ given $\Omega_{\nu, {\rm m}}$. This is achieved via the relationship (see, e.g., \citealt{Lesgourgues2012})
\begin{equation}
\sum m_\nu = 93.14 {\rm eV} \times \Omega_{\nu, {\rm m}}
\label{summnu_om}
\end{equation}
and then by splitting $\sum m_\nu$ into three neutrinos masses using the convention given by the user (the default being the normal mass hierarchy).

\section{Validation}
\label{sec:validation}

Our goal in building \ccl was to ensure that all outputs are validated to a well-established high level of numerical accuracy. We described the core of our validation procedure in Section \ref{sec:intro}. Validation was achieved by performing different types of tests of \ccl outputs. When possible, we established the accuracy of \ccl against known analytic solutions. As there are few cases of observables and cosmologies for which there is such an analytic prediction, this is often not sufficient for our purposes. In one specific case (Section \ref{ss:cosmicemu}), we can compare the \ccl outputs against numerical simulations and there is a specific threshold of accuracy that needs to be achieved. Most commonly, we compare \ccl outputs against one or multiple independent implementations obtained for the same cosmology. For such cases, we occasionally know the independent implementation to be more accurate. When this is not the case, we describe the differences between the implementations made by \ccl and the independent benchmark code. Those independent implementations are provided and within the \ccl repository together with our main library. For each feature, we define and quantify a numerical accuracy parameter, $\mathcal{A}$, in the following sub-sections, which describes the relative or absolute difference between the \ccl prediction and the independent one.

The ultimate goal was to guarantee that any numerical uncertainty in the predictions for correlation functions are within a fraction of the expected statistical uncertainty for LSST. Moreover, we ensure that any prediction of the matter power spectrum, necessary for predicting cross-correlations between probes, has a well-established numerical accuracy. In this section, we document the numerical accuracy achieved for each observable and demonstrate that our overall goal has been achieved. We emphasise that the hereby presented tests pertain to {\it numerical} accuracy alone, while details of the {\it physical} accuracy of each model are provided in Section \ref{sec:models}. 

There are two cases where \ccl is calling external codes to perform the computations. {\tt CLASS} and the Cosmic Emulator, described in Section \ref{sec:models}, are used by \ccl in making power spectrum predictions. In doing so, and to improve on the speed of the code, power spectra from these codes are tabulated and interpolated. To ensure that this procedure does not introduce any significant deviations compared to the direct outputs of those codes, we compare \ccl power spectra outputs to {\tt CLASS} in sub-sections \ref{ss:classval} and  \ref{ss:classval2}, and to the simulated power spectra used to calibrate the Cosmic Emulator in sub-section \ref{ss:cosmicemu}.

\begin{sidewaystable*}[!htp]
  \vskip 4cm
  \centering
  \begin{tabular}{ l|c c c c c}
    \hline
    Quantity & Equation/ & Cosmologies & Range & Agreement with & Figure \\
    & Reference &  & & benchmarks, $\mathcal{A}$ & \\
    \hline
    Comoving radial distance, $\chi$ & (\ref{eq:comrdist}) & CCL1-5,7-11 & $0.01 \leq z\leq 1000$ &  $5\times 10^{-7}$ & Fig. \ref{fig:distancegrow}\\
    Growth factor, $D$ & (\ref{eq:growth}) & CCL1-5 &  $0.01 \leq z\leq 1000 $ &  $6\times 10^{-6}$ & Fig. \ref{fig:distancegrow}\\
    $\sigma(M)$ (BBKS) & (\ref{eq:sigR}) & CCL1-3 &  $10^{10}\leq M/{\rm M}_\odot\leq 10^{16}$ &  $3\times 10^{-5}$ & Fig. \ref{fig:hmf}\\
    $\log[\sigma^{-1}(M)]$ (BBKS) & (\ref{eq:tildesig}) & CCL1 &  $10^{10}\leq M/{\rm M}_\odot\leq 10^{16}$ &  $10^{-3}$ & Fig. \ref{fig:hmf}\\
    $\mathcal H \equiv \log[(M^2/\bar{\rho}_m)dn/dM]$  & (\ref{eq:newhmf}), \citet{Tinker2010} & CCL1 & $10^{10}\leq M/{\rm M}_\odot\leq 10^{16}$ \& $z=0$ & $5\times 10^{-5}$ & Fig. \ref{fig:hmf}\\
    $P(k)$ (BBKS) & (\ref{eq:bbks}) & CCL1-3 & $10^{-3}\leq k/(h/{\rm Mpc})\leq 10$ \& $0\leq z\leq 5$ &  $10^{-5}$ & -\\
    $P(k)$ (Eisenstein \& Hu) & \citet{1998ApJ...496..605E}  & CCL1 & $10^{-3}\leq k/(h/{\rm Mpc})\leq 10$ \& $z=0$ & $10^{-5}$ & -\\
    $P(k)$ ({\tt CLASS} linear \& HaloFit)$^\star$ & \citet{CLASS_halofit}  & see Table 5 & $10^{-3}\leq k/{\rm Mpc}\leq 20$ \& $z=\{0,2\}$  &$\sim 10^{-3}$ & Figs. \ref{fig:NLextrapol} , \ref{fig:power_nu}, \ref{fig:power_paramspace} \& \ref{fig:power_paramspace_z2} \\
    $P(k)$ (CosmicEmu $w$CDM)$^\dag$ & \citet{Lawrence17} & M1,M3,M & $10^{-3}\leq k/{\rm Mpc}^{-1}\leq 5$ \& $z=0$  & $10^{-2}$ & Fig. \ref{fig:emuacc} \\
    & & M6,M8,M10 & & & (left panel)\\
    $P(k)$ (CosmicEmu $\nu$CDM)$^\dag$ & \citet{Lawrence17} & M38,M39,M40 & $10^{-3}\leq k/{\rm Mpc}^{-1}\leq 5$ \& $z=0$ & $3\times 10^{-2}$ & Fig. \ref{fig:emuacc} \\
    &  & M42 &  &  & (right panel)\\
    $P(k)$ (Halo model) & \citet{Cooray2002} & CCL1, {\it WMAP7} & $10^{-4}\leq k/h{\rm Mpc}^{-1}\leq 10^{2}$ \& $z=0,1$ & $10^{-3}$ & Fig. \ref{fig:halo_model_benchmark}\\
     &  & {\it Planck} 2013 &  & \\
    $P(k)$ (baryonic) & (\ref{eq:bcm}), \citet{Schneider15} &  - & $10^{-5}\leq k/h{\rm Mpc}^{-1}\leq 10$ \& $z=0$ & $10^{-12}$ & -\\
    $C_\ell$ clustering & (\ref{eq:cls}),(\ref{eq:transfer_nc})& CCL6 &$2 \leq \ell\leq 3000$ &  $0.1\sigma_\ell$  & Fig. \ref{fig:cls_limber}\\
    $C_\ell$ weak lensing & (\ref{eq:cls}),(\ref{eq:transfer_lensing})& CCL6 &$2 \leq \ell\leq 3000$ &  $0.1\sigma_\ell$  & Fig. \ref{fig:cls_limber}\\
    $C_\ell$ gxy-gxy lensing & (\ref{eq:cls}),(\ref{eq:transfer_nc}),(\ref{eq:transfer_lensing})& CCL6 &$2 \leq \ell\leq 3000$ &  $0.1\sigma_\ell$ & Fig. \ref{fig:cls_limber}\\
    $C_\ell$ intrinsic alignments & (\ref{eq:cls}),(\ref{eq:transfer_ia})& CCL6 &$2 \leq \ell\leq 3000$ &  $0.1\sigma_\ell$  & -\\
    $C_\ell$ CMB lensing auto &(\ref{eq:cls}),(\ref{eq:cmblens}) & CCL6 & $2 \leq \ell\leq 3000$& $0.1\sigma_l$  & Fig. \ref{fig:cls_cmblens}\\
    $C_\ell$ CMB lensing cross &(\ref{eq:cls}),(\ref{eq:transfer_nc}),(\ref{eq:transfer_lensing}),(\ref{eq:cmblens}) & CCL6 & $2 \leq \ell\leq 3000$& $0.1\sigma_\ell$  & Fig. \ref{fig:cls_cmblens}\\
    $\xi_{\pm},\xi_{gg},\xi_{ggl}$ & (\ref{eq:xi22flat}),(\ref{eq:xi02flat}),(\ref{eq:xi00flat}) & CCL6 & $0.01< \theta/{\rm deg}< 5$&  $0.5\sigma_{\rm LSST}$ & Figs. \ref{fig:corrval} and \ref{fig:corrval2}\\
    3D correlation$^\S$, $\xi$ & (\ref{eq:xi3d}) & CCL1-3 & $0.1<r/{\rm Mpc}<250$ \& $0 \leq z \leq 5$& $4\times 10^{-2}$ & Figs. \ref{fig:benchmark_xi} and \ref{fig:analytic_xi} \\
    $C_\ell$ clustering {\tt non-Limber} &  (\ref{eq:cls}),(\ref{eq:transfer_nc}),(\ref{eq:transfer_rsd}) & CCL1 & $500 \leq \ell < 1000$ & $2\times 10^{-2}$ & - \\
    $C_\ell$ clustering {\tt Angpow} & (\ref{eq:cls}),(\ref{eq:transfer_nc}),(\ref{eq:transfer_rsd}) & CCL1 & $2 \leq \ell < 1000$ & $3\times 10^{-3}$  & Fig. \ref{fig:angpow} (right panel)\\
    \hline
  \end{tabular}
  \caption{Summary of \ccl validation tests and the level of agreement achieved with respect to the benchmarks ($\mathcal{A}$). These tests can be reproduced by the user and are integrated into the \ccl repository. The $C_\ell$ accuracy is set to $10\%$ of the expected uncertainty due to cosmic variance, $\sigma_\ell$, given in Eq. (\ref{eq:sigmaell}). In the case of intrinsic alignments, we validate auto-spectra as well as cross-spectra with galaxy shear and positions. For this case, $\sigma_\ell$ includes the lensing contribution as well. Notice that the last row of the table compares the {\tt Angpow} output for the clustering $C_\ell$ to an independent non-Limber implementation. The row immediately above demonstrates that the non-Limber method can also reproduce the Limber case at high $\ell$ with sufficient accuracy compared to the expected cosmic variance. For the BCM case, we compared the fractional impact of baryons on the matter power spectrum by dividing the $P(k)$ prediction by the dark-matter-only case. Hence, the choice of cosmology becomes irrelevant in this case. Cosmologies are documented in Tables \ref{tab:cosmologies} and \ref{tab:cosmologies_nu} for the ``CCL'' case, and in \citet{Lawrence17} for the ``M'' cosmologies.\\
    $^\star$ indicates the accuracy was established against a known higher precision implementation.\\
    $^\dag$ indicates a given level of accuracy was required in comparison to simulations.\\
    $^\S$ indicates at least one test was performed against an analytical solution.
  }
  \label{tab:tests}
\end{sidewaystable*}

Table \ref{tab:tests} summarises all the \ccl validation tests discussed in this section. All plots presented in this section can be reproduced by means of a {\tt python} notebook available in the public repository. Accuracy checks can also be run automatically upon installation of the software. All the independent scripts used to generate the predictions used to validate \ccl are also released\footnote{A list of the scripts available can be found in the \ccl wiki: \url{https://github.com/LSSTDESC/CCL/wiki/Benchmarks}}.

In the following sub-sections, we also comment on potential discrepancies in the implementation of cosmological predictions between \ccl and the benchmarks that could be responsible for the level of agreement achieved.

\subsection{Background quantities \& growth of perturbations}

Comoving radial distances, the growth factor and distance moduli were compared against independently produced benchmarks for redshifts between $z = 0.01$ and $z = 1000$. These comparisons were performed for the cosmologies listed in Table \ref{tab:cosmologies} and for the cosmologies with massive neutrinos listed in Table \ref{tab:cosmologies_nu}. (Notice that the growth function with massive neutrinos is not supported by \ccl because it is scale-dependent and therefore ill-defined in our framework. Hence, no tests are provided for the growth function in those cosmologies.) The accuracy metric was defined as the fractional difference between the prediction made by \ccl and by an independent implementation (labeled $i$), i.e., for the growth factor,
\begin{equation}
  \mathcal{A} \equiv \frac{|D_{\tt CCL}(z)-D_{i}(z)|}{D_i(z)}
\end{equation}
and analogously defined for the comoving radial distance and distance moduli.

\begin{figure*}
  \centering
  \includegraphics[width=0.49\textwidth]{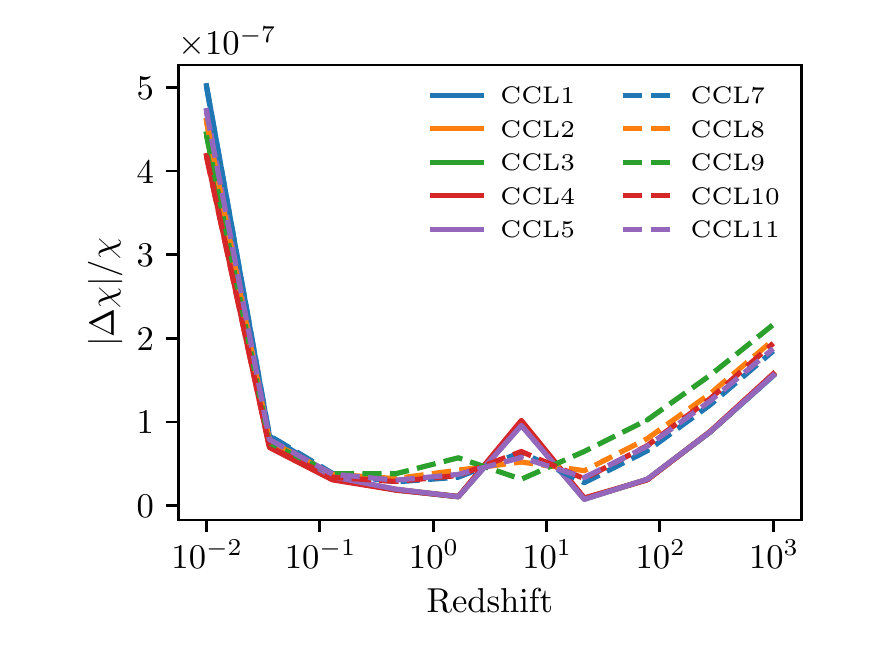}
  \includegraphics[width=0.49\textwidth]{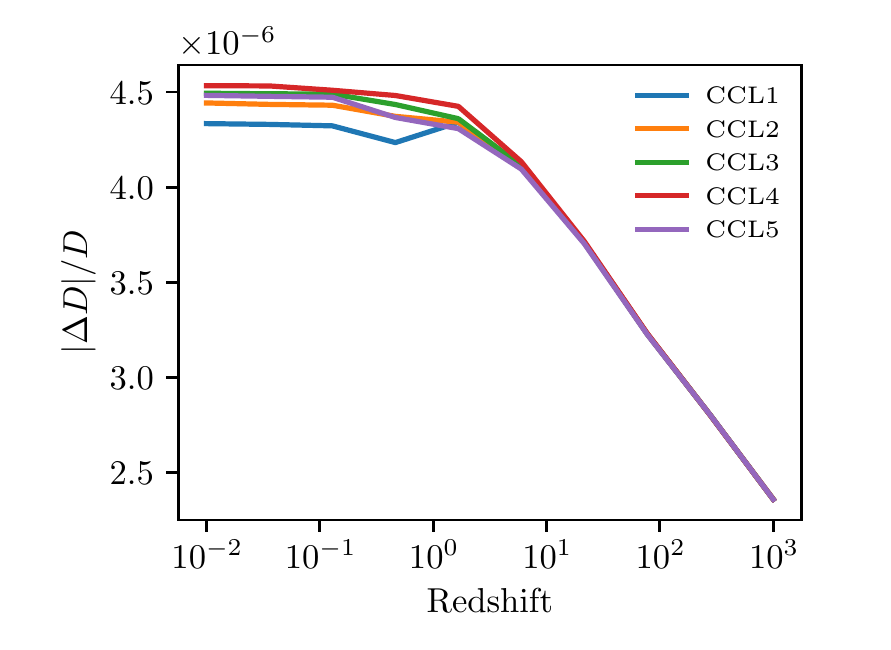}
  \caption{Accuracy achieved by \ccl in the prediction of background quantities. Left panel: fractional difference between the predictions of the comoving radial distance by \ccl and the benchmark for models CCL1--5 documented in Table \ref{tab:cosmologies} (solid lines) and models CCL7--11 with massive neutrinos documented in Table \ref{tab:cosmologies_nu} (dashed lines). Right panel: fractional difference between the predictions of the growth factor by \ccl and the benchmark for models CCL1--5. The growth factor in cosmologies with massive neutrinos is scale-dependent and not supported by \ccl.}
  \label{fig:distancegrow}
\end{figure*}

Figure \ref{fig:distancegrow} summarizes our results. The left panel shows the distance accuracy achieved for different cosmological models (curves of different thickness) as a function of redshift, which is always better than $5\times 10^{-7}$. Distance comparisons are made against benchmarks produced with the {\tt CosmoMAD} package\footnote{\url{https://github.com/damonge/CosmoMAD}} as well as the python version for {\tt CLASS}\footnote{{\tt classy},\url{https://github.com/lesgourg/class_public}}. For speed, \ccl relies on an intermediate instance where we adopt a specific grid for interpolating the comiving radial distance as a function of the scale factor for a given input cosmology. This instance is not there in {\tt CosmoMAD} and can introduce additional uncertainty. \class similarly interpolates background quantities from a pre-computed grid, but no efforts have been made to match the interpolation method or grid. Distance moduli in \ccl are also obtained from the interpolated comoving radial distance.

Similarly, the growth function is predicted with better than $6\times 10^{-6}$ accuracy in the right panel of Figure \ref{fig:distancegrow}. The growth function is obtained by solving the differential Eq. (\ref{eq:growth}) by means of a Runge-Kutta Cash-Karp algorithm. \ccl then adopts a specific grid for interpolating the growth function with the scale factor. {\tt CosmoMAD}, used for benchmarking the growth function produced by \ccl, implements a similar algorithm. Additional tests against independent codes (e.g., {\tt CosmoLike}\footnote{\url{https://github.com/CosmoLike}} \citealt{krause17}, {\tt cosmosis}\footnote{\url{https://bitbucket.org/joezuntz/cosmosis/wiki/Home}} \citealt{Zuntz15}) for a wide range of wCDM models yielded agreement to $10^{-3}$.

We validate the implementation of the modified growth function described in Eq. \ref{eq:mgrowth} against an analytical prediction. In particular, we verify that, by setting $\Delta f(a)=k\,a$ for a constant $k$, the growth factor computed by CCL is compatible with the analytical solution $D(a)=D_0(a)\,\exp[k\,(a-1)]$, where $D_0(a)$ is the solution for $\Delta f(a)=0$, to better than one part in $10^{5}$. 

Additional independent distance benchmarks were obtained from {\tt astropy} \citep{astropy}. We find that the agreement between \ccl and {\tt astropy} is only at the $10^{-3}$ level for cosmologies with massive neutrinos, in contrast to the much better agreement shown in left panel of Figure~\ref{fig:distancegrow}, which relies on benchmarks obtained from {\tt CLASS}. We believe that this is due to the fact that \ccl uses the full phase-space integral in computing the massive neutrino density as defined in Eq. (\ref{phasespacenu}), while {\tt astropy} uses a fitting function which is itself only accurate at a $10^{-3}$ level. Hence, true accuracy is probably better than quoted and closer to the value reported in the first row of Table \ref{tab:tests}, as shown by Figure \ref{fig:distancegrow}.

\begin{table*}[t]
  \centering
  \begin{tabular}{ l c | c c c c c c c c }
    \hline
    \multicolumn{10}{|c|}{Cosmological models with massless neutrinos} \\
    \hline
    \hline
    Acronym & Model & $\Omega_m$ & $\Omega_b$ & $\Omega_\Lambda$ & $h_0$ & $\sigma_8$ & $n_s$ & $w_0$ & $w_a$ \\
    \hline
    CCL1 & flat $\Lambda$CDM & 0.3 & 0.05 & 0.7 & 0.7 & 0.8 & 0.96 & -1 & 0 \\
    CCL2 & $w$CDM & 0.3 & 0.05 & 0.7 & 0.7 & 0.8 & 0.96 & -0.9 & 0  \\
    CCL3 & $w$CDM & 0.3 & 0.05 & 0.7 & 0.7 & 0.8 & 0.96 & -0.9 & 0.1  \\
    CCL4 & open $w$CDM & 0.3 & 0.05 & 0.65 & 0.7 & 0.8 & 0.96 & -0.9 & 0.1  \\
    CCL5 & closed $w$CDM & 0.3 & 0.05 & 0.75 & 0.7 & 0.8 & 0.96 & -0.9 & 0.1  \\
    CCL6 & flat $\Lambda$CDM & 0.3 & 0 & 0.7 & 0.7 & 0.8 & 0.96 & -1 & 0 \\
    WMAP7 & flat $\Lambda$CDM & 0.272 & 0.0455 & 0.728 & 0.704 & 0.810 & 0.967 & -1 & 0 \\
    Planck 2013 & flat $\Lambda$CDM & 0.318 & 0.0490 & 0.682 & 0.671 & 0.834 & 0.962 & -1 & 0 \\
    \hline
  \end{tabular}
  \caption{Cosmological models with massless neutrinos used in testing \ccl against independently produced benchmarks.}
  \label{tab:cosmologies}
\end{table*}

\begin{table*}[t]
  \centering
  \begin{tabular}{ l c | c c c c c c c c c c }
    \hline
    \multicolumn{12}{|c|}{Cosmological models with massive neutrinos} \\
    \hline
    \hline
    Acronym & Model & $\Omega_m$ & $\Omega_b$ & $\Omega_\Lambda$ & $h_0$ & $\sigma_8$ & $n_s$ & $w_0$ & $w_a$ & $N_{\rm eff}$ & $m_\nu$ (eV) \\
    \hline
    CCL7 & flat $\Lambda$CDM, $m_\nu$ & 0.3 & 0.05 & 0.7 & 0.7 & 0.8 & 0.96 & -1 & 0 & 3.013 & \{0.04, 0, 0\} \\
    CCL8 & $w$CDM, $m_\nu$ & 0.3 & 0.05 & 0.7 & 0.7 & 0.8 & 0.96 & -0.9 & 0 & 3.026 & \{0.05, 0.01, 0\} \\
    CCL9 & $w$CDM, $m_\nu$ & 0.3 & 0.05 & 0.7 & 0.7 & 0.8 & 0.96 & -0.9 & 0.1 & 3.040 & \{0.03, 0.02, 0.04\} \\
    CCL10 & open $w$CDM, $m_\nu$ & 0.3 & 0.05 & 0.65 & 0.7 & 0.8 & 0.96 & -0.9 & 0.1 & 3.013 & \{0.05, 0, 0\}  \\
    CCL11 & closed $w$CDM, $m_\nu$ & 0.3 & 0.05 & 0.75 & 0.7 & 0.8 & 0.96 & -0.9 & 0.1 & 3.026 &\{0.03, 0.02, 0\} \\
    \hline
  \end{tabular}
  \caption{Cosmological models with massive neutrinos used in testing \ccl against independently produced benchmarks. We calculate $N_{\rm eff}$ according to Eq.~(\ref{Nnurel}), based on the number of massless and massive neutrino species.
}
  \label{tab:cosmologies_nu}
\end{table*}

\subsection{Matter power spectra}

\subsubsection{Analytic expressions}
\label{ss:bbksval}

As discussed in Section \ref{sec:matterps}, several power spectrum methods are implemented in \ccl. Two of them, the BBKS \citep{BBKS} and the \citet{1998ApJ...496..605E} methods are implemented for validation purposes only and feed into the tests for observables such as angular power spectra and correlation functions, as we will see in subsequent sections. These two implementations have been validated against independent implementations. The accuracy in this case was defined as the absolute fractional difference between the \ccl and the independent predictions, $i$, at any given $k$ and $z$:
\begin{equation}
  \mathcal{A}\equiv \frac{|P_{\tt CCL}(k,z)-P_{i}(k,z)|}{P_i(k,z)}.
  \label{eq:pkacc}
\end{equation}
For BBKS, this test was performed at $0\leq z \leq 5$ in the wavenumber range $10^{-3} \leq k \leq 10 h\,\text{Mpc}^{-1}$ with 10 bins per decade, and yielded an accuracy level of $10^{-5}$.\footnote{We noticed that there are 2 typographical errors for the BBKS transfer function in ``Modern Cosmology'' \citep{DodelsonBook} compared to the original BBKS paper. The quadratic term should be $(16.1q)^2$ and the cubic term should be $(5.46q)^3$. The BBKS equation is correct in \citet{PeacockBook}. Using the wrong equation can give differences in the results above the $10^{-4}$ level.}
For the \citet{1998ApJ...496..605E} matter power spectrum, we obtained similar accuracy at $z=0$ for the same wavenumbers. The cosmologies for which the tests were implemented are specified in Table \ref{tab:tests}.

For both BBKS and the \citet{1998ApJ...496..605E} matter power spectra, the comparisons with \ccl were performed using {\tt CosmoMAD}. As in \ccl, {\tt CosmoMAD} implements analytical functions to produce these predictions and then creates an interpolation of the result with logarithmic wavenumber. The level of agreement between \ccl and the benchmarks is sensitive to the choice of interpolation scheme and resolution for the power spectrum in $k$ and redshift.

The BCM implementation for the impact of baryons on the matter power spectrum, described in Section \ref{sec:matterps} is also analytical. Following Eq.~(\ref{eq:pkacc}), we found it to be accurate to $10^{-12}$. In this case, we expect no sources of discrepancy between the independent implementation and \ccl other than the numerical precision of the variables involved in the computation.

\subsubsection{Validation of interpolation schemes}
\label{ss:classval}

In its default configuration, \ccl adopts the \halofit \citep{CLASS_halofit} implementation by interpolating \class power spectra outputs to model the matter power spectrum. The computation of the power spectrum from \class can be significantly sped up by interpolating the matter power spectra in the range {\tt K$\_$MIN$\_$SPLINE}~$<k<$~{\tt K$\_$MAX$\_$SPLINE} and extrapolating beyond it, as described in Section \ref{sec:implement}. In this section, we describe the loss of accuracy due to this method. The tests presented are performed in a flat $\Lambda$CDM cosmology similar to CCL1, but with a normalization of the power spectrum set by $A_s=2.1\times10^{-9}$ rather than $\sigma_8$.

\begin{figure*}
\centering
  \includegraphics[width=0.49\textwidth]{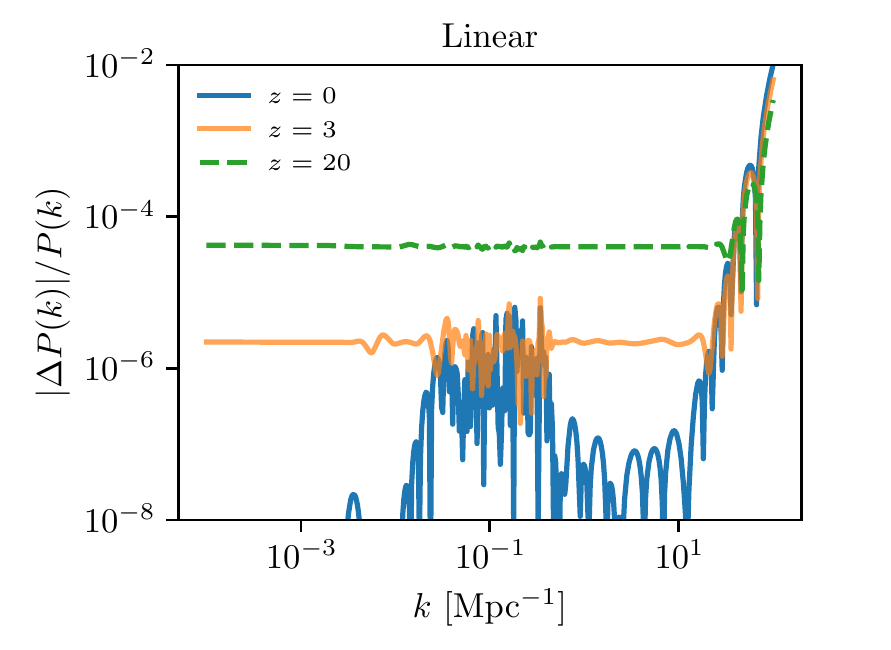}
  \includegraphics[width=0.49\textwidth]{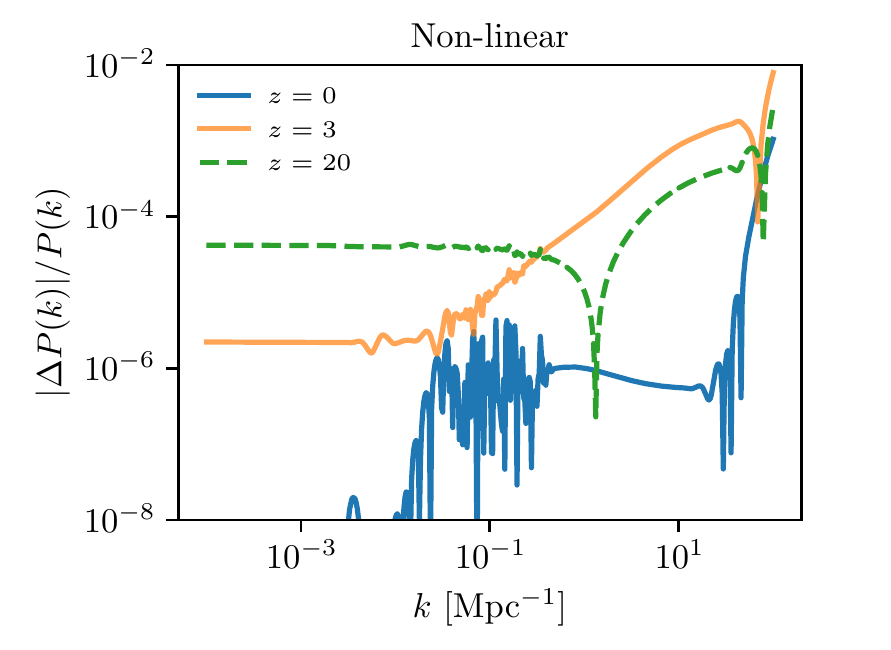}
\caption{The relative error compared to power spectra produced with high values of the power spectrum splines, $P_{fid}$, produced by splining the matter power spectrum up to {\tt K$\_$MAX$\_$SPLINE}$=50\,\text{Mpc}^{-1}$ and extrapolating beyond this value with a second order Taylor expansion the natural logarithm of the matter power spectrum. The left panel shows the relative errors for the linear matter power spectrum at $z=0$, $z=3$ and $z=20$. The right panel shows the results for the non-linear matter power spectrum at the same redshifts. While the relative error increases substantially at high $k$, we note that it is still well below the uncertainty from baryonic physics at these scales, which is $\sim 10\%$ at $k=1\,\text{Mpc}^{-1}$ \citep{Schneider15}.}
\label{fig:NLextrapol}
\end{figure*}

The accuracy of this approximation is shown in Figure \ref{fig:NLextrapol} for redshifts $z=0$, $z=3$ and $z=20$. We compare the non-linear matter power spectrum at these redshifts, computed with the previously described approximation, to the matter power spectrum obtained by setting the power spectrum splines to high-accuracy values. We find that for typical values of $\Delta \ln k=10^{-2}$ and {\tt K$\_$MAX$\_$SPLINE}$=50\,\text{Mpc}^{-1}$, $\ln P$ has converged to an accuracy that surpasses the expected impact of baryonic effects on the matter power spectrum at $k>10\,\text{Mpc}^{-1}$. (For an estimate of the impact of baryons on the total matter power spectrum, see \citealt{Schneider15}.) 

With the implementation described above, the power spectrum splines are initialized up to {\tt K$\_$MAX$\_$SPLINE}. This is also true for the linear matter power spectrum, which is used within \ccl in particular to obtain $\sigma_8$ (see Eq. \ref{eq:sigR}). We have tested how this procedure affects the convergence of the linear matter power spectrum. We compare the fiducial \ccl output to the case where we increase the precision of all spline parameters by an order of magnitude (i.e. we use 10 times larger sampling rates in $k$ and $a$, and extend the interpolation ranges in $k$ by 1 decade on either end). The result is shown in Figure \ref{fig:NLextrapol}. For some applications that use the linear power spectrum, the user might need to increase the value of {\tt K$\_$MAX$\_$SPLINE}, but overall the impact of the fiducial interpolation parameters is negligible for most applications.

\begin{figure}
\centering
\includegraphics[width=0.49\textwidth]{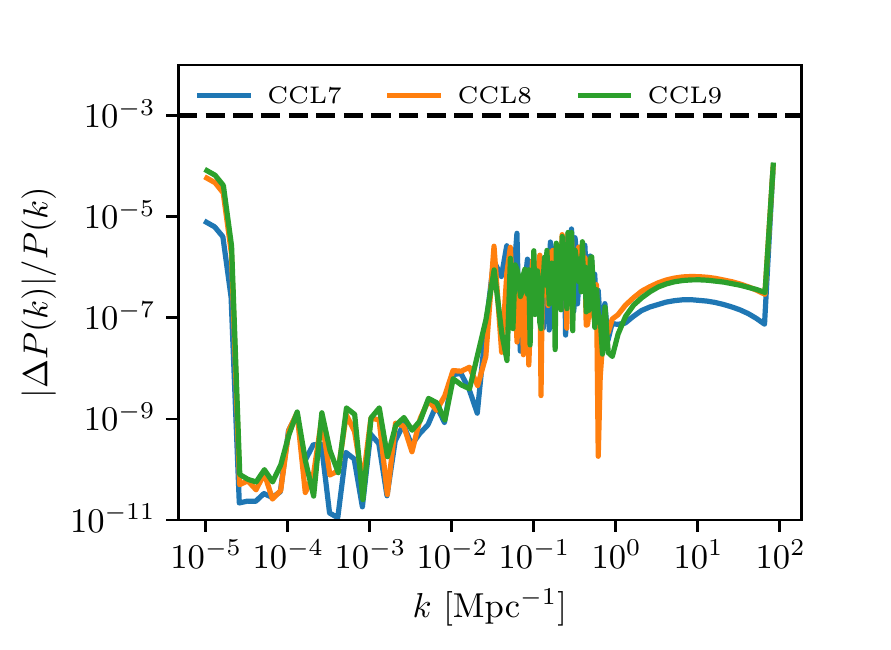}
\caption{Fractional difference between the non-linear matter power spectrum as computed directly via \class with that computed using \class via \ccl in cosmologies CCL7, CCL8, and CCL9 with massive neutrinos.}
\label{fig:power_nu}
\end{figure}

In addition to the above tests in $\Lambda$CDM cosmologies without massive neutrinos, we have checked the impact of using splines (at intermediate $k$) and extrapolation (at low and high $k$) in cosmologies CCL7, CCL8, and CCL9 with massive neutrinos, defined in Table \ref{tab:cosmologies_nu}. We compare the linear and non-linear matter power spectrum as computed directly via \class to that computed using \class via \ccl . We find that for $k$  between {\tt K$\_$MIN} and {\tt K$\_$MIN$\_$SPLINE}, the two power spectra agree to better than $10^{-4}$ in all models. For $k$ between {\tt K$\_$MAX$\_$SPLINE} and {\tt K$\_$MAX}, agreement is better than $10^{-3}$, which is sufficient given the significant physical uncertainties introduced at these small scales by effects such as galaxy formation \citep{vanDaalen11}. The fractional difference between the two non-linear power spectra is shown in Figure \ref{fig:power_nu}. 

\subsubsection{Generalized validation of the power spectrum over $\Lambda$CDM parameter space}
\label{ss:classval2}

While concentrating on individual points in cosmological parameter space allows us to perform detailed validation tests, as above, it is important for \ccl to also be validated across a wide range of cosmological parameter values, e.g., to ensure validity for MCMC analyses. In this section, we present a set of validation tests for the \ccl linear and non-linear matter power spectrum functions that spans a broad range of $\Lambda$CDM parameters.

Covering a full range of all 5 $\Lambda$CDM parameters on a regular grid would be prohibitively expensive, so an alternative method for fairly (but more sparsely) sampling the parameter space is needed. We use Latin Hypercube Sampling to determine a tractably-sized set of sample points. This splits the parameter space into a grid with $N$ bins per dimension. The sample points are then chosen by going through each dimension in turn and choosing a bin at random without replacement, so that a given bin in each dimension is only ever chosen once. This is repeated until all bins in each dimension contain a single sample (or until a maximum number of sample points has been reached). This has the effect of covering the space uniformly but sparsely, with only $N$ sample points chosen from the $N^5$ available positions on the grid. The exact location of the sample within each bin can be chosen from a uniform distribution within that bin, but for simplicity we put each sample at the bin center. We use $N=100$ sample points per dimension, with the ranges for each parameter given in Table~\ref{tab:paramranges}. These ranges were chosen to be significantly wider than those allowed by current observational constraints, to ensure that the full parameter range expected to be accessed by MCMC analyses is covered. For the purposes of this exercise, we allow only massless neutrinos ($N_{\rm eff} = 3.046$), and set $T_{\rm CMB}$ to the same value in \ccl and \class.

\begin{figure*}
\centering
\includegraphics[width=1\textwidth]{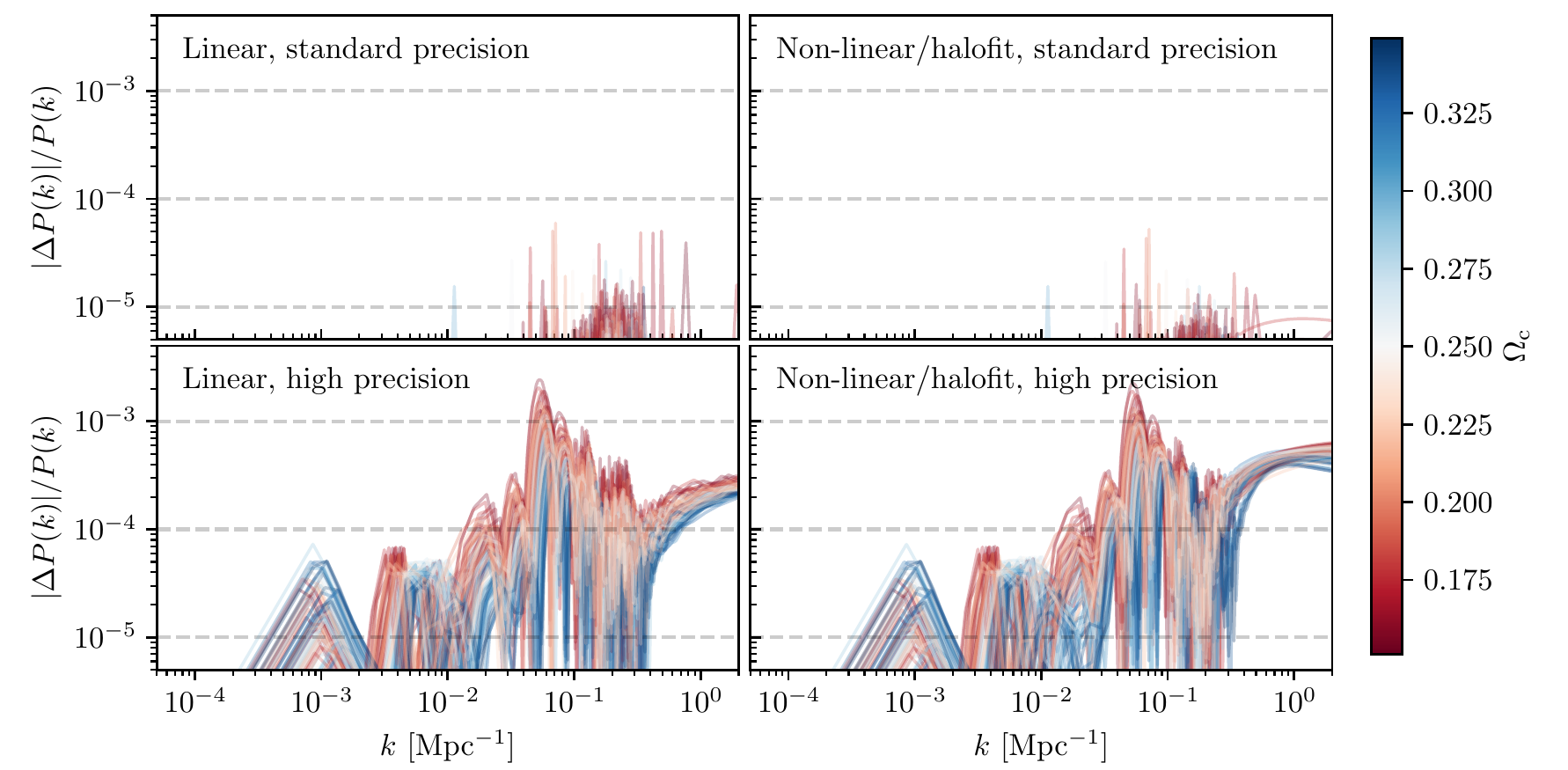}
\caption{Absolute fractional difference between the matter power spectra at $z=0$ calculated using \class via \ccl, and \class directly, for a range of cosmological parameter values, and different \class precision settings (standard vs. high-precision) and power spectrum types (linear vs. \halofit). The lines are colored according to the value of $\Omega_{\rm c}$ for each set of cosmological parameters (see Table~\ref{tab:paramranges} for the ranges of other parameters). }
\label{fig:power_paramspace}
\end{figure*}
%
\begin{table}[ht]
  \centering
  \begin{tabular}{l | l}
    \hline
    Parameter & Range \\
    \hline
    $h$ & $[0.55, 0.8]$ \\
    $\Omega_c$ & $[0.15, 0.35]$ \\
    $\Omega_b$ & $[0.018, 0.052]$ \\
    $A_s$ & $[1.5, 2.5] \times 10^{-9}$ \\
    $n_s$ & $[0.94, 0.98]$ \\
    \hline
  \end{tabular}
  \caption{Ranges of $\Lambda$CDM parameters used for the generalised \ccl validation tests of the matter power spectrum.}
  \label{tab:paramranges}
\end{table}

For each set of parameters, we then calculate the linear and non-linear (\halofit) power spectra using \ccl for a range of redshifts. A corresponding set of reference power spectra is then produced using \class directly, i.e. using a regular installation of \class ({\tt v2.6.3}). We run this with either default precision settings (`standard precision'), or settings intended to produce high-precision CMB results (`high precision'), taken from the {\tt pk\_ref.pre} precision file that is bundled with \class.

\begin{figure*}
\centering
\includegraphics[width=1\textwidth]{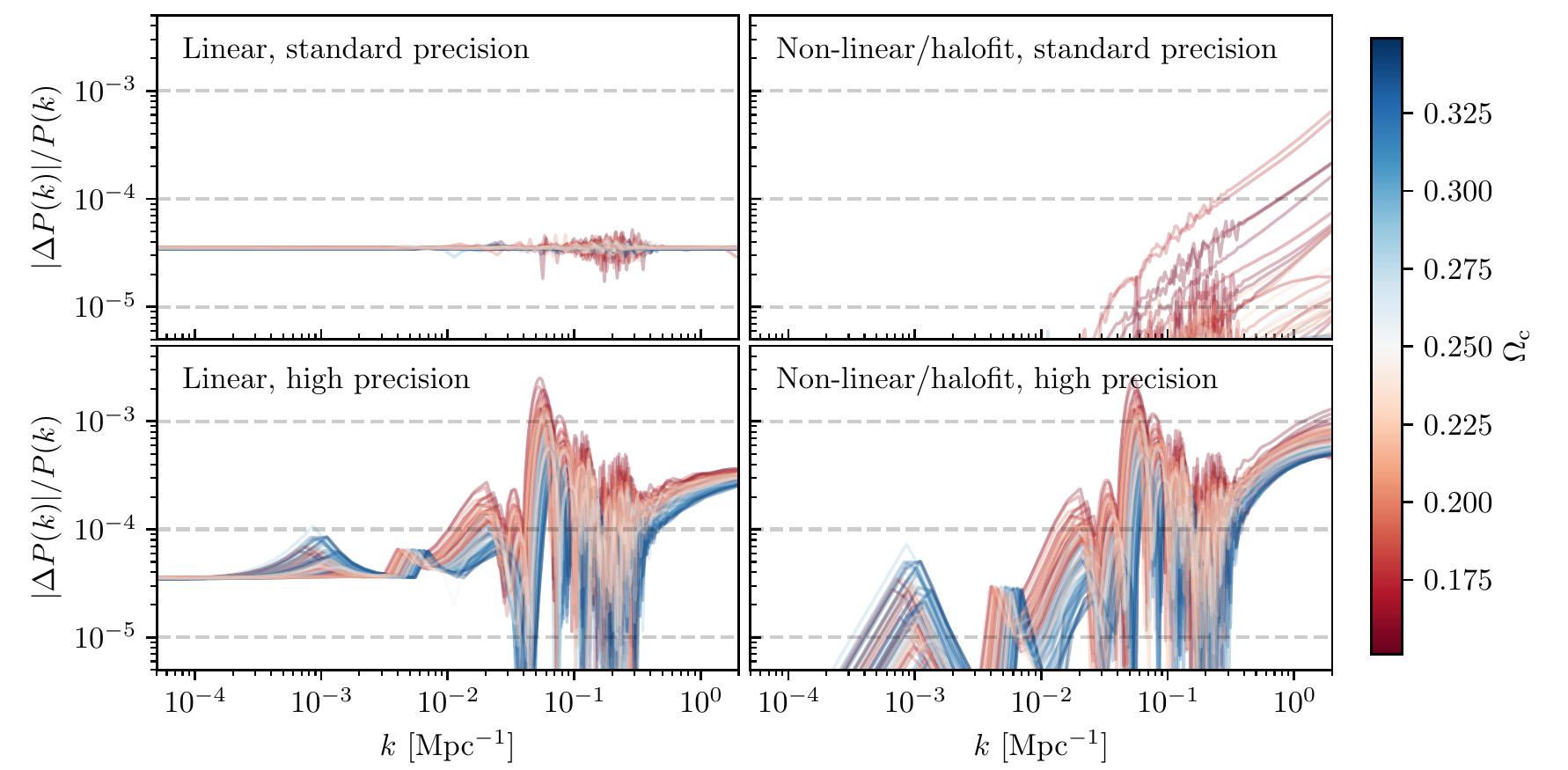}
\caption{Absolute fractional difference between \ccl and \class matter power spectra, as plotted in Figure~\ref{fig:power_paramspace}, but now at $z=2$.}
\label{fig:power_paramspace_z2}
\end{figure*}

Figure~\ref{fig:power_paramspace} shows the fractional difference between the \ccl and \class matter power power spectra at $z=0$ for 100 sample points over the parameter space, with each line colored according to the value of $\Omega_c$ for that sample. Results for different power spectrum types (linear vs. \halofit) and \class precision settings are shown for comparison.

As shown in the top two panels of Figure~\ref{fig:power_paramspace}, \ccl reproduces the standard \class results well across a broad range of parameter values, always remaining well within a fractional precision of $10^{-4}$. This demonstrates the robustness of our choice of spline parameters to different cosmological parameter values.

The lower panels in Figure~\ref{fig:power_paramspace} show the fractional deviation between \ccl (which always uses the `standard' \class precision) and \class with high precision settings. These deviations are more significant, especially around the wavenumbers where the baryon acoustic oscillation feature is most prominent. The precision is still generally better than $10^{-3}$ however, and is only worse than that for the very lowest values of $\Omega_c$.

Figure~\ref{fig:power_paramspace_z2} shows the same comparison, but now for $z=2$. The precision on the linear matter power spectrum is almost an order of magnitude worse than at $z=0$ for the standard precision settings (but still always better than $10^{-4}$), and only slightly worse than at $z=0$ for the high precision settings. The picture is slightly different for the \halofit power spectrum however, where moderate deviations are seen for the standard precision settings in models with small values of $\Omega_c$. This appears to be caused by a setting inside \class that switches off \halofit corrections when a redshift- and cosmology-dependent threshold is reached, and can be mitigated by increasing the value of the {\tt P\_k\_max\_1/Mpc} parameter (which is already set to a relatively high value of $50$ in \ccl by default). Larger values of $\Omega_c$ produce only slightly worse precision than at $z=0$, and the high precision \halofit results are also relatively unchanged.

These results show that the \class-based \ccl power spectrum calculations are robust across a broad range of cosmological parameters, especially for the linear power spectrum, but that some caution must be taken when using the \halofit power spectrum in MCMC studies that involve higher redshifts for example.\footnote{Note that the {\tt K\_MAX\_SPLINE} setting in the {\tt ccl\_params.ini} file can be used to change the value of {\tt P\_k\_max\_1/Mpc} used for the \halofit calculation, so this issue can be avoided at the expense of an increase in runtime.}

\subsubsection{Validation of the Cosmic Emulator implementation}
\label{ss:cosmicemu}

The matter power spectrum emulation procedure from \citet{Lawrence17} has an intrinsic accuracy compared to the simulated results used for its construction. It effectively provides a fitting scheme which allows interpolation between the simulation results. As a consequence, the method itself has some limitations in how well it can reproduce the simulation results. \ccl takes the emulator predictions and interpolates between the wavenumber and scale-factor notes in the emulator output. To validate the final power spectra coming out of \ccl, we compared them directly to the simulated spectra from \citet{Lawrence17} for a subset of the cosmologies adopted in that work. In this section, we quantify the accuracy of the CCL predictions by estimating
\begin{equation}
  \mathcal{A}\equiv\frac{|P_{\tt CCL}(k,z)-P_{\rm L17}(k,z)|}{P_{\rm L17}(k,z)}
  \label{eq:pkaccemu}
\end{equation}
where the label L17 refers to the smoothed simulated power spectra from \citet{Lawrence17}. Notice that the emulator is intrinsically accurate to $1\%$ for cosmologies without massive neutrinos, and to $3\%$ for cosmologies with massive neutrinos. In other words, replacing $P_{\tt CCL}(k,z)$ in Eq. (\ref{eq:pkaccemu}) by the direct emulator output would yield $\mathcal{A}$ of $0.01$ and $0.03$ for the two different families of cosmologies. In the validation test presented in this section, we focus on ensuring that \ccl does not deviate from that overall level of accuracy.

\begin{figure*}
  \centering
  \includegraphics[width=0.49\textwidth]{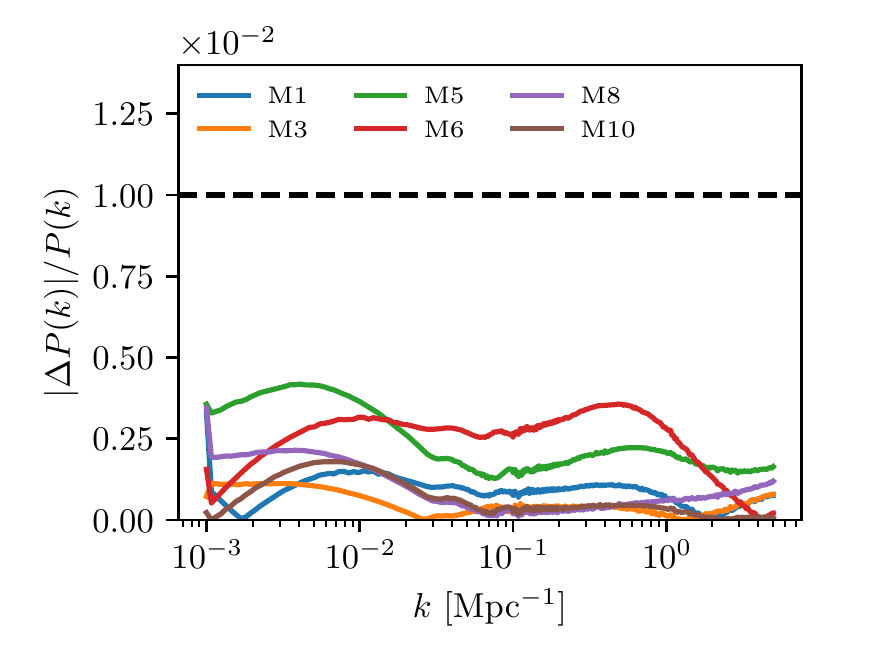}
  \includegraphics[width=0.49\textwidth]{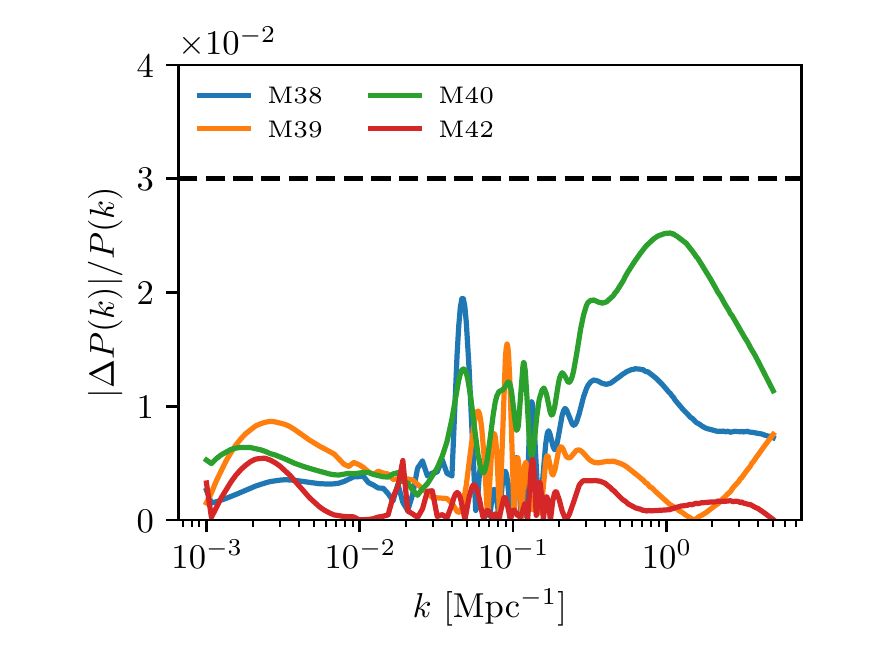}
  \caption{Absolute fractional accuracy in the matter power spectra, Eq. (\ref{eq:pkaccemu}), obtained by calling the Cosmic Emulator from \ccl and the smoothed simulated spectra from \citet{Lawrence17}. The left panel shows the results for cosmologies without neutrinos; the right panel, results for cosmologies with neutrinos. The dashed line in both panels represents our target accuracy, based on the claimed accuracy of the emulator by \citet{Lawrence17}.}
  \label{fig:emuacc}
\end{figure*}

Our results are shown in Figure \ref{fig:emuacc}. For cosmologies without neutrinos, we required the matter power spectrum at $z=0$ to be within $1\%$ of the smoothed simulated power spectrum from \citet{Lawrence17} (see their Figure 6). Similarly, we required $3\%$ accuracy for cosmologies with neutrinos (their Figure 5). The cosmologies that were tested are the ones listed in Table \ref{tab:cosmologies}, whose parameter values are specified in \citet{Lawrence17}. In both cases, we find that the \ccl implementation falls below the target accuracy for the full range of scales tested. This is not surprising, as \ccl directly incorporates the publicly available emulator prediction code\footnote{\url{https://github.com/lanl/CosmicEmu}}. The essential difference in the implementation is that the predictions of the public code for the matter power spectrum are interpolated in wavenumber and scale factor as we described in Section \ref{ssec:mpspec}.

\subsection{Halo bias and halo mass function}

The accuracy of the halo mass function calculation was checked against benchmarks produced by {\tt CosmoMAD} for power spectra obtained using the BBKS approximation, which allows us to isolate the numerical inaccuracies associated to the mass function calculation from those arising from the power spectrum calculation. For the halo mass function, we compare the value of $\sigma$,
\begin{equation}
  \tilde\sigma\equiv\log[\sigma^{-1}(M)],\label{eq:tildesig}
\end{equation}
and the value of the halo mass function in the form used in \citet{Tinker2008},
\begin{equation}
  \mathcal{H}\equiv \log[(M^2/\bar{\rho}_m)dn/dM].
  \label{eq:newhmf}
\end{equation}
We define three new accuracy metrics:
\begin{eqnarray}
  \mathcal{A}_{hmf1}&\equiv&\frac{|\sigma_{\tt CCL}-\sigma_i|}{\sigma_i},\\
  \mathcal{A}_{hmf2}&\equiv&\frac{|\tilde\sigma_{\tt CCL}-\tilde\sigma_i|}{\tilde\sigma_i},\\
  \mathcal{A}_{hmf3}&\equiv&\frac{|\mathcal{H}_{\tt CCL}-\mathcal{H}_i|}{\mathcal{H}_i}.
\end{eqnarray}
Note that for $\sigma(M)$, it is important to set the desired precision level correctly for the numerical integrator. As the integral yields $\sigma^2(M)$, this becomes the relevant concern for numerical accuracy. 

\begin{figure*}
\includegraphics[width=0.32\textwidth]{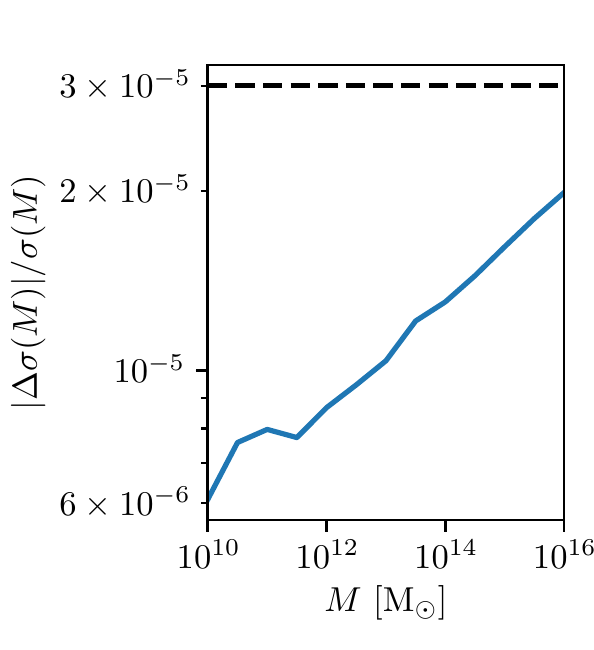}
\includegraphics[width=0.32\textwidth]{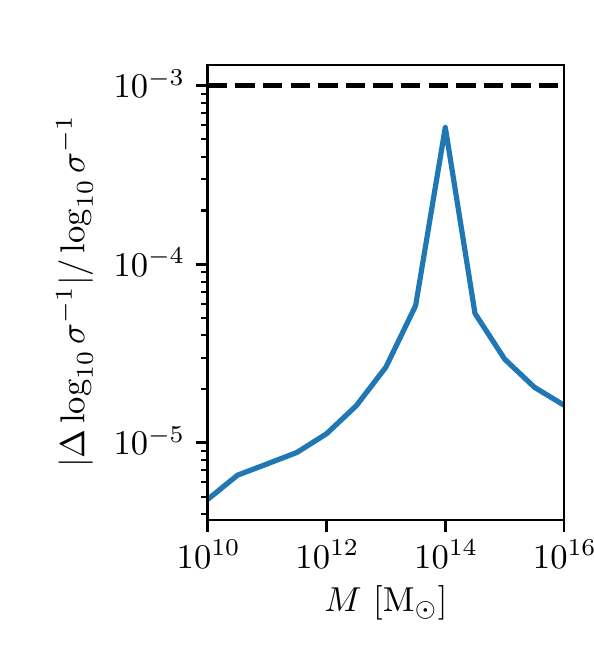}
\includegraphics[width=0.32\textwidth]{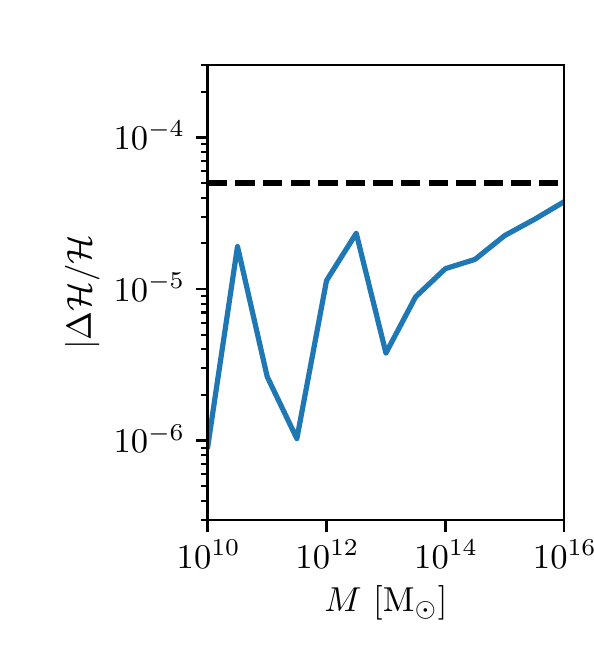}
\caption{Three different numerical tests of the halo mass function calculation. In each line, the blue line is the fractional error in the function, while the black dashed line represents our error tolerance. The first panel demonstrates the robust calculation of $\sigma(M)$. The second panel demonstrates a numerical quirk in our spline treatment that is currently not addressed, but does reduce the numerical accuracy in returning the log inverse of $\sigma(M)$. We note that this does not significantly impact the error in the halo mass function in the final panel.}
\label{fig:hmf}
\end{figure*}

For $\mathcal{A}_{hmf1}$ and $\mathcal{A}_{hmf3}$ we achieve accuracies of $3\times 10^{-5}$ and $5 \times 10^{-5}$, respectively. For $\mathcal{A}_{hmf2}$, the accuracy degrades to a value of $10^{-3}$. These accuracy levels are acceptable, as it is significantly better than the physical accuracy of current halo mass function models. This is demonstrated in Figure~\ref{fig:hmf}, where this calculation has been run for a single cosmology using the \citet{Tinker2010} halo mass function\footnote{A single cosmology is used for this analysis as the Tinker fitting parameters do not vary with cosmology.}. While there is a degradation in accuracy due to our spline treatment of the log inverse of $\sigma(M)$, we note that it does not significantly degrade our halo mass function determination. While improvement on this remains a task for the future, the halo mass function varies between fitting functions significantly more than this remaining error. As of this time, we do not have independent implementations for the halo bias function, though it should be noted that this calculation does not involve any additional functions beyond $\sigma(M)$ and should not exceed a $10^{-4}$ tolerance level. Some deviation may exist between \ccl and other implementations due to our choice of spline interpolation between \citet{Tinker2010} fitting parameters as a function of $\Delta_\mathrm{v}$; we use an Akima interpolation between those provided in \citet{Tinker2010} for the halo mass function and this may lead to mild numerical change. This approach is motivated by the fact that the difference in parameters at $\Delta_\mathrm{v} = 200$ between the fitting formula result and the tabulated best-fit parameters leading to a greater than $10^{-4}$ error from our calculated benchmarks in the halo mass function. As the tabulated version is in common use in the literature, having higher accuracy for the tabulated points was prioritized.

\subsection{Halo model}
\label{sec:halo_model_verification}

\begin{figure}
\includegraphics[width=0.49\textwidth]{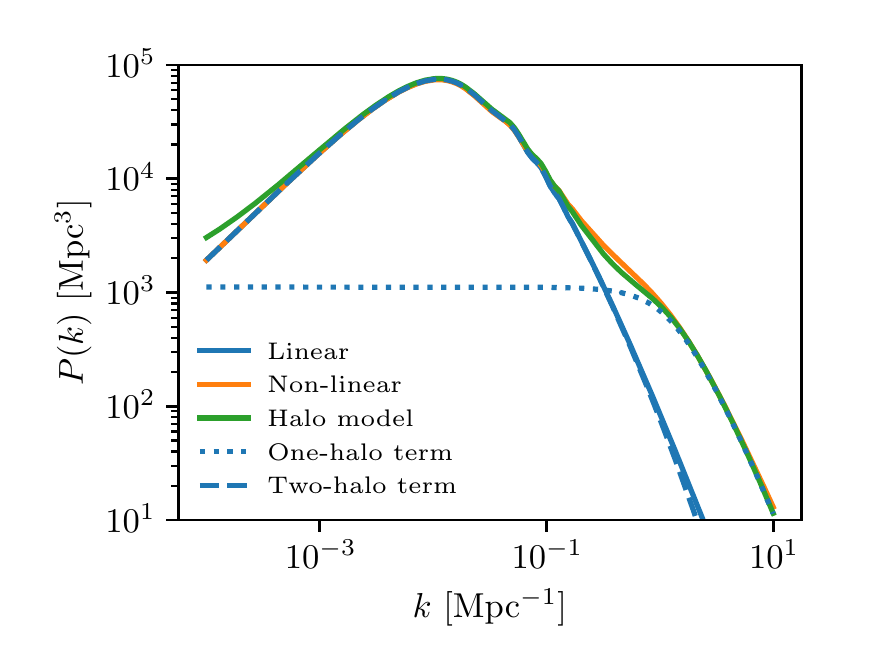}
\caption{The matter power spectrum computed according to linear theory, \halofit and the \ccl halo model for the CCL1 cosmology. The halo model two- and one-halo terms are also shown, their sum is the total halo model prediction. \halofit is accurate compared to $N$-body simulations at the $\simeq5\%$ level for the scales shown. The halo model prediction deviates from \halofit at the $\simeq30\%$ level in the transition region when both the two- and one-halo terms are important ($k\simeq0.5\,\mathrm{Mpc}^{-1}$) but shows better agreement at smaller scales.}
\label{fig:halo_model}
\end{figure}

In Figure~\ref{fig:halo_model} we show the power spectrum computed by the \ccl halo model compared to that from \halofit and to the linear matter spectrum for the CCL1 cosmology from Table~\ref{tab:cosmologies} at $z=0$. The halo model predictions show the correct general trend for the non-linear power spectrum but differ in details. Compared to simulations they are only accurate at the $\sim 30\%$ level.

\begin{figure}
\includegraphics[width=0.49\textwidth]{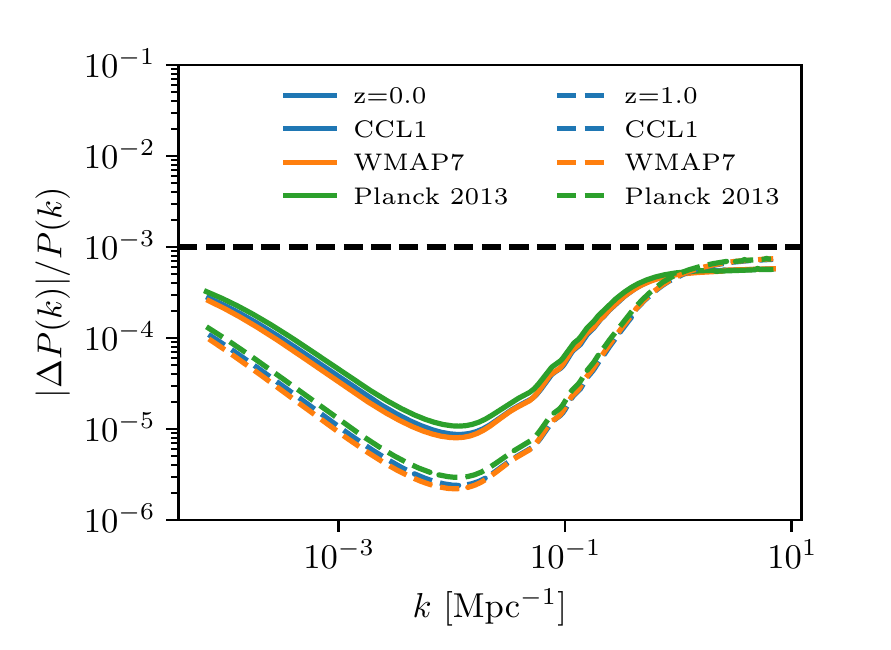}
\caption{The relative accuracy of the halo model power spectrum calculation compared to our benchmarks. We achieve a precision of $10^{-3}$ for the range of scales shown. Solid lines show $z=0$ while dashed lines show $z=1$. Different colors show different cosmological models (CCL1, {\it WMAP7}, {\it Planck} 2013).}
\label{fig:halo_model_benchmark}
\end{figure}

In Figure~\ref{fig:halo_model_benchmark} we show the accuracy of the halo model power spectrum compared to our independently produced benchmark. The benchmark was generated using a standalone {\tt python} script that uses the CCL power spectrum as input, but includes independent implementations of the halo profiles and concentration-mass relations. The script produced predictions for the 1-halo and 2-halo contributions to the power spectrum solving the corresponding mass integrals using a trapezium rule and {\tt scipy}'s {\tt quad} method. We define an accuracy criterion $\mathcal{A}$ as the ratio of power from \ccl compared to that from the independent code released with the \ccl repository. With this definition, we achieve an accuracy of $10^{-3}$ across scales from $10^{-4}\,\mathrm{Mpc}^{-1}<k<10\,\mathrm{Mpc}^{-1}$ for three different cosmological models (CCL1, {\it WMAP7}, {\it Planck} 2013) at both $z=0$ and $z=1$. The mass range for the halo model integrations are identical for the benchmark and the \ccl implementation, as are the mass function, halo bias and halo profiles. Therefore we suspect that the residual differences are due to the differing integration schemes between the benchmark and \ccl. The benchmark shown here was produced using a non-adaptive trapezium rule, so this level of difference is not surprising.

\subsection{Two-point statistics}

Validation tests for two-point statistics relied on the BBKS linear matter power spectrum. This choice of method was intended to remove any potential discrepancies between the \ccl implementation and the independent one with regards to the matter power spectrum. By using BBKS, we are relying on predictions that we know to be fast and which we have already validated to known numerical accuracy (Section \ref{ss:bbksval}).

\begin{figure}
\centering
\includegraphics[width=0.49\textwidth]{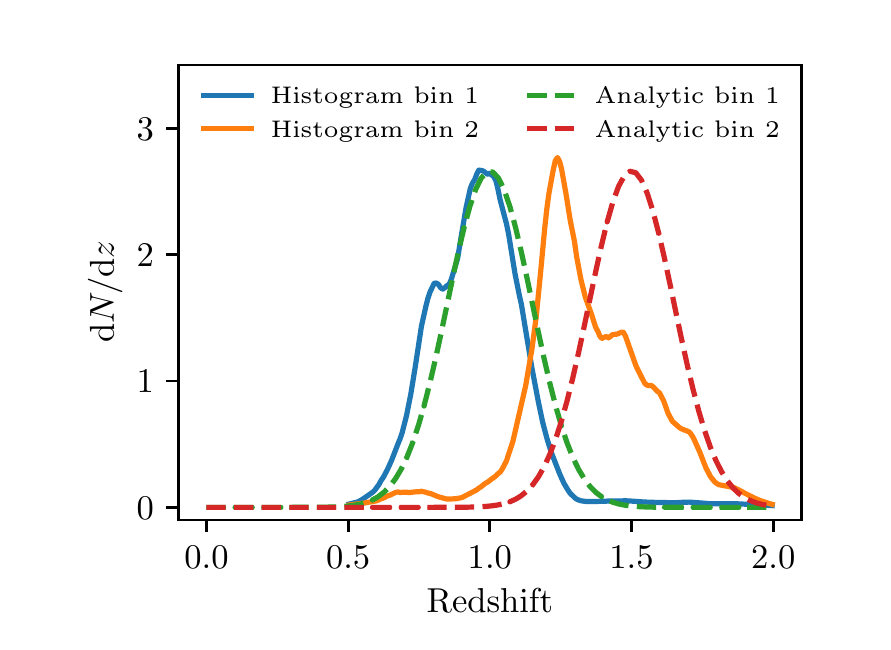}
\caption{Redshift distributions used for validating the computation of angular power spectra and correlation functions.}
\label{fig:zhistos}
\end{figure}

We thus used the BBKS linear matter power spectrum to compare two-point statistics for two redshift bins, resulting in four tomographic combinations, labelled 1-1, 1-2, 2-1 and 2-2. The validation tests were performed for two kinds of redshift distributions: analytic and binned ones. The goal of defining these two sets was to capture any numerical deviation produced by the interpolation of the binned distribution. We adopted the following analytic redshift distributions: a Gaussian with $\sigma = 0.15$, centered at $z_1 = 1$; and another Gaussian with the same dispersion but centered at $z_2 = 1.5$. In the case of the binned distributions, we adopted the two redshift distribution histograms shown in Figure \ref{fig:zhistos}.

For both types of distributions, we computed the following quantities:
\begin{itemize}
\item Number counts angular power spectra: density term only (no magnification, RSD, etc.) with non-evolving linear bias $b(z) = 1$, in the range $2 \leq \ell \leq 3000$.
\item Lensing $E$-mode angular power spectra: leading order term only (no magnification), on the same scales.
\item The cross-power spectrum between galaxy positions and galaxy shear (galaxy-galaxy lensing), on the same scales.
\item Intrinsic alignments $E$-mode angular power spectra, cross-spectra with galaxy shear and galaxy positions, on the same scales, with $b_{\rm IA}(z)$ set to correspond to the commonly used parameterization of alignment amplitude \citep[e.g.][]{Joudaki18} with $A_{\rm IA}=1$ and $f_{\rm IA}(z)=1$.
\item The cross-power spectrum between number counts and CMB lensing, on the same scales.
\item The cross-power spectrum between galaxy weak lensing and CMB weak lensing, on the same scales.
\item Number counts angular correlation functions in the range $0.01 \deg < \theta < 8 \deg$, using 5 bins per decade, and
\item Lensing shear angular correlation functions ($\xi_+$,$\xi_-$), similarly to above.
\item The cross-correlation between both quantities (galaxy-galaxy lensing), as above.
\item The full shape-shape and shape-position observables for both angular power spectra and correlation functions, where the intrinsic alignment contributions to the observables are included.
\end{itemize}
Notice that RSD and magnification predictions are not currently validated. We do not include real-space (correlation function) benchmarks involving CMB lensing, since they are functionally the same as number counts (i.e. a spin-0 quantity). Angular power spectrum and correlation function benchmarks for the full `$3 \times 2{\rm{pt}}$' calculation of galaxy clustering and weak lensing observables was created internally by {\tt CosmoLSS}\footnote{\url{https://github.com/sjoudaki/cosmolss}} and other independent codes.

\begin{figure}
\includegraphics[width=0.49\textwidth]{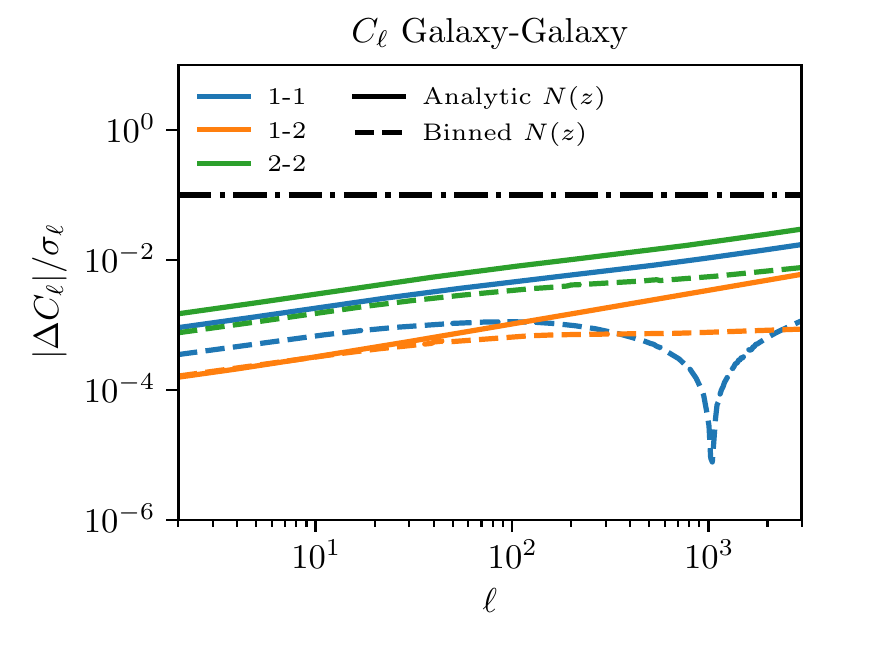}
\includegraphics[width=0.49\textwidth]{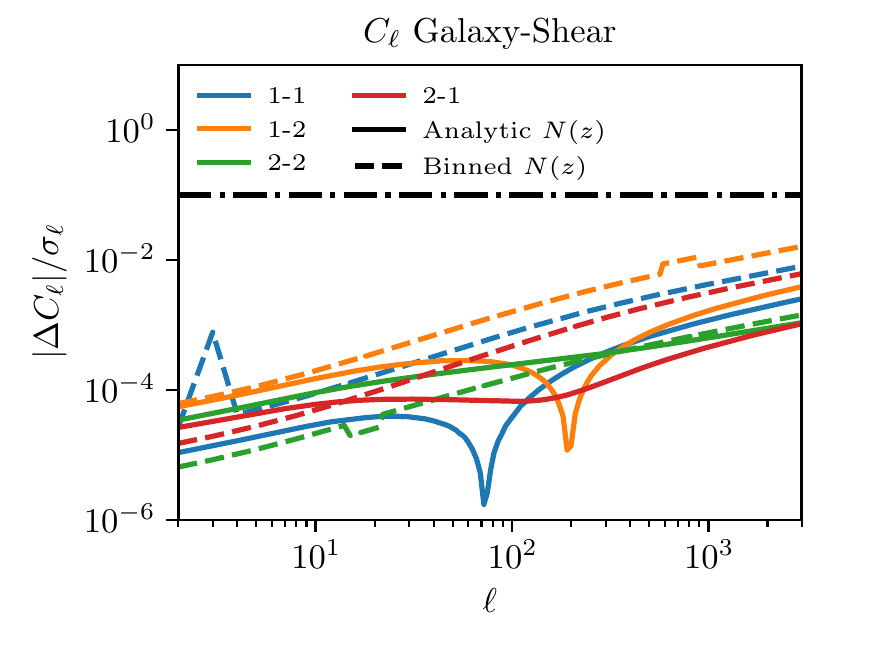}
\includegraphics[width=0.49\textwidth]{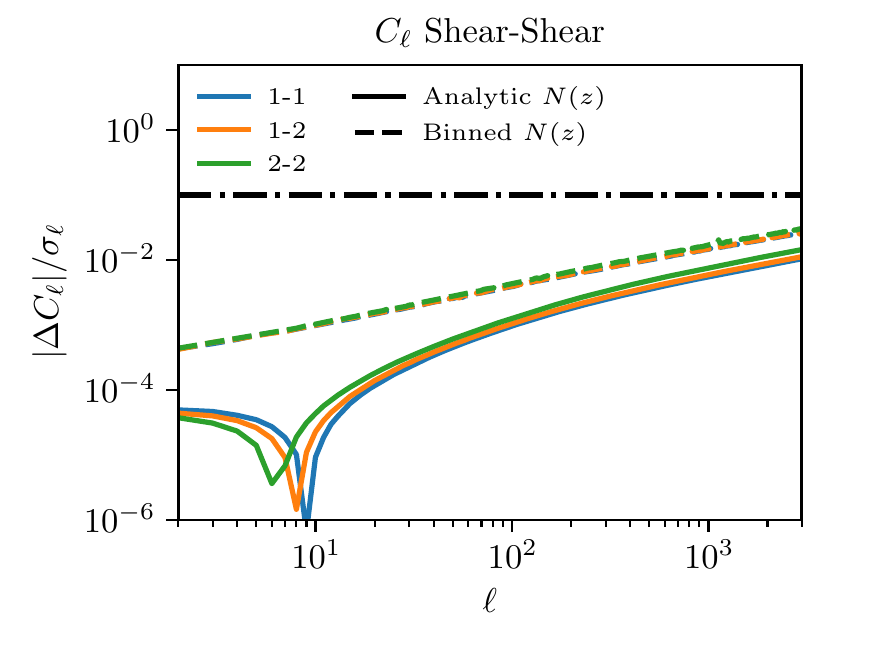}
\caption{Tests of the angular power spectrum accuracy. The black dot-dashed line indicates the target numerical accuracy of $0.1\sigma_\ell$. {\sl Top:} benchmark comparisons for power spectra between number counts in pairs of redshift bins. Results are shown for analytic (solid) and binned (dashed) redshift distributions, for the bin pairs 1-1 (blue), 1-2 (orange), 2-2 (green). {\sl Middle:} same as the top panel for cross-correlations between number counts and weak lensing, this time including the tomographic bin combination 2-1 (red). {\sl Bottom:} same as the top panel, but for the weak lensing power spectra.}
\label{fig:cls_limber}
\end{figure}
\begin{figure}
\includegraphics[width=0.49\textwidth]{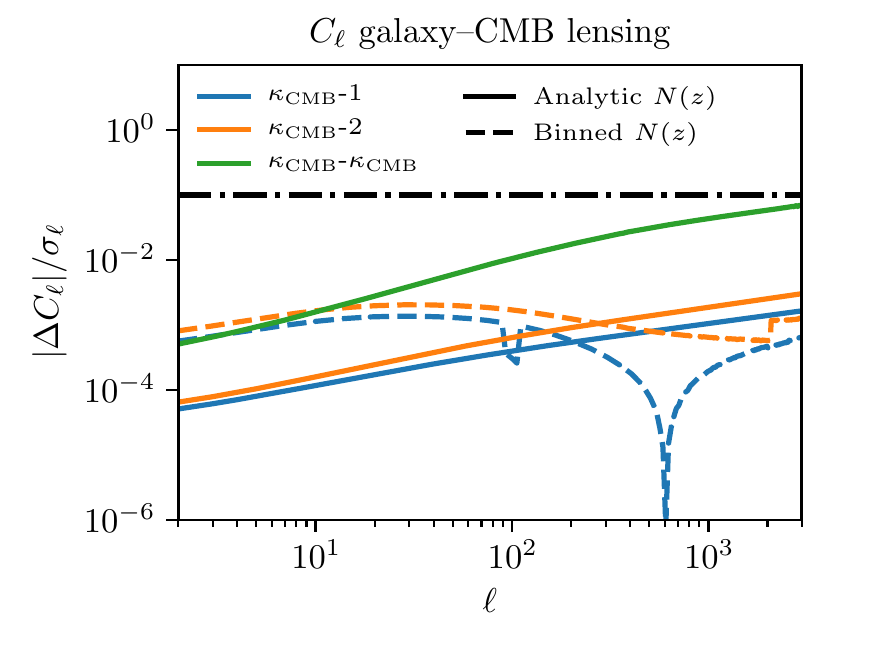}
\includegraphics[width=0.49\textwidth]{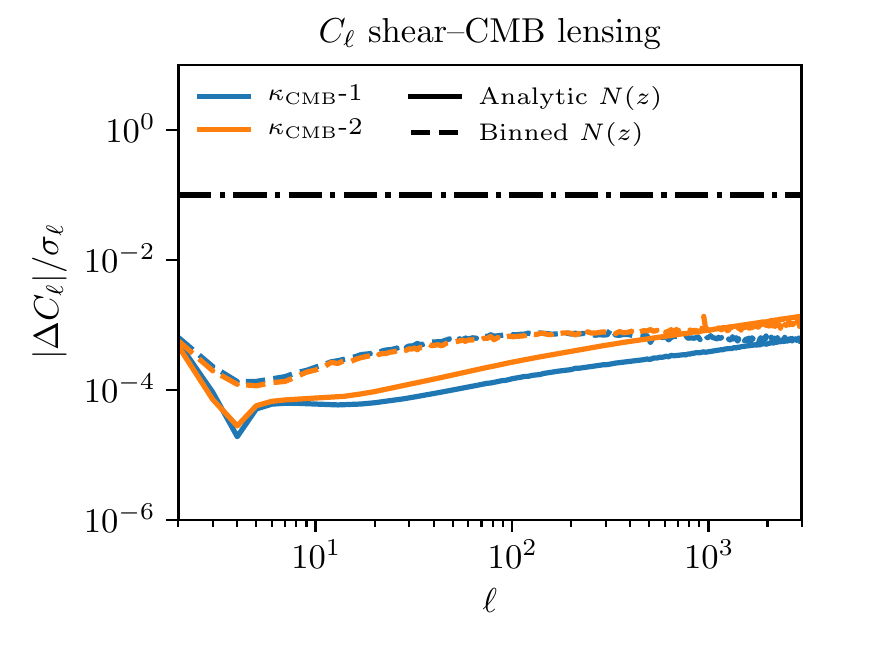}
\caption{Same as Figure \ref{fig:cls_limber} for angular power spectra involving CMB lensing. The black dot-dashed line indicates the target numerical accuracy of $0.1\sigma_\ell$. {\sl Top:} cross-correlations with number counts and CMB auto-correlation. {\sl Bottom:} cross-correlations with weak lensing.}
\label{fig:cls_cmblens}
\end{figure}

For $C_\ell$ computations, we define as our accuracy metric the absolute value of the difference between \ccl and an independent realisation, $i$, as a fraction of the cosmic-variance-limit uncertainties:
\begin{equation}
  \mathcal{A}=\left|\frac{C_\ell^{\tt CCL}-C_\ell^{(i)}}{\sigma_\ell}\right|.
  \label{eq:errorCl}
\end{equation}
For the power spectrum $C^{ab}_\ell$ between two fields $a$ and $b$, the cosmic variance errors are given by
\begin{equation}
  \sigma_\ell^2=\frac{C^{aa}_\ell C^{bb}_\ell+(C^{ab}_\ell)^2}{2\ell+1}.
  \label{eq:sigmaell}
\end{equation}
Our accuracy requirement, for all auto- and cross-correlations with analytic and binned redshift distributions is $\mathcal{A}<0.1$ (i.e. differences must be smaller than one tenth of the cosmic-variance errors). Notice that Eq. (\ref{eq:sigmaell}) assumes a full-sky survey. As a consequence, our requirement is conservative compared to a survey covering $40\%$ of the sky, as expected for LSST, where the uncertainty from cosmic variance would be $\sim 60\%$ larger.

The results for the auto- and cross-correlations between number counts and weak lensing are shown in Figure \ref{fig:cls_limber}, while those involving CMB lensing are shown in Figure \ref{fig:cls_cmblens}. In addition, we have verified that the intrinsic alignment auto-spectra and cross-spectra with shear and galaxy positions satisfy our accuracy requirement (intrinsic-intrinsic, shear-intrinsic, galaxy-intrinsic, and the full observables, i.e, II, GI, gI, GG+II+GI, gG+gI). In this case, the denominator in Eq. (\ref{eq:errorCl}) includes the lensing contribution as well.

The differences between the \ccl~results and the benchmarks are mostly due to the integration and interpolation methods. {\tt CosmoLSS} is written in Fortran (compared to \ccl written in C) independently of any other existing cosmic shear, galaxy-galaxy lensing, and galaxy clustering code. It obtains the expansion history and comoving distance with redshift from \camb (in general \camb also provides the matter power spectrum, which is however taken to be the same BBKS $P(k)$ as \ccl here). It uses a cubic spline interpolation for the redshift distributions, it performs the integral of the lensing kernel using Romberg's method, and for the outer integral to obtain the angular power spectra it uses the trapezoidal rule with 370 logarithmically spaced bins over the full redshift range (it also allows the user to choose Romberg's method for this outer integral). The integral from angular power spectra to correlation functions is performed through a direct summation over all integer multipoles in the range $1<\ell<59000$ (other choices of $\ell_{\rm max}$ and integration methods have been explored and found to yield consistent results). For purposes of benchmarking \ccl, while the impact is negligible, the angular power spectra are not interpolated in this process but directly computed at each of the multipoles.

The independent code that produces additional galaxy-galaxy lensing benchmarks is written in {\tt python} and has a number of differences with respect to \ccl. Not only are redshift distributions interpolated prior to the computation of the angular power spectra, but also lensing kernels. Integration over redshift is performed using the {\tt quad} routine in {\tt python} with a pre-specified relative accuracy threshold of $10^{-7}$. The angular power spectrum is then interpolated between $0<\ell<6\times 10^4$ and integrated to obtain the angular correlation function for this observable.

In the case of CMB lensing, we note that the results are particularly sensitive to numerical errors in the computation of the distance to the last scattering surface. The independent code that provides predictions for auto- and cross-correlations of CMB lensing also adopts a different integration strategy from \ccl. In this case, integration over redshift is performed via direct summation with a default number of $10$ redshift bins. The independent code relies on {\tt astropy} for constants and background computations, and this difference in implementation can contribute to the discrepancies with \ccl.

Cosmological constraints from current weak lensing surveys are also derived from correlation functions. As we discussed in Section \ref{sec:models}, the correlation functions are modeled by Eq.(\ref{eq:cl_xi}) and obtained by \ccl through numerical integration of predicted angular power spectra. We require that the absolute difference between the \ccl prediction and an independent one is smaller than our expected error bars:
\begin{equation}
  \mathcal{A}=\left|\xi^{\tt CCL}-\xi^{(i)}\right|<0.5\sigma_{\rm LSST}.
\end{equation}
where $\sigma_{\rm LSST}$ is the expected statistical uncertainty of any given correlation function between tracers. The choice of an absolute tolerance criterion here (compared to fractional ones in the previous sub-sections) is driven by the fact that the correlation function approaches zero at large scales.

To obtain realistic targets for the convergence of projected correlation function computations for LSST analyses, we calculated the expected statistical uncertainty of the clustering and lensing correlation functions of the LSST gold sample \citep{LSSTSB} assuming an effective source galaxy density of $n_\mathrm{eff} = 26\,\mathrm{gal/sq\,arcmin}$ for galaxy shape distortions \citep{Chang13}, and galaxy density of $n_\mathrm{gold} = 45\,\mathrm{gal/sq\,arcmin}$ for number counts. Specifically, we calculated the Gaussian covariance of angular correlation functions following the formalism of \citet{2008A&A...477...43J}, and note that leaving out the non-Gaussian covariance terms makes our accuracy criterion more conservative. We split the galaxy samples into 10 tomography bins, defined to contain equal numbers of galaxies.

\begin{figure*}
\centering
\includegraphics[width=0.49\textwidth]{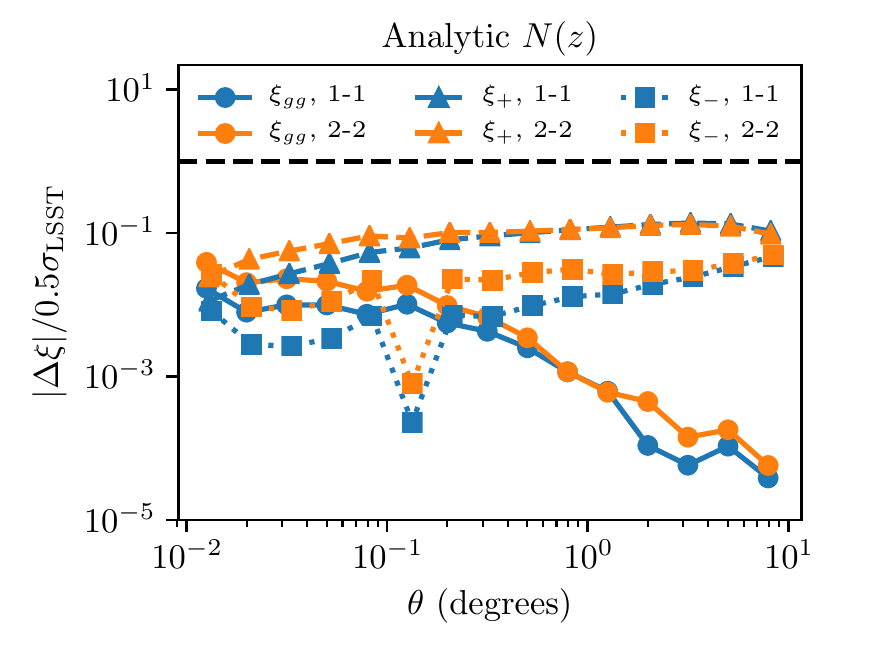} 
\includegraphics[width=0.49\textwidth]{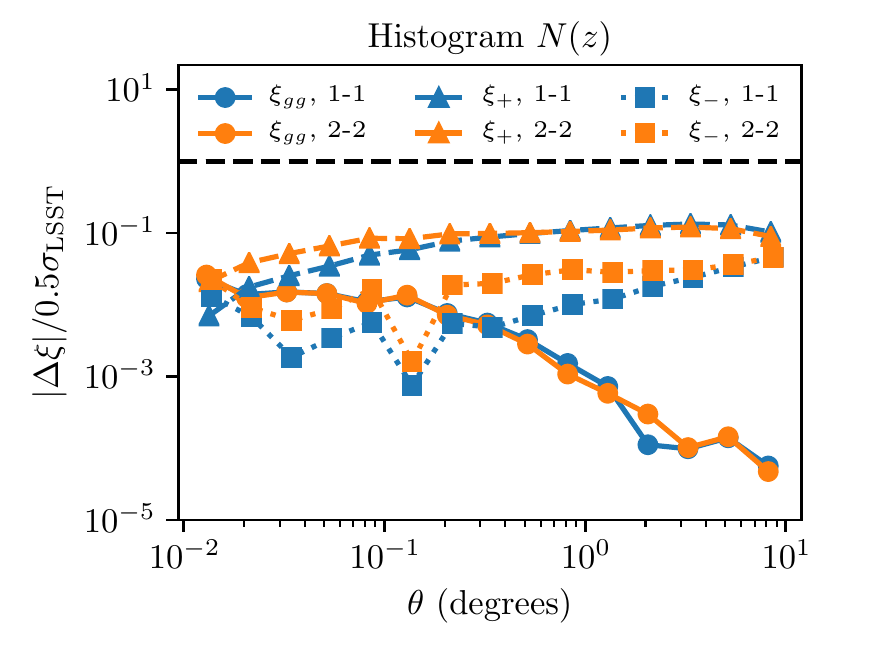} 
\caption{Comparison between the predicted projected correlation functions and the expected uncertainties for LSST. The black dot-dashed line indicates the target numerical accuracy of $0.5\sigma_{\rm LSST}$. The left panel shows predictions for the analytic redshift distributions, while the right panel shows the case of the redshift histograms. The different markers and colors indicate clustering ($\xi_{gg}$, Eq. \ref{eq:xi00flat}, filled circles) or lensing ($\xi_{\pm}$, Eq. \ref{eq:xi22flat}, filled triangles and squares) auto-correlations of the $1$-$1$ or $2$-$2$ redshift bin combinations.}
\label{fig:corrval}
\end{figure*}

\begin{figure}
\centering
\includegraphics[width=0.49\textwidth]{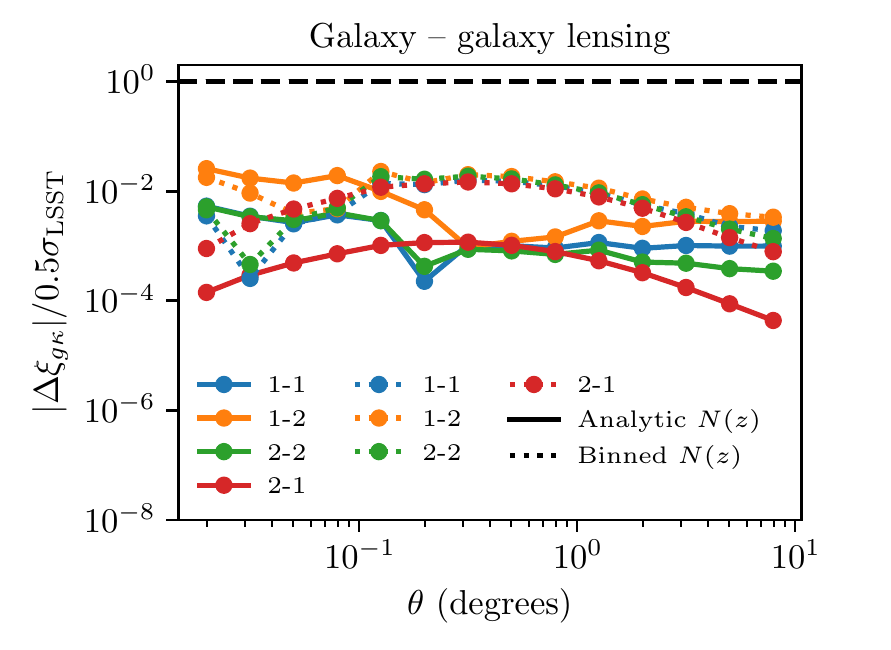}
\caption{Comparison between the predicted galaxy-galaxy lensing projected correlation function and the expected uncertainty for LSST for both analytic and histogram redshift distributions.}
\label{fig:corrval2}
\end{figure}

We compared the difference between \ccl and the benchmarks for the cosmic shear, galaxy-galaxy lensing, and galaxy clustering correlations, for all tomographic bin combinations and both redshift distributions. Specifically, we took the value of the covariance in the bins centered at $z=1$ and $z=1.5$ to compare to the benchmarks. The results of this validation procedure for the projected correlation function are shown in Figures \ref{fig:corrval} and \ref{fig:corrval2}. These suggest that the convergence between \ccl predictions and benchmarks is below the expected statistical uncertainty. Similar to the power spectra, we have verified that the target precision is achieved when including intrinsic alignments.


The three-dimensional spatial correlation function $\xi(r)$ predicted by \ccl was validated by comparing it with an independent, precise numerical transform\footnote{This independent implementation is based on the {\tt cluster toolkit} package, available at \url{http://cluster-toolkit.readthedocs.io/en/latest/}.}. We calculated $\xi(r)$  by transforming the \ccl non-linear \halofit power spectrum using this independent method for the five cosmologies listed in Table~\ref{tab:cosmologies} at redshifts $z = 0,1,2,3,4,5$.  We then compared it with the $\xi(r)$ from \ccl with a sampling of $P(k)$ equal to {\tt N$\_$K$\_$3DCOR} bins per decade. The accuracy metric is defined as
\begin{equation}
	\mathcal{A}=|\xi^{\tt CCL}(r)-\xi^{(i)}(r)| / \xi^{(i)}(r).
  \label{eq:accxi}
\end{equation}
The default value of {\tt N$\_$K$\_$3DCOR}~=~100,000 results in $\mathcal{A} < 2.5 \times 10^{-3}$ for $0.1 < r < 250$~Mpc and $z=0$. The agreement was better for higher redshifts.

\begin{figure*}
\centering
\includegraphics[width=0.49\textwidth]{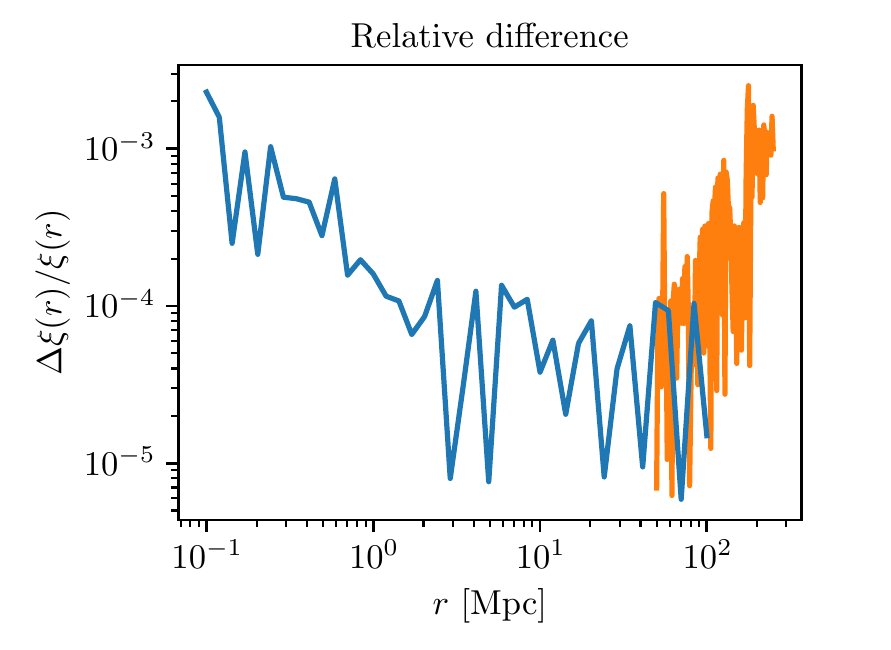}
\includegraphics[width=0.49\textwidth]{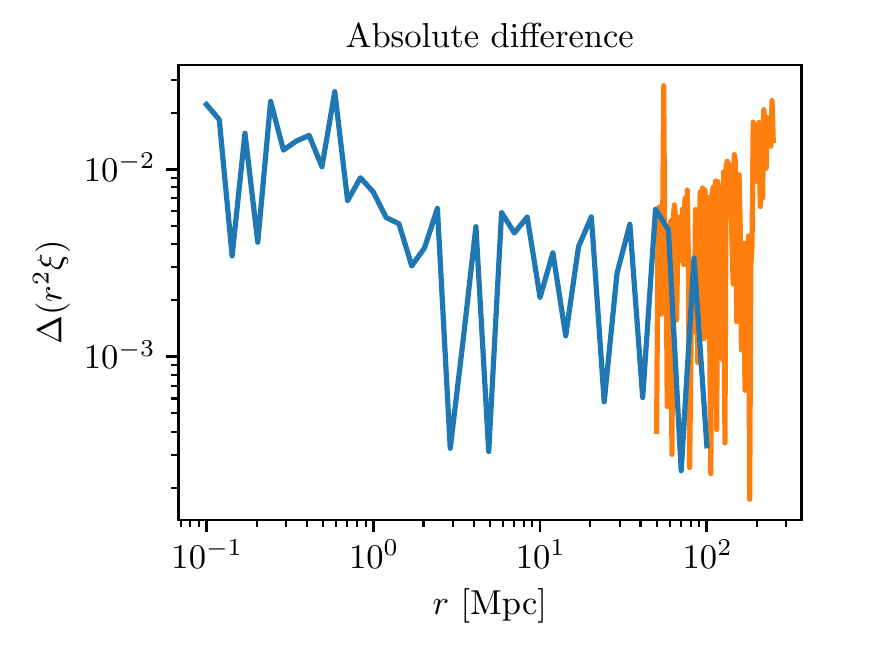} 
\caption{Comparison of the \ccl calculation of the three-dimensional spatial correlation function $\xi(r)$ with a precise, numerical transform of the \ccl non-linear \halofit power spectrum. The left panel shows the relative error $\Delta \xi(r) / \xi(r)$. The right panel shows the absolute error in $r^2 \xi(r)$. Both panels are for the CCL1 model of Table~\ref{tab:cosmologies} at redshift zero. The comparison made with 40 points in the range $r = 0.1 - 100$~Mpc (blue curve) and 100 points in the region $r = 50 - 250$~Mpc encompassing the baryon acoustic oscillation peak.}
\label{fig:benchmark_xi}
\end{figure*}

We also compared the absolute value of $r^2 \xi(r)$ and find a maximum difference of $\Delta (r^2 \xi(r)) < 3.0 \times 10^{-2}$ for the range $r = 0.1 - 250$~Mpc. This corresponds to approximately $0.08\%$ of the baryon acoustic oscillation peak value of $r^2 \xi(r)$. At the peak, the difference is only $9.0 \times 10^{-3}$, or 0.024\% of the peak height. The results are shown in Figure~\ref{fig:benchmark_xi}.

\begin{figure}[htbp]
\centering
\includegraphics[width=0.49\textwidth]{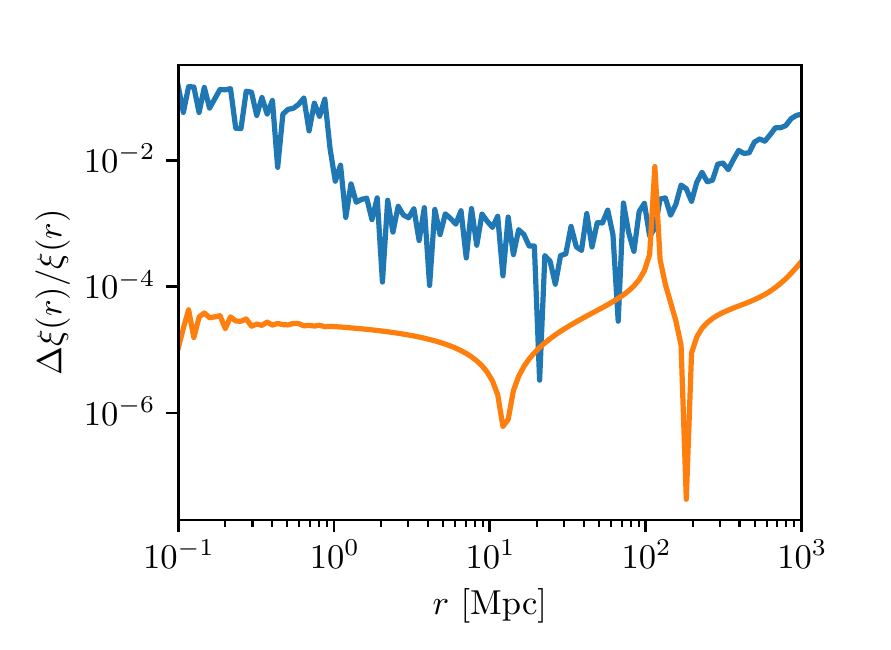}
\caption{The relative error in the three-dimensional spatial correlation function computed using the \ccl algorithm compared to an analytic function $\xi(r) = (r/r_0)^{-1.67}$ (blue curve) and using $P(k) = A k / (k_0 + k)^4$ (orange curve). Both functions have known analytic transforms, $\xi(r)$, but the second one has an asymptotic behavior that matches $P(k)\sim k$ at low $k$. In this validation test, the known $P(k)$ was transformed with the \ccl algorithm and compared to the known analytic result for $\xi(r)$.}
\label{fig:analytic_xi}
\end{figure}

To further validate the $P(k) \to \xi(r)$ transform we performed a test using an analytical function $\xi(r) = (r / r_0)^a$, whose inverse transform $P(k)$ has a known analytic form, $P(k) \propto k^{3+a}$. We used $r_0 = 5 h^{-1}$~Mpc$^{-1}$ and $a = -1.67$, which approximates the actual 3-dimensional correlation function.  We then compared the \ccl calculation of $\xi(r)$ to the known analytic result, defining an analogous metric to Eq. (\ref{eq:accxi}). This was found to be less than 0.4\% in the range $1 < r < 200$~Mpc rising to about 5\% at $r = 1000$~Mpc (see Figure~\ref{fig:analytic_xi}). For $r=0.1-0.8$~Mpc the relative difference is $\approx$8\%. The accuracy at low and high distances can be improved by increasing the range over which the power spectrum splines are evaluated. 

Although this function approximates the true three-dimensional correlation distribution over the range of interest in $r$, the transform $P(k)$ does not have the correct behavior at low $k$ where $P(k) \sim  k$. Therefore, a second test was performed with the function $P(k) = A k / (k_0 + k)^4$, where $A = 100$ and $k_0 = 0.045$~Mpc$^{-1}$, that approximates the true behavior of the power spectrum for all $k$.
The transform to $\xi(r)$ was performed using {\tt Mathematica}\footnote{\url{http://www.wolfram.com/mathematica/}.} and compared to the \ccl calculation. The results, shown in Figure~\ref{fig:analytic_xi} show agreement to within $10^{-4}$ over most of the range in $r$. Since this power spectrum results in a correlation function that turns negative at $r \approx 150$~Mpc, the accuracy metric is large near this value. 

The differences between the \ccl calculations and the benchmarks are primarily due to the method used to compute the transform of Eq.~(\ref{eq:xi3d}). In \ccl we use {\tt FFTLog}, while the benchmarks use either a slower precise numerical integration or an exact analytic expression, and therefore differences at the levels observed are not surprising.

\begin{figure*}[htbp]
\centering
\includegraphics[width=0.49\textwidth]{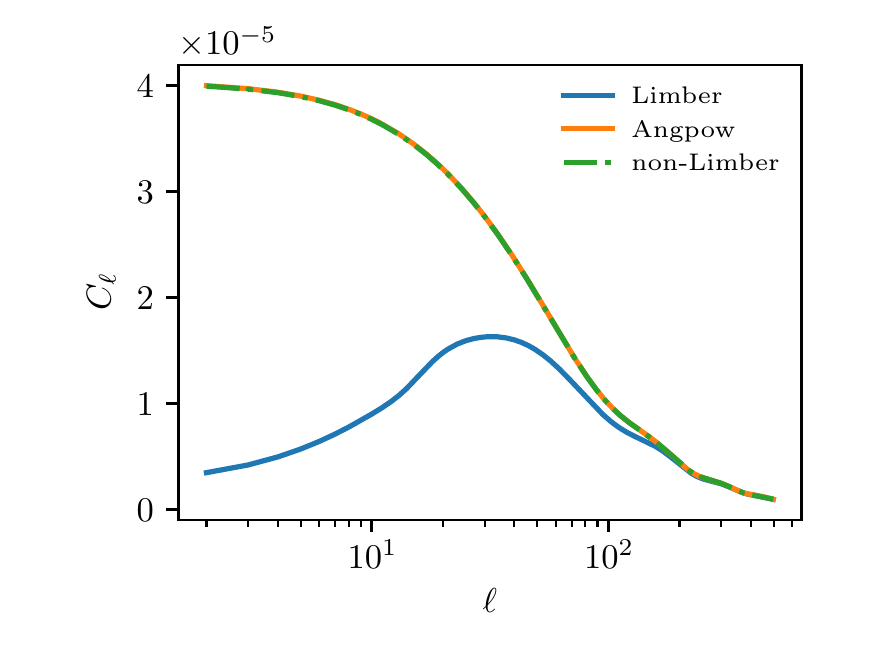}
\includegraphics[width=0.49\textwidth]{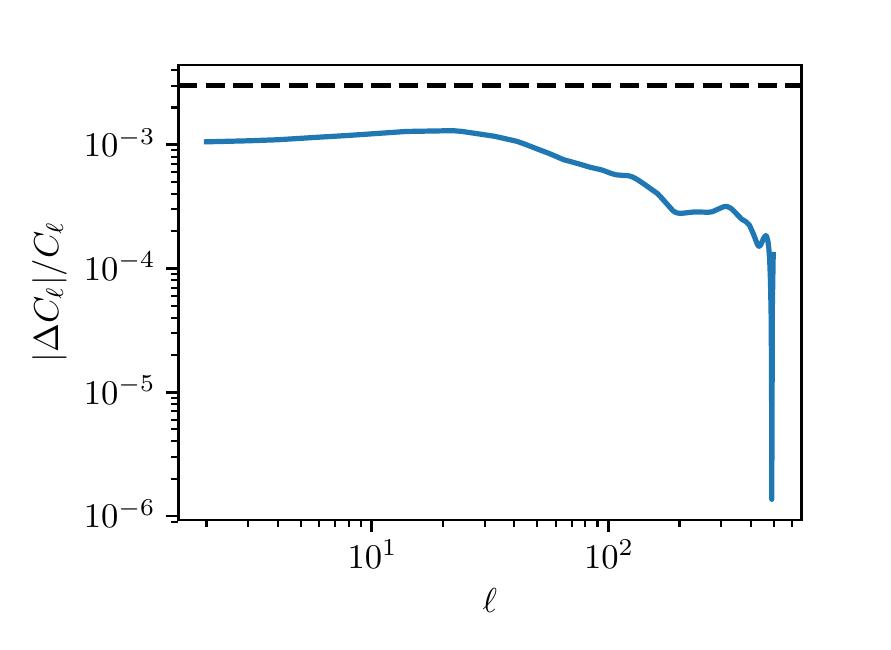}
\caption{Different predictions for the clustering angular power spectrum of a sample of galaxies with a Gaussian redshift distribution centered on $\langle z \rangle =1$. The left panel shows the $C_\ell$ predictions from the Limber case (green), the non-Limber case external to \ccl (yellow dot-dashed) and {\tt Angpow} (black). The right panel shows the fractional difference in the predicted clustering angular power spectrum between {\tt Angpow} and the extrernal brute-force non-Limber computation. The relative numerical difference between the non-Limber computations is lower than two orders of magnitude.}
\label{fig:angpow}
\end{figure*}

\ccl performs non-Limber computations of angular power spectra through the {\tt Angpow} library as detailed in Section \ref{sec:angpow}. The \texttt{Angpow} software \citep{2017A&A...602A..72C} was tested against \class and an external brute-force non-Limber implementation, and can perform the same computations approximately an order of magnitude faster ($\mathcal{O}(1s)$). The external code first carried out a brute-force integration of the transfer functions in Eqs. \ref{eq:transfer_nc}-\ref{eq:cmblens} using a simple trapezium rule, before computing the integral over $k$ in Eq. \ref{eq:cls} using an adaptive quadrature method. Its precision and speed parameters were optimised so that the relative numerical difference between the non-Limber computations is lower than two orders of magnitude, from $\ell=2$ to $\ell=1000$. We demonstrate this in Figure \ref{fig:angpow}, where we plot the angular clustering power spectrum for a sample of galaxies with $\langle z \rangle=1$ and a Gaussian redshift distribution that extends between $|z-\langle z \rangle|<5\sigma_z$, where $\sigma_z=0.02$, for a CCL1 cosmology. The non-Limber prediction deviates from the Limber case at low $l$ as expected. The right panel shows the fractional difference between the non-Limber curves, demonstrating the accuracy of the {\tt Angpow} prediction for our choice of precision and speed parameters. Also the external brute-force non-Limber computation and {\tt Angpow} were tested to recover the Limber approximated curve at high $\ell$ for a wide Gaussian window ($\sigma_z=0.1$). The relative errors with respect to the Limber result at high $\ell$ are also lower than two orders of magnitudes compared to the expected cosmic variance. The differences between the \ccl~results and benchmarks from \class can be due to the integration methods and in particular the choice of the integral cut-off for small scales. The {\tt Angpow} implementation in \ccl sets the $k$ integral cut-off automatically using the user-defined maximum multipole $\ell_{\rm max}$ as $k_{\rm max}=\pi \ell_{\rm max} / \chi(z_{\rm min})$ where $\chi(z_{\rm min})$ is the minimum comoving distance within the redshift shells, while the \class user has to set it appropriately.

\section{Usage}
\label{sec:usage}

\ccl is a public tool developed by the members of the LSST-DESC and can be downloaded from the collaboration's {\tt GitHub} repository\footnote{\url{https://github.com/LSSTDESC/CCL}}. Installation instructions are provided in a README file available in that same repository. In particular, \ccl is installable via {\tt pip} with minimal dependencies. These include {\tt cmake}\footnote{\url{http://cmake.org}}, the GNU Scientific Library\footnote{\url{https://www.gnu.org/software/gsl/}} and FFTW3\footnote{\url{http://www.fftw.org/}}. Instructions on how to generate a Docker\footnote{\url{https://www.docker.com/}} image are provided for portability to different architectures.

A suite of tests can be run to ensure installation was successful and all features perform normally. These comprise accuracy checks performed in C and unit tests available in {\tt python}. These are also run regularly with the {\tt Travis} continuous integration service\footnote{\url{https://travis-ci.org}}, ensuring that the code remains reliable as we continue to improve it. 

The steps to follow to perform a standard cosmological computation (for example, to obtain angular power spectra for galaxy clustering) in \ccl are the following:
\begin{itemize}
\item Set up a {\tt cosmology} object which contains all the information on the cosmological model. This will not only specify the values of the cosmological parameters but also the choice of algorithm for computing the matter power spectrum and information on whether to work under a linear approximation. This step already allows the user to compute quantities such as distances, the Hubble rate or growth functions.
\item In a second step, the user specifies a {\tt tracer} object, which contains all the information pertaining to the sample of galaxies to be modelled. For galaxy clustering, this includes information on the bias of the sample and its redshift distribution. The {\tt tracer} also contains information on how the clustering is to be modelled, e.g., taking into account magnification effects. 
\item Finally, the user can proceed to compute angular power spectra for a given set of multipoles via the function {\tt ccl\_angular\_cls}, by providing the {\tt tracer} object as input.
\end{itemize}
An example run corresponding to this case can be found in the {\tt 3x2demo} notebook or in the {\tt ccl\_sample\_run.c} example in the {\tt examples} folder within the repository.

\ccl is documented online\footnote{\url{https://readthedocs.org/projects/ccl/}} and through Doxygen\footnote{\url{http://www.doxygen.org/}} files released with the repository. The repository also includes multiple example files in C and several {\tt Jupyter} notebooks showing many common use cases.

\ccl is released under terms consistent with BSD 3-Clause licensing\footnote{\url{https://github.com/LSSTDESC/CCL/blob/master/LICENSE}}.

\section{Outlook}
\label{sec:conclusion}

Science software development to facilitate the cosmological inference from LSST data is one of the critical tasks of the DESC. Recent cosmological analyses of the Dark Energy Survey (DES) have relied on {\tt CosmoSIS} \citep{Zuntz14} and {\tt CosmoLike} \citep{krause17}, while analyses of the Kilo Degree Survey (KiDS) have relied on {\tt CosmoLSS} \citep{Joudaki18} (based on {\tt CosmoMC} \citep{Lewis02}) and {\tt Monte Python} \citep{Audren12}. All of these frameworks employ \class, \camb \citep{Challinor2005}, or the Cosmic Emulator to compute the density power spectra. Compared to the analyses of DES, KiDS and the Hyper-Suprime Cam Survey (HSC), future data sets (e.g. LSST, {\it Euclid} and {\it WFIRST}) have substantially higher demands on analysis frameworks. Analyses are becoming more complex in terms of cosmological physics that is included in the analyses (neutrinos, modified gravity, and dark matter models) and in terms of modeling astrophysical and observational systematics at the required precision. 

It is the primary goal of \ccl to become the back-bone of all cosmological analyses carried out by the LSST-DESC. \ccl can also have applications for analytic covariance calculations needed for future analyses of cosmological observables. This unified approach of a validated \ccl will ensure that LSST-DESC results are both consistent (in that they will all be based on the same theory framework) and accurate (in that this framework has undergone a rigorous numerical validation).

Within LSST-DESC, the implementation of \ccl in realistic analysis pipelines has already begun: all likelihood module prototypes under development use it as its back-bone, and the first of these, cosmological analysis of angular galaxy clustering cross-correlations, will serve as a model for the design of the joint-probes likelihood of the LSST-DESC. This work has allowed us to validate the performance of \ccl in a realistic analysis scenario, verifying its accuracy and efficiency in the context of computationally demanding Markov Chain Monte Carlo runs. 

Beyond its usefulness in the LSST-DESC, the flexible design of \ccl makes it an ideal tool for the analysis of other cosmological datasets, as well as for the cross-correlation of different experiments. To this end, and to allow a generic and flexible analysis of the LSST data, further functionality will be added to \ccl. Plans are in place to extend the range of standard and non-standard cosmological models covered by the code, including basic and more complex modified gravity parametrisations \citep{Silvestri2013, Bellini2014} and consistent treatment of the growth function and the matter power spectrum in modified gravity theories. Work is already underway to add predictions for cosmology with clusters \citep{McClintock18}. The simplified treatment of the galaxy-matter connection for galaxy clustering and intrinsic alignments will be improved by implementing generic perturbation-theory approaches to structure formation \citep{FASTPT}. A more complete implementation of all relevant cross-correlations between large-scale structure observables and other cosmological probes (e.g. CMB integrated Sachs-Wolfe effect ---\citealt{1967ApJ...147...73S}---, and other secondary anisotropies) will also soon be included. Likewise, \ccl is expected to provide consistent modeling of complex astrophysical and observational systematics across all probes, critical to LSST analyses.

In general, although this document presents the functionality and performance of \ccl shortly after its release, we expect the library to be a continuously evolving piece of software. In particular, we expect to study the trade-offs between numerical accuracy and speed in the future, as well as to propagate the uncertainties in theoretical predictions to a forecasting framework that can determine their impact in cosmological parameters. This will allow \ccl to satisfy the analysis needs of future large data sets, as well as more accurate and sophisticated models for a broad range of cosmological and astrophysical observables.

\section*{Acknowledgments}

\vskip 5pt
This paper has undergone internal review in the LSST Dark Energy Science Collaboration. We thank the reviewers: Mike Jarvis, Yao-Yuan Mao and Mariana Penna-Lima for comments that helped improved this manuscript and the \ccl library overall. We specially thank Mike Jarvis for performing the \ccl code review and Matt Becker for helping us address it.

The DESC acknowledges ongoing support from the Institut National de Physique Nucl\'eaire et de Physique des Particules in France; the Science \& Technology Facilities Council in the United Kingdom; and the Department of Energy, the National Science Foundation, and the LSST Corporation in the United States.  DESC uses resources of the IN2P3 Computing Center (CC-IN2P3--Lyon/Villeurbanne - France) funded by the Centre National de la Recherche Scientifique; the National Energy Research Scientific Computing Center, a DOE Office of Science User Facility supported by the Office of Science of the U.S.\ Department of Energy under Contract No.\ DE-AC02-05CH11231; STFC DiRAC HPC Facilities, funded by UK BIS National E-infrastructure capital grants; and the UK particle physics grid, supported by the GridPP Collaboration. This work was performed in part under DOE Contract DE-AC02-76SF00515. 

We would like to thank the organisers of the DESC meetings and hack weeks in the period 2015-2018, where this work was partly developed. We would also like to acknowledge the contribution of the participants of the Theory and Joint Probes Code Comparison Project, some of whom are among the \ccl contributors, for providing the benchmarks for testing CCL. We also acknowledge Louis Penafiel and Elizabeth Kimura, who developed the {\tt VARRIC} code to compare and visualize power spectra calculated by \ccl and {\tt CLASS}. We thank Pedro Ferreira and Eric Linder for feedback on this manuscript. We are grateful for the feedback received from other working groups of DESC, including Strong Lensing, Supernovae, Clusters and Photometric Redshifts. We are grateful to Katrin Heitmann and Earl Lawrence for discussions concerning the Cosmic Emulator. We are also thankful to the {\tt CLASS} authors and to Andrew Hamilton for making their codes available and allowing us to use them in this work. We thank Peter Williams for making his ApJ bibstyle file available\footnote{\url{https://github.com/pkgw/tex-stuff/blob/master/yahapj.bst}.}.  

DA is supported by the Science and Technology Facilities Council (STFC) through an Ernest Rutherford Fellowship, grant reference ST/P004474/1. NEC acknowledges support from a Beecroft fellowship and a Royal Astronomical Society Research Fellowship. TT acknowledges funding from the European Union's Horizon 2020 research and innovation programme under the Marie Sk{l}odowska-Curie grant agreement No.\ 797794. MI acknowledges that this material is based upon work supported in part by NSF under grant AST-1517768 and the U.S. Department of Energy, Office of Science, under Award Number DE-SC0019206.

\vskip 5pt
Author contributions are listed below. \\
Nora Elisa Chisari: Co-led project, coordinated hack projects \& communication, contributed to: correlation function \& power spectrum implementation, documentation, and comparisons with benchmarks. \\
David Alonso: Co-led project; developed structure for angular power spectra; implemented autotools; integrated into LSS pipeline; contributed to: background, power spectrum, mass function, documentation and benchmarks; reviewed code \\
Elisabeth Krause: Initiated and co-led project; developed CLASS interface and error handling; contributed to other code; reviewed pull requests. \\
C. Danielle Leonard: Wrote and tested code for LSST specifications, user-defined photo-z interface, and support of neutrinos; reviewed other code; wrote text for this note. \\
Philip Bull: Implemented and documented the Python wrapper; extensive infrastructure work, including maintenance of the build system and unit tests, bug fixes, and code review; enhanced the error handling system; performed power spectrum comparison across parameter space; various architecture/design contributions. \\
J\'er\'emy Neveu: Contributed to Angpow, built the interface with CCL and the comparisons with benchmarks. \\
Antonio Villarreal: Contributed to initial benchmarking, halo mass function code, and general code and issues review. \\
Sukhdeep Singh: Contributed to the correlation functions code and weak lensing benchmarks. \\
Thomas McClintock: Wrote Python and doxygen documentation. \\
John Ellison: Implemented the 3d correlation function; documentation of 3d correlation function. \\
Zilong Du: Implemented the 3d correlation function. \\
Joe Zuntz: Wrote initial infrastructure, C testing setup, and reviewed code. \\
Alexander Mead: Wrote halo-model code and documentation \\
Shahab Joudaki: Performed benchmarking of angular power spectra and correlation functions for galaxy clustering, galaxy-galaxy lensing, and cosmic shear (including intrinsic alignments); reviewed all CCL benchmarks; contributed to implementation of background functions and documentation. \\
Christiane S. Lorenz: Contributed to accurate high-redshift cosmological background quantities and benchmarked background splines. \\
Tilman Tr\"oster: Wrote code for user-changable precision parameters and distance tests, found and fixed bugs. \\
Javier Sanchez: Modified setup.py to allow pip installation and uninstall. \\
Francois Lanusse: Contributed to improving installation options. \\
Mustapha Ishak: Contributed to planning of code capabilities and structure; reviewed code; identified bugs and wrote code to fix them; wrote text for this note. \\
Ren\'ee Hlozek: Contributed initial code for error handling structures, reviewed other code edits. \\
Jonathan Blazek: Planning capabilities and structure; documentation, review, and testing. \\
Jean-Eric Campagne: Angpow main builder. \\
Husni Almoubayyed: Reviewed code/contributed to example notebooks and issues. \\
Tim Eifler: Reviewed/tested code. \\
Matthew Kirby: Performed comparison of physical constants. \\
David Kirkby: Writing, testing and reviewing code. Asking questions. \\
St\'ephane Plaszczynski: Contributed to Angpow. \\
An\v{z}e Slosar: Wrote and reviewed code. \\
Michal Vrastil: Wrote documentation and example code, reviewed code. \\
Erika L. Wagoner: Set up logarithmic binning in wave number. \\

\bibliography{main}

\begin{thebibliography}{}
\providecommand\natexlab[1]{#1}
\providecommand\JournalTitle[1]{#1}

\bibitem[{{Afshordi} {et~al.}(2004){Afshordi}, {Loh}, \&
  {Strauss}}]{2004PhRvD..69h3524A}
{Afshordi}, N., {Loh}, Y.-S., \& {Strauss}, M.~A. 2004,
  \href{http://dx.doi.org/10.1103/PhysRevD.69.083524}{\JournalTitle{\prd}, 69,
  083524}

\bibitem[{{Allison} {et~al.}(2015){Allison}, {Caucal}, {Calabrese}, {Dunkley},
  \& {Louis}}]{Allison15}
{Allison}, R., {Caucal}, P., {Calabrese}, E., {Dunkley}, J., \& {Louis}, T.
  2015,
  \href{http://dx.doi.org/10.1103/PhysRevD.92.123535}{\JournalTitle{\prd}, 92,
  123535}

\bibitem[{{Alonso} {et~al.}(2015){Alonso}, {Bull}, {Ferreira}, {Maartens}, \&
  {Santos}}]{GReffects}
{Alonso}, D., {Bull}, P., {Ferreira}, P.~G., {Maartens}, R., \& {Santos}, M.~G.
  2015,
  \href{http://dx.doi.org/10.1088/0004-637X/814/2/145}{\JournalTitle{ApJ}, 814,
  145}

\bibitem[{{Alonso} \& {Ferreira}(2015)}]{Alonso15}
{Alonso}, D., \& {Ferreira}, P.~G. 2015,
  \href{http://dx.doi.org/10.1103/PhysRevD.92.063525}{\JournalTitle{\prd}, 92,
  063525}

\bibitem[{{Angulo} {et~al.}(2012){Angulo}, {Springel}, {White}, {Jenkins},
  {Baugh}, \& {Frenk}}]{Angulo2012}
{Angulo}, R.~E., {Springel}, V., {White}, S.~D.~M., {et~al.} 2012,
  \href{http://dx.doi.org/10.1111/j.1365-2966.2012.21830.x}{\JournalTitle{\mnras},
  426, 2046}

\bibitem[{{Astropy Collaboration} {et~al.}(2013){Astropy Collaboration},
  {Robitaille}, {Tollerud}, {Greenfield}, {Droettboom}, {Bray}, {Aldcroft},
  {Davis}, {Ginsburg}, {Price-Whelan}, {Kerzendorf}, {Conley}, {Crighton},
  {Barbary}, {Muna}, {Ferguson}, {Grollier}, {Parikh}, {Nair}, {Unther},
  {Deil}, {Woillez}, {Conseil}, {Kramer}, {Turner}, {Singer}, {Fox}, {Weaver},
  {Zabalza}, {Edwards}, {Azalee Bostroem}, {Burke}, {Casey}, {Crawford},
  {Dencheva}, {Ely}, {Jenness}, {Labrie}, {Lim}, {Pierfederici}, {Pontzen},
  {Ptak}, {Refsdal}, {Servillat}, \& {Streicher}}]{astropy}
{Astropy Collaboration}, {Robitaille}, T.~P., {Tollerud}, E.~J., {et~al.} 2013,
  \href{http://dx.doi.org/10.1051/0004-6361/201322068}{\JournalTitle{\aap},
  558, A33}

\bibitem[{{Audren} {et~al.}(2013){Audren}, {Lesgourgues}, {Benabed}, \&
  {Prunet}}]{Audren12}
{Audren}, B., {Lesgourgues}, J., {Benabed}, K., \& {Prunet}, S. 2013,
  \href{http://dx.doi.org/10.1088/1475-7516/2013/02/001}{\JournalTitle{\jcap},
  2, 001}

\bibitem[{{Bardeen} {et~al.}(1986){Bardeen}, {Bond}, {Kaiser}, \&
  {Szalay}}]{BBKS}
{Bardeen}, J.~M., {Bond}, J.~R., {Kaiser}, N., \& {Szalay}, A.~S. 1986,
  \href{http://dx.doi.org/10.1086/164143}{\JournalTitle{\apj}, 304, 15}

\bibitem[{{Bartelmann} \& {Schneider}(2001)}]{Bartelmann01}
{Bartelmann}, M., \& {Schneider}, P. 2001,
  \href{http://dx.doi.org/10.1016/S0370-1573(00)00082-X}{\JournalTitle{\physrep},
  340, 291}

\bibitem[{{Bellini} \& {Sawicki}(2014)}]{Bellini2014}
{Bellini}, E., \& {Sawicki}, I. 2014,
  \href{http://dx.doi.org/10.1088/1475-7516/2014/07/050}{\JournalTitle{\jcap},
  7, 050}

\bibitem[{Beringer {et~al.}(2012)}]{Beringer:1900zz}
Beringer, J., {et~al.} 2012,
  \href{http://dx.doi.org/10.1103/PhysRevD.86.010001}{\JournalTitle{Phys.
  Rev.}, D86, 010001}

\bibitem[{{Blas} {et~al.}(2011){Blas}, {Lesgourgues}, \& {Tram}}]{class}
{Blas}, D., {Lesgourgues}, J., \& {Tram}, T. 2011, {CLASS: Cosmic Linear
  Anisotropy Solving System}, Astrophysics Source Code Library,
  \href{http://arxiv.org/abs/1106.020}{{\sffamily ascl:1106.020}}

\bibitem[{{Blazek} {et~al.}(2017){Blazek}, {MacCrann}, {Troxel}, \&
  {Fang}}]{Blazek17}
{Blazek}, J., {MacCrann}, N., {Troxel}, M.~A., \& {Fang}, X. 2017,
  \JournalTitle{ArXiv e-prints},
  \href{http://arxiv.org/abs/1708.09247}{{\sffamily arXiv:1708.09247}}

\bibitem[{{Bonvin} \& {Durrer}(2011)}]{2011PhRvD..84f3505B}
{Bonvin}, C., \& {Durrer}, R. 2011,
  \href{http://dx.doi.org/10.1103/PhysRevD.84.063505}{\JournalTitle{\prd}, 84,
  063505}

\bibitem[{Bryan \& Norman(1998)}]{Bryan1998}
Bryan, G.~L., \& Norman, M.~L. 1998,
  \href{http://dx.doi.org/10.1086/305262}{\JournalTitle{\apj}, 495, 80}

\bibitem[{{Campagne} {et~al.}(2017{\natexlab{a}}){Campagne}, {Neveu}, \&
  {Plaszczynski}}]{2017A&A...602A..72C}
{Campagne}, J.-E., {Neveu}, J., \& {Plaszczynski}, S. 2017{\natexlab{a}},
  \href{http://dx.doi.org/10.1051/0004-6361/201730399}{\JournalTitle{\aap},
  602, A72}

\bibitem[{{Campagne} {et~al.}(2017{\natexlab{b}}){Campagne}, {Plaszczynski}, \&
  {Neveu}}]{2017ApJ...845...28C}
{Campagne}, J.-E., {Plaszczynski}, S., \& {Neveu}, J. 2017{\natexlab{b}},
  \href{http://dx.doi.org/10.3847/1538-4357/aa7cf8}{\JournalTitle{\apj}, 845,
  28}

\bibitem[{{Carroll}(2001)}]{Carroll2001CC}
{Carroll}, S.~M. 2001,
  \href{http://dx.doi.org/10.12942/lrr-2001-1}{\JournalTitle{Living Reviews in
  Relativity}, 4, 1}

\bibitem[{{Castorina} {et~al.}(2015){Castorina}, {Carbone}, {Bel}, {Sefusatti},
  \& {Dolag}}]{Castorina15}
{Castorina}, E., {Carbone}, C., {Bel}, J., {Sefusatti}, E., \& {Dolag}, K.
  2015,
  \href{http://dx.doi.org/10.1088/1475-7516/2015/07/043}{\JournalTitle{\jcap},
  7, 043}

\bibitem[{{Catelan} {et~al.}(2001){Catelan}, {Kamionkowski}, \&
  {Blandford}}]{Catelan01}
{Catelan}, P., {Kamionkowski}, M., \& {Blandford}, R.~D. 2001,
  \href{http://dx.doi.org/10.1046/j.1365-8711.2001.04105.x}{\JournalTitle{\mnras},
  320, L7}

\bibitem[{{Challinor} \& {Lewis}(2005)}]{Challinor2005}
{Challinor}, A., \& {Lewis}, A. 2005,
  \href{http://dx.doi.org/10.1103/PhysRevD.71.103010}{\JournalTitle{\prd}, 71,
  103010}

\bibitem[{{Challinor} \& {Lewis}(2011)}]{2011PhRvD..84d3516C}
---. 2011,
  \href{http://dx.doi.org/10.1103/PhysRevD.84.043516}{\JournalTitle{\prd}, 84,
  043516}

\bibitem[{{Chang} {et~al.}(2013){Chang}, {Jarvis}, {Jain}, {Kahn}, {Kirkby},
  {Connolly}, {Krughoff}, {Peng}, \& {Peterson}}]{Chang13}
{Chang}, C., {Jarvis}, M., {Jain}, B., {et~al.} 2013,
  \href{http://dx.doi.org/10.1093/mnras/stt1156}{\JournalTitle{\mnras}, 434,
  2121}

\bibitem[{{Chevallier} \& {Polarski}(2001)}]{Chevallier01}
{Chevallier}, M., \& {Polarski}, D. 2001,
  \href{http://dx.doi.org/10.1142/S0218271801000822}{\JournalTitle{International
  Journal of Modern Physics D}, 10, 213}

\bibitem[{{Chisari} {et~al.}(2018){Chisari}, {Richardson}, {Devriendt},
  {Dubois}, {Schneider}, {Le Brun}, {Beckmann}, {Peirani}, {Slyz}, \&
  {Pichon}}]{Chisari18}
{Chisari}, N.~E., {Richardson}, M.~L.~A., {Devriendt}, J., {et~al.} 2018,
  \href{http://dx.doi.org/10.1093/mnras/sty2093}{\JournalTitle{\mnras}, 480,
  3962}

\bibitem[{{Chon} {et~al.}(2004){Chon}, {Challinor}, {Prunet}, {Hivon}, \&
  {Szapudi}}]{2004MNRAS.350..914C}
{Chon}, G., {Challinor}, A., {Prunet}, S., {Hivon}, E., \& {Szapudi}, I. 2004,
  \href{http://dx.doi.org/10.1111/j.1365-2966.2004.07737.x}{\JournalTitle{\mnras},
  350, 914}

\bibitem[{Cooray \& Sheth(2002)}]{Cooray2002}
Cooray, A., \& Sheth, R. 2002,
  \href{http://dx.doi.org/10.1016/S0370-1573(02)00276-4}{\JournalTitle{Physics
  Reports}, 372, 1}

\bibitem[{{Copeland} {et~al.}(2006){Copeland}, {Sami}, \&
  {Tsujikawa}}]{Copeland2006}
{Copeland}, E.~J., {Sami}, M., \& {Tsujikawa}, S. 2006,
  \href{http://dx.doi.org/10.1142/S021827180600942X}{\JournalTitle{International
  Journal of Modern Physics D}, 15, 1753}

\bibitem[{{Croton} {et~al.}(2007){Croton}, {Gao}, \& {White}}]{Croton2007}
{Croton}, D.~J., {Gao}, L., \& {White}, S.~D.~M. 2007,
  \href{http://dx.doi.org/10.1111/j.1365-2966.2006.11230.x}{\JournalTitle{\mnras},
  374, 1303}

\bibitem[{{Dalal} {et~al.}(2008){Dalal}, {Dor{\'e}}, {Huterer}, \&
  {Shirokov}}]{Dalal08}
{Dalal}, N., {Dor{\'e}}, O., {Huterer}, D., \& {Shirokov}, A. 2008,
  \href{http://dx.doi.org/10.1103/PhysRevD.77.123514}{\JournalTitle{\prd}, 77,
  123514}

\bibitem[{{DES Collaboration} {et~al.}(2017){DES Collaboration}, {Abbott},
  {Abdalla}, {Alarcon}, {Aleksi{\'c}}, {Allam}, {Allen}, {Amara}, {Annis},
  {Asorey}, {Avila}, {Bacon}, {Balbinot}, {Banerji}, {Banik}, {Barkhouse},
  {Baumer}, {Baxter}, {Bechtol}, {Becker}, {Benoit-L{\'e}vy}, {Benson},
  {Bernstein}, {Bertin}, {Blazek}, {Bridle}, {Brooks}, {Brout}, {Buckley-Geer},
  {Burke}, {Busha}, {Capozzi}, {Carnero Rosell}, {Carrasco Kind}, {Carretero},
  {Castander}, {Cawthon}, {Chang}, {Chen}, {Childress}, {Choi}, {Conselice},
  {Crittenden}, {Crocce}, {Cunha}, {D'Andrea}, {da Costa}, {Das}, {Davis},
  {Davis}, {De Vicente}, {DePoy}, {DeRose}, {Desai}, {Diehl}, {Dietrich},
  {Dodelson}, {Doel}, {Drlica-Wagner}, {Eifler}, {Elliott}, {Elsner},
  {Elvin-Poole}, {Estrada}, {Evrard}, {Fang}, {Fernandez}, {Fert{\'e}},
  {Finley}, {Flaugher}, {Fosalba}, {Friedrich}, {Frieman},
  {Garc{\'{\i}}a-Bellido}, {Garcia-Fernandez}, {Gatti}, {Gaztanaga}, {Gerdes},
  {Giannantonio}, {Gill}, {Glazebrook}, {Goldstein}, {Gruen}, {Gruendl},
  {Gschwend}, {Gutierrez}, {Hamilton}, {Hartley}, {Hinton}, {Honscheid},
  {Hoyle}, {Huterer}, {Jain}, {James}, {Jarvis}, {Jeltema}, {Johnson},
  {Johnson}, {Kacprzak}, {Kent}, {Kim}, {King}, {Kirk}, {Kokron}, {Kovacs},
  {Krause}, {Krawiec}, {Kremin}, {Kuehn}, {Kuhlmann}, {Kuropatkin}, {Lacasa},
  {Lahav}, {Li}, {Liddle}, {Lidman}, {Lima}, {Lin}, {MacCrann}, {Maia},
  {Makler}, {Manera}, {March}, {Marshall}, {Martini}, {McMahon}, {Melchior},
  {Menanteau}, {Miquel}, {Miranda}, {Mudd}, {Muir}, {M{\"o}ller}, {Neilsen},
  {Nichol}, {Nord}, {Nugent}, {Ogando}, {Palmese}, {Peacock}, {Peiris},
  {Peoples}, {Percival}, {Petravick}, {Plazas}, {Porredon}, {Prat}, {Pujol},
  {Rau}, {Refregier}, {Ricker}, {Roe}, {Rollins}, {Romer}, {Roodman},
  {Rosenfeld}, {Ross}, {Rozo}, {Rykoff}, {Sako}, {Salvador}, {Samuroff},
  {S{\'a}nchez}, {Sanchez}, {Santiago}, {Scarpine}, {Schindler}, {Scolnic},
  {Secco}, {Serrano}, {Sevilla-Noarbe}, {Sheldon}, {Smith}, {Smith}, {Smith},
  {Soares-Santos}, {Sobreira}, {Suchyta}, {Tarle}, {Thomas}, {Troxel},
  {Tucker}, {Tucker}, {Uddin}, {Varga}, {Vielzeuf}, {Vikram}, {Vivas},
  {Walker}, {Wang}, {Wechsler}, {Weller}, {Wester}, {Wolf}, {Yanny}, {Yuan},
  {Zenteno}, {Zhang}, {Zhang}, \& {Zuntz}}]{DEScombined}
{DES Collaboration}, {Abbott}, T.~M.~C., {Abdalla}, F.~B., {et~al.} 2017,
  \JournalTitle{ArXiv e-prints},
  \href{http://arxiv.org/abs/1708.01530}{{\sffamily arXiv:1708.01530}}

\bibitem[{{Desjacques} {et~al.}(2016){Desjacques}, {Jeong}, \&
  {Schmidt}}]{2016arXiv161109787D}
{Desjacques}, V., {Jeong}, D., \& {Schmidt}, F. 2016, \JournalTitle{ArXiv
  e-prints}, \href{http://arxiv.org/abs/1611.09787}{{\sffamily
  arXiv:1611.09787}}

\bibitem[{{Desjacques} \& {Seljak}(2010)}]{Desjacques10}
{Desjacques}, V., \& {Seljak}, U. 2010,
  \href{http://dx.doi.org/10.1088/0264-9381/27/12/124011}{\JournalTitle{Classical
  and Quantum Gravity}, 27, 124011}

\bibitem[{{Dodelson}(2004)}]{DodelsonBook}
{Dodelson}, S. 2004, {Modern Cosmology}, Vol.~57, 60

\bibitem[{{Doux} {et~al.}(2018){Doux}, {Penna-Lima}, {Vitenti}, {Tr{\'e}guer},
  {Aubourg}, \& {Ganga}}]{numcosmo}
{Doux}, C., {Penna-Lima}, M., {Vitenti}, S.~D.~P., {et~al.} 2018,
  \href{http://dx.doi.org/10.1093/mnras/sty2160}{\JournalTitle{\mnras}},
  \href{http://arxiv.org/abs/1706.04583}{{\sffamily arXiv:1706.04583}}

\bibitem[{{Duffy} {et~al.}(2008){Duffy}, {Schaye}, {Kay}, \& {Dalla
  Vecchia}}]{Duffy2008}
{Duffy}, A.~R., {Schaye}, J., {Kay}, S.~T., \& {Dalla Vecchia}, C. 2008,
  \href{http://dx.doi.org/10.1111/j.1745-3933.2008.00537.x}{\JournalTitle{\mnras},
  390, L64}

\bibitem[{{Durrer}(2008)}]{2008cmb..book.....D}
{Durrer}, R. 2008, {The Cosmic Microwave Background} (Cambridge University
  Press)

\bibitem[{{Eifler} {et~al.}(2015){Eifler}, {Krause}, {Dodelson}, {Zentner},
  {Hearin}, \& {Gnedin}}]{Eifler15}
{Eifler}, T., {Krause}, E., {Dodelson}, S., {et~al.} 2015,
  \href{http://dx.doi.org/10.1093/mnras/stv2000}{\JournalTitle{\mnras}, 454,
  2451}

\bibitem[{{Eisenstein} \& {Hu}(1998)}]{1998ApJ...496..605E}
{Eisenstein}, D.~J., \& {Hu}, W. 1998,
  \href{http://dx.doi.org/10.1086/305424}{\JournalTitle{\apj}, 496, 605}

\bibitem[{{Feng}(2010)}]{Feng10}
{Feng}, J.~L. 2010,
  \href{http://dx.doi.org/10.1146/annurev-astro-082708-101659}{\JournalTitle{\araa},
  48, 495}

\bibitem[{{Frigo} \& {Johnson}(2012)}]{FFTW}
{Frigo}, M., \& {Johnson}, S.~G. 2012, {FFTW: Fastest Fourier Transform in the
  West}, Astrophysics Source Code Library,
  \href{http://arxiv.org/abs/1201.015}{{\sffamily ascl:1201.015}}

\bibitem[{{Gao} {et~al.}(2005){Gao}, {Springel}, \& {White}}]{Gao2005}
{Gao}, L., {Springel}, V., \& {White}, S.~D.~M. 2005,
  \href{http://dx.doi.org/10.1111/j.1745-3933.2005.00084.x}{\JournalTitle{\mnras},
  363, L66}

\bibitem[{{Ghosh} {et~al.}(2018){Ghosh}, {Durrer}, \& {Sellentin}}]{ghosh18}
{Ghosh}, B., {Durrer}, R., \& {Sellentin}, E. 2018, \JournalTitle{ArXiv
  e-prints}, \href{http://arxiv.org/abs/1801.02518}{{\sffamily
  arXiv:1801.02518}}

\bibitem[{{Giocoli} {et~al.}(2010){Giocoli}, {Bartelmann}, {Sheth}, \&
  {Cacciato}}]{Giocoli2010}
{Giocoli}, C., {Bartelmann}, M., {Sheth}, R.~K., \& {Cacciato}, M. 2010,
  \href{http://dx.doi.org/10.1111/j.1365-2966.2010.17108.x}{\JournalTitle{\mnras},
  408, 300}

\bibitem[{{Green} {et~al.}(2011){Green}, {Schechter}, {Baltay}, {Bean},
  {Bennett}, {Brown}, {Conselice}, {Donahue}, {Gaudi}, {Lauer}, {Perlmutter},
  {Rauscher}, {Rhodes}, {Roellig}, {Stern}, {Sumi}, {Tanner}, {Wang}, {Wright},
  {Gehrels}, {Sambruna}, \& {Traub}}]{green11}
{Green}, J., {Schechter}, P., {Baltay}, C., {et~al.} 2011, \JournalTitle{ArXiv
  e-prints}, \href{http://arxiv.org/abs/1108.1374}{{\sffamily arXiv:1108.1374
  [astro-ph.IM]}}

\bibitem[{{Hamilton}(2000)}]{Hamilton2000}
{Hamilton}, A.~J.~S. 2000,
  \href{http://dx.doi.org/10.1046/j.1365-8711.2000.03071.x}{\JournalTitle{\mnras},
  312, 257}

\bibitem[{{Heitmann} {et~al.}(2016){Heitmann}, {Bingham}, {Lawrence},
  {Bergner}, {Habib}, {Higdon}, {Pope}, {Biswas}, {Finkel}, {Frontiere}, \&
  {Bhattacharya}}]{Heitmann16}
{Heitmann}, K., {Bingham}, D., {Lawrence}, E., {et~al.} 2016,
  \href{http://dx.doi.org/10.3847/0004-637X/820/2/108}{\JournalTitle{\apj},
  820, 108}

\bibitem[{{Hellwing} {et~al.}(2016){Hellwing}, {Schaller}, {Frenk}, {Theuns},
  {Schaye}, {Bower}, \& {Crain}}]{Hellwing16}
{Hellwing}, W.~A., {Schaller}, M., {Frenk}, C.~S., {et~al.} 2016,
  \href{http://dx.doi.org/10.1093/mnrasl/slw081}{\JournalTitle{\mnras}, 461,
  L11}

\bibitem[{{Hildebrandt} {et~al.}(2017){Hildebrandt}, {Viola}, {Heymans},
  {Joudaki}, {Kuijken}, {Blake}, {Erben}, {Joachimi}, {Klaes}, {Miller},
  {Morrison}, {Nakajima}, {Verdoes Kleijn}, {Amon}, {Choi}, {Covone}, {de
  Jong}, {Dvornik}, {Fenech Conti}, {Grado}, {Harnois-D{\'e}raps}, {Herbonnet},
  {Hoekstra}, {K{\"o}hlinger}, {McFarland}, {Mead}, {Merten}, {Napolitano},
  {Peacock}, {Radovich}, {Schneider}, {Simon}, {Valentijn}, {van den Busch},
  {van Uitert}, \& {Van Waerbeke}}]{Hildebrandt17}
{Hildebrandt}, H., {Viola}, M., {Heymans}, C., {et~al.} 2017,
  \href{http://dx.doi.org/10.1093/mnras/stw2805}{\JournalTitle{\mnras}, 465,
  1454}

\bibitem[{{Hirata} {et~al.}(2007){Hirata}, {Mandelbaum}, {Ishak}, {Seljak},
  {Nichol}, {Pimbblet}, {Ross}, \& {Wake}}]{2007MNRAS.381.1197H}
{Hirata}, C.~M., {Mandelbaum}, R., {Ishak}, M., {et~al.} 2007,
  \href{http://dx.doi.org/10.1111/j.1365-2966.2007.12312.x}{\JournalTitle{\mnras},
  381, 1197}

\bibitem[{{Hirata} \& {Seljak}(2004)}]{2004PhRvD..70f3526H}
{Hirata}, C.~M., \& {Seljak}, U. 2004,
  \href{http://dx.doi.org/10.1103/PhysRevD.70.063526}{\JournalTitle{\prd}, 70,
  063526}

\bibitem[{{Ishak}(2007)}]{Ishak2007}
{Ishak}, M. 2007,
  \href{http://dx.doi.org/10.1007/s10701-007-9175-z}{\JournalTitle{Foundations
  of Physics}, 37, 1470}

\bibitem[{{Joachimi} \& {Bridle}(2010)}]{Joachimi10}
{Joachimi}, B., \& {Bridle}, S.~L. 2010,
  \href{http://dx.doi.org/10.1051/0004-6361/200913657}{\JournalTitle{\aap},
  523, A1}

\bibitem[{{Joachimi} {et~al.}(2008){Joachimi}, {Schneider}, \&
  {Eifler}}]{2008A&A...477...43J}
{Joachimi}, B., {Schneider}, P., \& {Eifler}, T. 2008,
  \href{http://dx.doi.org/10.1051/0004-6361:20078400}{\JournalTitle{\aap}, 477,
  43}

\bibitem[{{Joudaki} {et~al.}(2018){Joudaki}, {Blake}, {Johnson}, {Amon},
  {Asgari}, {Choi}, {Erben}, {Glazebrook}, {Harnois-D{\'e}raps}, {Heymans},
  {Hildebrandt}, {Hoekstra}, {Klaes}, {Kuijken}, {Lidman}, {Mead}, {Miller},
  {Parkinson}, {Poole}, {Schneider}, {Viola}, \& {Wolf}}]{Joudaki18}
{Joudaki}, S., {Blake}, C., {Johnson}, A., {et~al.} 2018,
  \href{http://dx.doi.org/10.1093/mnras/stx2820}{\JournalTitle{\mnras}, 474,
  4894}

\bibitem[{{Kamionkowski} \& {Spergel}(1994)}]{1994ApJ...432....7K}
{Kamionkowski}, M., \& {Spergel}, D.~N. 1994,
  \href{http://dx.doi.org/10.1086/174543}{\JournalTitle{\apj}, 432, 7}

\bibitem[{{Kitching} \& {Heavens}(2017)}]{2017PhRvD..95f3522K}
{Kitching}, T.~D., \& {Heavens}, A.~F. 2017,
  \href{http://dx.doi.org/10.1103/PhysRevD.95.063522}{\JournalTitle{\prd}, 95,
  063522}

\bibitem[{{Krause} {et~al.}(2016){Krause}, {Eifler}, \& {Blazek}}]{Krause15}
{Krause}, E., {Eifler}, T., \& {Blazek}, J. 2016,
  \href{http://dx.doi.org/10.1093/mnras/stv2615}{\JournalTitle{\mnras}, 456,
  207}

\bibitem[{{Krause} {et~al.}(2017){Krause}, {Eifler}, {Zuntz}, {Friedrich},
  {Troxel}, {Dodelson}, {Blazek}, {Secco}, {MacCrann}, {Baxter}, {Chang},
  {Chen}, {Crocce}, {DeRose}, {Ferte}, {Kokron}, {Lacasa}, {Miranda}, {Omori},
  {Porredon}, {Rosenfeld}, {Samuroff}, {Wang}, {Wechsler}, {Abbott}, {Abdalla},
  {Allam}, {Annis}, {Bechtol}, {Benoit-Levy}, {Bernstein}, {Brooks}, {Burke},
  {Capozzi}, {Carrasco Kind}, {Carretero}, {D'Andrea}, {da Costa}, {Davis},
  {DePoy}, {Desai}, {Diehl}, {Dietrich}, {Evrard}, {Flaugher}, {Fosalba},
  {Frieman}, {Garcia-Bellido}, {Gaztanaga}, {Giannantonio}, {Gruen}, {Gruendl},
  {Gschwend}, {Gutierrez}, {Honscheid}, {James}, {Jeltema}, {Kuehn},
  {Kuhlmann}, {Lahav}, {Lima}, {Maia}, {March}, {Marshall}, {Martini},
  {Menanteau}, {Miquel}, {Nichol}, {Plazas}, {Romer}, {Rykoff}, {Sanchez},
  {Scarpine}, {Schindler}, {Schubnell}, {Sevilla-Noarbe}, {Smith},
  {Soares-Santos}, {Sobreira}, {Suchyta}, {Swanson}, {Tarle}, {Tucker},
  {Vikram}, {Walker}, \& {Weller}}]{krause17}
{Krause}, E., {Eifler}, T.~F., {Zuntz}, J., {et~al.} 2017, \JournalTitle{ArXiv
  e-prints}, \href{http://arxiv.org/abs/1706.09359}{{\sffamily
  arXiv:1706.09359}}

\bibitem[{{Lattanzi} \& {Gerbino}(2017)}]{Gerbino2017}
{Lattanzi}, M., \& {Gerbino}, M. 2017, \JournalTitle{ArXiv e-prints},
  \href{http://arxiv.org/abs/1712.07109}{{\sffamily arXiv:1712.07109}}

\bibitem[{{Laureijs} {et~al.}(2011){Laureijs}, {Amiaux}, {Arduini},
  {Augu{\`e}res}, {Brinchmann}, {Cole}, {Cropper}, {Dabin}, {Duvet}, {Ealet},
  \& et~al.}]{Laureijs11}
{Laureijs}, R., {Amiaux}, J., {Arduini}, S., {et~al.} 2011, \JournalTitle{ArXiv
  e-prints}, \href{http://arxiv.org/abs/1110.3193}{{\sffamily arXiv:1110.3193
  [astro-ph.CO]}}

\bibitem[{{Lawrence} {et~al.}(2017){Lawrence}, {Heitmann}, {Kwan}, {Upadhye},
  {Bingham}, {Habib}, {Higdon}, {Pope}, {Finkel}, \& {Frontiere}}]{Lawrence17}
{Lawrence}, E., {Heitmann}, K., {Kwan}, J., {et~al.} 2017,
  \href{http://dx.doi.org/10.3847/1538-4357/aa86a9}{\JournalTitle{\apj}, 847,
  50}

\bibitem[{{Lesgourgues} \& {Pastor}(2012{\natexlab{a}})}]{Lesgourgues12}
{Lesgourgues}, J., \& {Pastor}, S. 2012{\natexlab{a}}, \JournalTitle{ArXiv
  e-prints}, \href{http://arxiv.org/abs/1212.6154}{{\sffamily arXiv:1212.6154
  [hep-ph]}}

\bibitem[{{Lesgourgues} \& {Pastor}(2012{\natexlab{b}})}]{Lesgourgues2012}
---. 2012{\natexlab{b}}, \JournalTitle{ArXiv e-prints},
  \href{http://arxiv.org/abs/1212.6154}{{\sffamily arXiv:1212.6154 [hep-ph]}}

\bibitem[{Lewis \& Bridle(2002)}]{camb}
Lewis, A., \& Bridle, S. 2002,
  \href{http://dx.doi.org/10.1103/PhysRevD.66.103511}{\JournalTitle{\prd}, 66,
  103511}

\bibitem[{{Lewis} \& {Bridle}(2002)}]{Lewis02}
{Lewis}, A., \& {Bridle}, S. 2002,
  \href{http://dx.doi.org/10.1103/PhysRevD.66.103511}{\JournalTitle{\prd}, 66,
  103511}

\bibitem[{{Limber}(1954)}]{1954ApJ...119..655L}
{Limber}, D.~N. 1954,
  \href{http://dx.doi.org/10.1086/145870}{\JournalTitle{\apj}, 119, 655}

\bibitem[{{Linder}(2003)}]{Linder03}
{Linder}, E.~V. 2003,
  \href{http://dx.doi.org/10.1103/PhysRevLett.90.091301}{\JournalTitle{Physical
  Review Letters}, 90, 091301}

\bibitem[{{LSST Dark Energy Science Collaboration}(2012)}]{DESCWhite}
{LSST Dark Energy Science Collaboration}. 2012, \JournalTitle{ArXiv e-prints},
  \href{http://arxiv.org/abs/1211.0310}{{\sffamily arXiv:1211.0310
  [astro-ph.CO]}}

\bibitem[{{LSST Science Collaboration} {et~al.}(2009){LSST Science
  Collaboration}, {Abell}, {Allison}, {Anderson}, {Andrew}, {Angel}, {Armus},
  {Arnett}, {Asztalos}, {Axelrod}, \& et~al.}]{LSSTSB}
{LSST Science Collaboration}, {Abell}, P.~A., {Allison}, J., {et~al.} 2009,
  \JournalTitle{ArXiv e-prints},
  \href{http://arxiv.org/abs/0912.0201}{{\sffamily arXiv:0912.0201
  [astro-ph.IM]}}

\bibitem[{{Mamajek} {et~al.}(2015){Mamajek}, {Prsa}, {Torres}, {Harmanec},
  {Asplund}, {Bennett}, {Capitaine}, {Christensen-Dalsgaard}, {Depagne},
  {Folkner}, {Haberreiter}, {Hekker}, {Hilton}, {Kostov}, {Kurtz}, {Laskar},
  {Mason}, {Milone}, {Montgomery}, {Richards}, {Schou}, \& {Stewart}}]{IAU15}
{Mamajek}, E.~E., {Prsa}, A., {Torres}, G., {et~al.} 2015, \JournalTitle{ArXiv
  e-prints}, \href{http://arxiv.org/abs/1510.07674}{{\sffamily arXiv:1510.07674
  [astro-ph.SR]}}

\bibitem[{{Mao} {et~al.}(2018){Mao}, {Zentner}, \& {Wechsler}}]{Mao2018}
{Mao}, Y.-Y., {Zentner}, A.~R., \& {Wechsler}, R.~H. 2018,
  \href{http://dx.doi.org/10.1093/mnras/stx3111}{\JournalTitle{\mnras}, 474,
  5143}

\bibitem[{{McClintock} {et~al.}(2018){McClintock}, {Varga}, {Gruen}, {Rozo},
  {Rykoff}, {Shin}, {Melchior}, {DeRose}, {Seitz}, {Dietrich}, {Sheldon},
  {Zhang}, {von der Linden}, {Jeltema}, {Mantz}, {Romer}, {Allen}, {Becker},
  {Bermeo}, {Bhargava}, {Costanzi}, {Everett}, {Farahi}, {Hamaus}, {Hartley},
  {Hollowood}, {Hoyle}, {Israel}, {Li}, {MacCrann}, {Morris}, {Palmese},
  {Plazas}, {Pollina}, {Rau}, {Simet}, {Soares-Santos}, {Troxel}, {Vergara
  Cervantes}, {Wechsler}, {Zuntz}, {Abbott}, {Abdalla}, {Allam}, {Annis},
  {Avila}, {Bridle}, {Brooks}, {Burke}, {Carnero Rosell}, {Carrasco Kind},
  {Carretero}, {Castander}, {Crocce}, {Cunha}, {D'Andrea}, {da Costa}, {Davis},
  {De Vicente}, {Diehl}, {Doel}, {Drlica-Wagner}, {Evrard}, {Flaugher},
  {Fosalba}, {Frieman}, {Garc{\'{\i}}a-Bellido}, {Gaztanaga}, {Gerdes},
  {Giannantonio}, {Gruendl}, {Gutierrez}, {Honscheid}, {James}, {Kirk},
  {Krause}, {Kuehn}, {Lahav}, {Li}, {Lima}, {March}, {Marshall}, {Menanteau},
  {Miquel}, {Mohr}, {Nord}, {Ogando}, {Roodman}, {Sanchez}, {Scarpine},
  {Schindler}, {Sevilla-Noarbe}, {Smith}, {Smith}, {Sobreira}, {Suchyta},
  {Swanson}, {Tarle}, {Tucker}, {Vikram}, {Walker}, \& {Weller}}]{McClintock18}
{McClintock}, T., {Varga}, T.~N., {Gruen}, D., {et~al.} 2018,
  \JournalTitle{ArXiv e-prints},
  \href{http://arxiv.org/abs/1805.00039}{{\sffamily arXiv:1805.00039}}

\bibitem[{{McEwen} {et~al.}(2016){McEwen}, {Fang}, {Hirata}, \&
  {Blazek}}]{FASTPT}
{McEwen}, J.~E., {Fang}, X., {Hirata}, C.~M., \& {Blazek}, J.~A. 2016,
  \href{http://dx.doi.org/10.1088/1475-7516/2016/09/015}{\JournalTitle{\jcap},
  9, 015}

\bibitem[{{Mead} {et~al.}(2015){Mead}, {Peacock}, {Heymans}, {Joudaki}, \&
  {Heavens}}]{Mead2015}
{Mead}, A.~J., {Peacock}, J.~A., {Heymans}, C., {Joudaki}, S., \& {Heavens},
  A.~F. 2015,
  \href{http://dx.doi.org/10.1093/mnras/stv2036}{\JournalTitle{\mnras}, 454,
  1958}

\bibitem[{{Mo} \& {White}(1996)}]{Mo1996}
{Mo}, H.~J., \& {White}, S.~D.~M. 1996, \JournalTitle{\mnras}, 282, 347

\bibitem[{{Mohammed} \& {Gnedin}(2017)}]{Mohammed17}
{Mohammed}, I., \& {Gnedin}, N.~Y. 2017, \JournalTitle{ArXiv e-prints},
  \href{http://arxiv.org/abs/1707.02332}{{\sffamily arXiv:1707.02332}}

\bibitem[{{Mohammed} \& {Seljak}(2014)}]{Mohammed14}
{Mohammed}, I., \& {Seljak}, U. 2014,
  \href{http://dx.doi.org/10.1093/mnras/stu1972}{\JournalTitle{\mnras}, 445,
  3382}

\bibitem[{{Mohr} {et~al.}(2016){Mohr}, {Newell}, \& {Taylor}}]{CODATA14}
{Mohr}, P.~J., {Newell}, D.~B., \& {Taylor}, B.~N. 2016,
  \href{http://dx.doi.org/10.1103/RevModPhys.88.035009}{\JournalTitle{Reviews
  of Modern Physics}, 88, 035009}

\bibitem[{{Nakamura} \& {Suto}(1997)}]{Nakamura1997}
{Nakamura}, T.~T., \& {Suto}, Y. 1997,
  \href{http://dx.doi.org/10.1143/PTP.97.49}{\JournalTitle{Progress of
  Theoretical Physics}, 97, 49}

\bibitem[{Navarro {et~al.}(1997)Navarro, Frenk, \& White}]{Navarro1997}
Navarro, J.~F., Frenk, C.~S., \& White, S. D.~M. 1997,
  \href{http://dx.doi.org/10.1086/304888}{\JournalTitle{\apj}, 490, 493}

\bibitem[{{Ng} \& {Liu}(1999)}]{Ng1999}
{Ng}, K.-W., \& {Liu}, G.-C. 1999,
  \href{http://dx.doi.org/10.1142/S0218271899000079}{\JournalTitle{International
  Journal of Modern Physics D}, 8, 61}

\bibitem[{{Padmanabhan}(2003)}]{Padmanabhan2003}
{Padmanabhan}, T. 2003,
  \href{http://dx.doi.org/10.1016/S0370-1573(03)00120-0}{\JournalTitle{\physrep},
  380, 235}

\bibitem[{{Paranjape}(2014)}]{Paranjape2014}
{Paranjape}, A. 2014,
  \href{http://dx.doi.org/10.1103/PhysRevD.90.023520}{\JournalTitle{\prd}, 90,
  023520}

\bibitem[{{Parfrey} {et~al.}(2011){Parfrey}, {Hui}, \& {Sheth}}]{Parfrey2011}
{Parfrey}, K., {Hui}, L., \& {Sheth}, R.~K. 2011,
  \href{http://dx.doi.org/10.1103/PhysRevD.83.063511}{\JournalTitle{\prd}, 83,
  063511}

\bibitem[{{Peacock}(1999)}]{PeacockBook}
{Peacock}, J.~A. 1999, {Cosmological Physics}, 704

\bibitem[{{Peacock} \& {Smith}(2000)}]{Peacock2000}
{Peacock}, J.~A., \& {Smith}, R.~E. 2000,
  \href{http://dx.doi.org/10.1046/j.1365-8711.2000.03779.x}{\JournalTitle{\mnras},
  318, 1144}

\bibitem[{{Peebles} \& {Ratra}(2003)}]{Peebles2003}
{Peebles}, P.~J., \& {Ratra}, B. 2003,
  \href{http://dx.doi.org/10.1103/RevModPhys.75.559}{\JournalTitle{Reviews of
  Modern Physics}, 75, 559}

\bibitem[{{Porter} {et~al.}(2011){Porter}, {Johnson}, \& {Graham}}]{Porter11}
{Porter}, T.~A., {Johnson}, R.~P., \& {Graham}, P.~W. 2011,
  \href{http://dx.doi.org/10.1146/annurev-astro-081710-102528}{\JournalTitle{\araa},
  49, 155}

\bibitem[{{Reinecke}(2011)}]{2011A&A...526A.108R}
{Reinecke}, M. 2011,
  \href{http://dx.doi.org/10.1051/0004-6361/201015906}{\JournalTitle{\aap},
  526, A108}

\bibitem[{{Sachs} \& {Wolfe}(1967)}]{1967ApJ...147...73S}
{Sachs}, R.~K., \& {Wolfe}, A.~M. 1967,
  \href{http://dx.doi.org/10.1086/148982}{\JournalTitle{\apj}, 147, 73}

\bibitem[{{Schneider} \& {Teyssier}(2015)}]{Schneider15}
{Schneider}, A., \& {Teyssier}, R. 2015,
  \href{http://dx.doi.org/10.1088/1475-7516/2015/12/049}{\JournalTitle{\jcap},
  12, 049}

\bibitem[{{Schulz} \& {White}(2006)}]{Schulz2006}
{Schulz}, A.~E., \& {White}, M. 2006,
  \href{http://dx.doi.org/10.1016/j.astropartphys.2005.11.007}{\JournalTitle{Astroparticle
  Physics}, 25, 172}

\bibitem[{{Seljak}(2000)}]{Seljak2000}
{Seljak}, U. 2000,
  \href{http://dx.doi.org/10.1046/j.1365-8711.2000.03715.x}{\JournalTitle{\mnras},
  318, 203}

\bibitem[{{Semboloni} {et~al.}(2013){Semboloni}, {Hoekstra}, \&
  {Schaye}}]{Semboloni13}
{Semboloni}, E., {Hoekstra}, H., \& {Schaye}, J. 2013,
  \href{http://dx.doi.org/10.1093/mnras/stt1013}{\JournalTitle{\mnras}, 434,
  148}

\bibitem[{{Semboloni} {et~al.}(2011){Semboloni}, {Hoekstra}, {Schaye}, {van
  Daalen}, \& {McCarthy}}]{Semboloni11}
{Semboloni}, E., {Hoekstra}, H., {Schaye}, J., {van Daalen}, M.~P., \&
  {McCarthy}, I.~G. 2011,
  \href{http://dx.doi.org/10.1111/j.1365-2966.2011.19385.x}{\JournalTitle{\mnras},
  417, 2020}

\bibitem[{Sheth {et~al.}(2001)Sheth, Mo, \& Tormen}]{Sheth2001}
Sheth, R.~K., Mo, H.~J., \& Tormen, G. 2001,
  \href{http://dx.doi.org/10.1046/j.1365-8711.2001.04006.x}{\JournalTitle{\mnras},
  323, 1}

\bibitem[{Sheth \& Tormen(1999)}]{Sheth1999}
Sheth, R.~K., \& Tormen, G. 1999,
  \href{http://dx.doi.org/10.1046/j.1365-8711.1999.02692.x}{\JournalTitle{\mnras},
  308, 119}

\bibitem[{{Silvestri} {et~al.}(2013){Silvestri}, {Pogosian}, \&
  {Buniy}}]{Silvestri2013}
{Silvestri}, A., {Pogosian}, L., \& {Buniy}, R.~V. 2013,
  \href{http://dx.doi.org/10.1103/PhysRevD.87.104015}{\JournalTitle{\prd}, 87,
  104015}

\bibitem[{{Singh} {et~al.}(2015){Singh}, {Mandelbaum}, \& {More}}]{Singh15}
{Singh}, S., {Mandelbaum}, R., \& {More}, S. 2015,
  \href{http://dx.doi.org/10.1093/mnras/stv778}{\JournalTitle{\mnras}, 450,
  2195}

\bibitem[{{Smith} \& {Markovic}(2011)}]{Smith2011}
{Smith}, R.~E., \& {Markovic}, K. 2011,
  \href{http://dx.doi.org/10.1103/PhysRevD.84.063507}{\JournalTitle{\prd}, 84,
  063507}

\bibitem[{{Smith} {et~al.}(2007){Smith}, {Scoccimarro}, \& {Sheth}}]{Smith2007}
{Smith}, R.~E., {Scoccimarro}, R., \& {Sheth}, R.~K. 2007,
  \href{http://dx.doi.org/10.1103/PhysRevD.75.063512}{\JournalTitle{\prd}, 75,
  063512}

\bibitem[{{Smith} {et~al.}(2003){Smith}, {Peacock}, {Jenkins}, {White},
  {Frenk}, {Pearce}, {Thomas}, {Efstathiou}, \& {Couchman}}]{Smith2003}
{Smith}, R.~E., {Peacock}, J.~A., {Jenkins}, A., {et~al.} 2003,
  \href{http://dx.doi.org/10.1046/j.1365-8711.2003.06503.x}{\JournalTitle{\mnras},
  341, 1311}

\bibitem[{{Springel} {et~al.}(2017){Springel}, {Pakmor}, {Pillepich},
  {Weinberger}, {Nelson}, {Hernquist}, {Vogelsberger}, {Genel}, {Torrey},
  {Marinacci}, \& {Naiman}}]{Springel17}
{Springel}, V., {Pakmor}, R., {Pillepich}, A., {et~al.} 2017,
  \JournalTitle{ArXiv e-prints},
  \href{http://arxiv.org/abs/1707.03397}{{\sffamily arXiv:1707.03397}}

\bibitem[{{Sugiyama}(1995)}]{Sugiyama95}
{Sugiyama}, N. 1995,
  \href{http://dx.doi.org/10.1086/192220}{\JournalTitle{\apjs}, 100, 281}

\bibitem[{{Sunayama} {et~al.}(2016){Sunayama}, {Hearin}, {Padmanabhan}, \&
  {Leauthaud}}]{Sunayama2016}
{Sunayama}, T., {Hearin}, A.~P., {Padmanabhan}, N., \& {Leauthaud}, A. 2016,
  \href{http://dx.doi.org/10.1093/mnras/stw332}{\JournalTitle{\mnras}, 458,
  1510}

\bibitem[{{Takahashi} {et~al.}(2012){Takahashi}, {Sato}, {Nishimichi},
  {Taruya}, \& {Oguri}}]{CLASS_halofit}
{Takahashi}, R., {Sato}, M., {Nishimichi}, T., {Taruya}, A., \& {Oguri}, M.
  2012,
  \href{http://dx.doi.org/10.1088/0004-637X/761/2/152}{\JournalTitle{\apj},
  761, 152}

\bibitem[{Talman(2009)}]{Talman2009}
Talman, J. 2009,
  \href{http://dx.doi.org/http://dx.doi.org/10.1016/j.cpc.2008.10.003}{\JournalTitle{Computer
  Physics Communications}, 180, 332 }

\bibitem[{{Tinker} {et~al.}(2008){Tinker}, {Kravtsov}, {Klypin}, {Abazajian},
  {Warren}, {Yepes}, {Gottl{\"o}ber}, \& {Holz}}]{Tinker2008}
{Tinker}, J., {Kravtsov}, A.~V., {Klypin}, A., {et~al.} 2008,
  \href{http://dx.doi.org/10.1086/591439}{\JournalTitle{\apj}, 688, 709}

\bibitem[{{Tinker} {et~al.}(2010){Tinker}, {Robertson}, {Kravtsov}, {Klypin},
  {Warren}, {Yepes}, \& {Gottl{\"o}ber}}]{Tinker2010}
{Tinker}, J.~L., {Robertson}, B.~E., {Kravtsov}, A.~V., {et~al.} 2010,
  \href{http://dx.doi.org/10.1088/0004-637X/724/2/878}{\JournalTitle{\apj},
  724, 878}

\bibitem[{{Troxel} \& {Ishak}(2015)}]{Troxel14}
{Troxel}, M.~A., \& {Ishak}, M. 2015,
  \href{http://dx.doi.org/10.1016/j.physrep.2014.11.001}{\JournalTitle{\physrep},
  558, 1}

\bibitem[{{Upadhye} {et~al.}(2014){Upadhye}, {Biswas}, {Pope}, {Heitmann},
  {Habib}, {Finkel}, \& {Frontiere}}]{Upadhye14}
{Upadhye}, A., {Biswas}, R., {Pope}, A., {et~al.} 2014,
  \href{http://dx.doi.org/10.1103/PhysRevD.89.103515}{\JournalTitle{\prd}, 89,
  103515}

\bibitem[{{van Daalen} {et~al.}(2011){van Daalen}, {Schaye}, {Booth}, \& {Dalla
  Vecchia}}]{vanDaalen11}
{van Daalen}, M.~P., {Schaye}, J., {Booth}, C.~M., \& {Dalla Vecchia}, C. 2011,
  \href{http://dx.doi.org/10.1111/j.1365-2966.2011.18981.x}{\JournalTitle{\mnras},
  415, 3649}

\bibitem[{{van Uitert} {et~al.}(2018){van Uitert}, {Joachimi}, {Joudaki},
  {Amon}, {Heymans}, {K{\"o}hlinger}, {Asgari}, {Blake}, {Choi}, {Erben},
  {Farrow}, {Harnois-D{\'e}raps}, {Hildebrandt}, {Hoekstra}, {Kitching},
  {Klaes}, {Kuijken}, {Merten}, {Miller}, {Nakajima}, {Schneider}, {Valentijn},
  \& {Viola}}]{vanUitert18}
{van Uitert}, E., {Joachimi}, B., {Joudaki}, S., {et~al.} 2018,
  \href{http://dx.doi.org/10.1093/mnras/sty551}{\JournalTitle{\mnras}},
  \href{http://arxiv.org/abs/1706.05004}{{\sffamily arXiv:1706.05004}}

\bibitem[{{Villarreal} {et~al.}(2017){Villarreal}, {Zentner}, {Mao}, {Purcell},
  {van den Bosch}, {Diemer}, {Lange}, {Wang}, \& {Campbell}}]{Villarreal2017}
{Villarreal}, A.~S., {Zentner}, A.~R., {Mao}, Y.-Y., {et~al.} 2017,
  \href{http://dx.doi.org/10.1093/mnras/stx2045}{\JournalTitle{\mnras}, 472,
  1088}

\bibitem[{{Vogelsberger} {et~al.}(2014){Vogelsberger}, {Genel}, {Springel},
  {Torrey}, {Sijacki}, {Xu}, {Snyder}, {Bird}, {Nelson}, \&
  {Hernquist}}]{Illustris}
{Vogelsberger}, M., {Genel}, S., {Springel}, V., {et~al.} 2014,
  \href{http://dx.doi.org/10.1038/nature13316}{\JournalTitle{\nat}, 509, 177}

\bibitem[{{Watson} {et~al.}(2013){Watson}, {Iliev}, {D'Aloisio}, {Knebe},
  {Shapiro}, \& {Yepes}}]{Watson2013}
{Watson}, W.~A., {Iliev}, I.~T., {D'Aloisio}, A., {et~al.} 2013,
  \href{http://dx.doi.org/10.1093/mnras/stt791}{\JournalTitle{\mnras}, 433,
  1230}

\bibitem[{{Weinberg} {et~al.}(2013){Weinberg}, {Mortonson}, {Eisenstein},
  {Hirata}, {Riess}, \& {Rozo}}]{Weinberg13}
{Weinberg}, D.~H., {Mortonson}, M.~J., {Eisenstein}, D.~J., {et~al.} 2013,
  \href{http://dx.doi.org/10.1016/j.physrep.2013.05.001}{\JournalTitle{\physrep},
  530, 87}

\bibitem[{{Wong}(2011)}]{Wong11}
{Wong}, Y.~Y.~Y. 2011,
  \href{http://dx.doi.org/10.1146/annurev-nucl-102010-130252}{\JournalTitle{Annual
  Review of Nuclear and Particle Science}, 61, 69}

\bibitem[{{Yoo}(2010)}]{2010PhRvD..82h3508Y}
{Yoo}, J. 2010,
  \href{http://dx.doi.org/10.1103/PhysRevD.82.083508}{\JournalTitle{\prd}, 82,
  083508}

\bibitem[{{Yoo} {et~al.}(2009){Yoo}, {Fitzpatrick}, \&
  {Zaldarriaga}}]{2009PhRvD..80h3514Y}
{Yoo}, J., {Fitzpatrick}, A.~L., \& {Zaldarriaga}, M. 2009,
  \href{http://dx.doi.org/10.1103/PhysRevD.80.083514}{\JournalTitle{\prd}, 80,
  083514}

\bibitem[{{Zaldarriaga} \& {Seljak}(1997)}]{1997PhRvD..55.1830Z}
{Zaldarriaga}, M., \& {Seljak}, U. 1997,
  \href{http://dx.doi.org/10.1103/PhysRevD.55.1830}{\JournalTitle{\prd}, 55,
  1830}

\bibitem[{{Zhao} {et~al.}(2013){Zhao}, {Saito}, {Percival}, {Ross},
  {Montesano}, {Viel}, {Schneider}, {Manera}, {Miralda-Escud{\'e}},
  {Palanque-Delabrouille}, {Ross}, {Samushia}, {S{\'a}nchez}, {Swanson},
  {Thomas}, {Tojeiro}, {Y{\`e}che}, \& {York}}]{2013MNRAS.436.2038Z}
{Zhao}, G.-B., {Saito}, S., {Percival}, W.~J., {et~al.} 2013,
  \href{http://dx.doi.org/10.1093/mnras/stt1710}{\JournalTitle{\mnras}, 436,
  2038}

\bibitem[{{Zumalac{\'a}rregui} {et~al.}(2017){Zumalac{\'a}rregui}, {Bellini},
  {Sawicki}, {Lesgourgues}, \& {Ferreira}}]{Zumalacarregui17}
{Zumalac{\'a}rregui}, M., {Bellini}, E., {Sawicki}, I., {Lesgourgues}, J., \&
  {Ferreira}, P.~G. 2017,
  \href{http://dx.doi.org/10.1088/1475-7516/2017/08/019}{\JournalTitle{\jcap},
  8, 019}

\bibitem[{{Zuntz} {et~al.}(2014){Zuntz}, {Paterno}, {Jennings}, {Rudd},
  {Manzotti}, {Dodelson}, {Bridle}, {Sehrish}, \& {Kowalkowski}}]{Zuntz14}
{Zuntz}, J., {Paterno}, M., {Jennings}, E., {et~al.} 2014, {CosmoSIS:
  Cosmological parameter estimation}, Astrophysics Source Code Library,
  \href{http://arxiv.org/abs/1409.012}{{\sffamily ascl:1409.012}}

\bibitem[{{Zuntz} {et~al.}(2015){Zuntz}, {Paterno}, {Jennings}, {Rudd},
  {Manzotti}, {Dodelson}, {Bridle}, {Sehrish}, \& {Kowalkowski}}]{Zuntz15}
---. 2015,
  \href{http://dx.doi.org/10.1016/j.ascom.2015.05.005}{\JournalTitle{Astronomy
  and Computing}, 12, 45}

\end{thebibliography}

\end{document}